\documentclass[aps,prd,twocolumn,groupedaddress]{revtex4-1}
\usepackage[french]{babel}
\usepackage[latin1]{inputenc}
\usepackage[T1]{fontenc}
\usepackage{amsmath}
\usepackage{textcomp}
\usepackage{etoolbox}
\usepackage{graphicx,rotating}
 \usepackage{epsfig}
  \usepackage{xcolor}

\definecolor{marron}{rgb}{.53,.29,.1}

\usepackage{graphicx,rotating}
 \usepackage{epsfig}
  \usepackage{xcolor}
\newcommand{\Dsm}{{D}_s^-}

\newcommand{\Bb}{\overline{{B}}}

\newcommand{\Dstarsm}{{D}^{\ast -}_s}

\newcommand{\pim}{\pi^{-}}

\newcommand{\Dstar}{{D}^{\ast}}

\newcommand{\Vcb}{\left | {\rm V}_{cb} \right |}

\newcommand{\sdce}{LLSWBi}
\newcommand{\sdcep}{LLSWB}

\newcommand{\Bob}{\overline{B}^0}

\newcommand{\Dun}{D_{1}}
\newcommand{\Dunp}{{D_1^+}}

\newcommand{\Ddestar}{{D_2^*}}
\newcommand{\Ddestarp}{{D_2^{*+}}}

\newcommand{\Dostar}{{D_0(2300)}}
\newcommand{\Dostarp}{{D_0(2300)^{+}}}

\newcommand{\Dunstar}{{D_1(2430)}}
\newcommand{\Dunstarp}{{D_1(2430)^{+}}}

\newcommand{\mumu}{\ifmmode {\mu^+\mu^-} \else ${\mu^+\mu^-} $ \fi}

\newcommand{\ba}{\begin{array}}
\newcommand{\ea}{\end{array}}
\newcommand{\bc}{\begin{center}}
\newcommand{\ec}{\end{center}}
\newcommand{\beq}{\begin{eqnarray}}
\newcommand{\eeq}{\end{eqnarray}}
\newcommand{\bes}{\begin{eqnarray*}}
\newcommand{\ees}{\end{eqnarray*}}
\newcommand{\Kz}{\ifmmode {\rm K^0_s} \else ${\rm K^0_s} $ \fi}
\newcommand{\Zz}{\ifmmode {\rm Z} \else ${\rm Z } $ \fi}
\newcommand{\qqbar}{\ifmmode {\rm q\bar{q}} \else ${\rm q\bar{q}} $ \fi}
\newcommand{\ccbar}{\ifmmode {\rm c\bar{c}} \else ${\rm c\bar{c}} $ \fi}
\newcommand{\bbbar}{\ifmmode {\rm b\bar{b}} \else ${\rm b\bar{b}} $ \fi}
\newcommand{\xxbar}{\ifmmode {\rm x\bar{x}} \else ${\rm x\bar{x}} $ \fi}
\newcommand{\rphi}{\ifmmode {\rm R\phi} \else ${\rm R\phi} $ \fi}
\makeatletter
\patchcmd{\maketitle}{\@fnsymbol}{\@alph}{}{}  % Footnote numbers from symbols to small letters
\makeatother
\begin{document}
%\preprint{}
%\preprint{}
%\preprint{}

\title{A model for  NL and SL decays by $\Bob \to D^{**}$ transitions with $BR(j=1/2) \ll BR(j=3/2)$ using the $LLSW$ scheme}

\author{Alain Le Yaouanc}\affiliation{Universit\'e Paris-Saclay, CNRS/IN2P3, IJCLab, 91405 Orsay, France}%\footnote{alain.le-yaouanc@ijclab.in2p3.fr}  
\author{Jean-Pierre Leroy}\affiliation{Universit\'e Paris-Saclay, CNRS/IN2P3, IJCLab, 91405 Orsay, France}%\footnote{jean-pierre.leroy@ijclab.in2p3.fr}

\author{Patrick Roudeau}\affiliation{Universit\'e Paris-Saclay, CNRS/IN2P3, IJCLab, 91405 Orsay, France}%\footnote{patrick.roudeau@ijclab.in2p3.fr}\footnote{Universit\'e Paris-Saclay, CNRS/IN2P3, IJCLab, 91405 Orsay, France}}
\vspace{0.3cm}
\date{\today}
%\maketitle

\begin{abstract}

We present a model for the vector and axial form factors of the transitions 
$\Bob \to D^{**}$ in good agreement with the presently available data and based on the present theoretical knowledge, combining a) the safe lattice QCD predictions at $m_Q=\infty$ and $w=1$ ; 
b) the predictions at general $w$ of a relativistic, covariant quark model at $m_Q=\infty$, including the well tested Godfrey and Isgur spectroscopic model and which agrees with lattice QCD at $w=1$ ; c) the constraint of Bjorken and Neubert relating Semi-leptonic (SL) and Class I  Non-leptonic (NL) decays, which shows that 
$\Bob \to \Dostarp \pi^-$  strongly constrains $\tau_{1/2}(w)$ to be much smaller than $\tau_{3/2}(w)$, in agreement with the theoretical expectation; d) the general HQET expansion which constrains the $1/m_Q$ corrections (cf \cite{LLSW}, denoted  hereafter as $LLSW$). 

An important element in the understanding of data is the large contribution of virtual $D^{(*)}_V$ to the broad structures seen in SL decays at low $D^{(*)} \pi$ masses - which makes it difficult to isolate  the broad resonances  denoted as $D_{1/2}$ in the following.
\end{abstract}

%\vskip 1cm
\maketitle

\section{Introduction}There is insufficient knowledge of both non-leptonic (NL) and semi-leptonic  (SL) transitions to charmed orbital excitations, generically termed as  $D^{**}$. The latter (SL) are especially interesting 

1) in themselves, 
\vskip 0.5cm
for a) striking theoretically expected features which contradict the naive idea of strong similarity between SL transitions to $j=1/2$ and $j=3/2$ $D^{**}$,
\vskip 0.5cm
and b) for the apparent contradiction between this expectation and the presently widespread interpretation of experimental data on broad states, a contradiction which we claim to have resolved by the interplay of the  possibly large $D^*_V$ background \cite{roudeau1}.
\vskip 0.5cm
2) for some applications, like the background to $\overline{B} \to D^{(*)} \ell^-(\tau^-) \overline{\nu}_{\ell(\tau)}$. Ultimately, one would like to have a description of the full system of semi-leptonic $\overline{B} \to D^{**}$ form factors, but this appears a hard task.

Indeed, there are no full calculations of the latter from first principles, as are provided in simpler cases by lattice QCD. Presently, the latter gives results only at $m_Q=\infty$ and $w=1$. The reasons are given below, Section \ref{theory}. 

On the other hand, the experimental data are scarce : total branching ratios, some points in the $d\Gamma/dw$ of certain transitions... Then, as explained 
in more details below, Sections \ref{theory} and \ref{sec:d32d12}, a more complicated path is to be followed, combining theoretical inputs and experience, the former being used in addition in several different ways.

We want indeed to use the very useful general ideas of the extensive analyses of \cite{ref:leibo, ref:ligeti1, ref:ligeti2} \footnote{From now on we shall refer to this set of analyses as $\sdcep$. When comparing such type of analyses with our model we will use the acronym $\sdce$ where the ``i'' means ``inspire'', implying slight modifications to $\sdcep$ required for a fair comparison with our own model, and which are explained in Appendix \protect\ref{app:second}}
relying on their HQET analysis, but to take  into account\\1) {\bf quantitative dynamical results at $m_Q=\infty$} of lattice QCD  (completed by quark models in the  Bakamjian-Thomas (BT) approach) 
 as well as \\2)  the {\bf very important experimental measurements of  $\Bob \to \Dostarp  \pi^-$} which constrain strongly the transition to $j=1/2$ to be small, 
{\color{black}  both  of them having been disregarded {in the $\sdcep$ analyses}}.

Finally we underline the very important role of the $D^*_V$ background in the SL decays, which we have already emphasized in \cite{roudeau1}, and which, we think,  has not been fully taken into account up to now. 

All these elements, taken together, change drastically the conclusions with respect to the ones of the above mentioned analyses, 
 especially the claim that the Isgur-Wise (IW) functions are roughly equal at $w=1$, as we explain now.
%This background, if not removed, almost completely obliterates the wide and flat $j=1/2$ states. 

The $LLSW$ approach  \cite{ref:leibo},  provides a framework to parameterize $1/m_Q$ corrections in corresponding hadronic form factors, and applies factorization to relate semi-leptonic and Class I non-leptonic decays. Such analyses  \cite{ref:ligeti1,ref:ligeti2} which take, as input, quoted values \cite{ref:hfag2016,ref:pdg2019} %[HFAG, PDG]
 for $\Bob \to D^{**,+} \ell^- \overline{\nu}_{\ell}$  conclude that production of narrow and broad $D^{**}$ states are similar. Meanwhile, to reach this conclusion, they have to discard the measurement of $\Bob \to \Dostarp \pi^-$ which implies, as a consequence of factorization, a very small value for the production of the $\Dostar$ meson in semi-leptonic decays, when compared with the rates measured for narrow states.  This  could seem to be justified because such a low value, evaluated  for $\Bob \to \Dostarp \ell^- \overline{\nu}_{\ell}$, appears to be in contradiction with the rates measured in experiments for this channel. However, on the opposite, we think that the non-leptonic data for the $D_{1/2}$ are much more trustable than the semi-leptonic ones. Indeed, the identification of the $\Dostar$ in the non-leptonic channel is supported by 
1) the extraction of the $D_V^*$  and 2) the measure of the phase shift, while no such work has been done in the semi-leptonic case (except, for the  $D_V^*$, the work  reported in \cite{ref:belle_dsstarl}). Moreover, from a theoretical point of view, the  smallness of the transitions to the $D_{1/2}$ states with respect to the  $D_{3/2}$ ones
 has been anticipated using, initially, quark models and, later on, by direct LQCD evaluations.

We have considered again this problem and propose a model, based also on the $LLSW$ parameterization and {which uses all the measured non-leptonic Class I decays} in the framework of factorization. This model agrees with LQCD and relativistic quark models (RQM) expectations. Because, in this model, production of broad $D^{**}$ mesons is expected to be much smaller than the production of narrow states, it is necessary to add another broad component to be able to explain the broad mass distributions measured in $\Bob \to D^{(*)}\pi \ell^- \overline{\nu}_{\ell}$ decays. We find that {$D^{(*)}_V$} decays can fill such a gap. In the following we detail our model and  provide comparisons with the $\sdce$  model in which the broad $D^{(*)}\pi$ mass distributions are explained by the contributions from $D^{**}$ decays alone, as done in previous analyses and where the measurement from  $\Bob \to \Dostarp \pi^-$ is not used.

Finally one is led to a solution with a $j=1/2$ rate much smaller than the $j = 3/2$ one.

\vspace{2cm}
\section{Theoretical inputs}\label{theory}

\vskip 1cm 
We speak of "inputs" because one lacks a systematical theoretical treatment : rather, one is led to use a mixture of procedures, including the very experimental data which are to be explained, and fitting.
%\vskip 0.5cm

In addition, several ingredients a)b)c)d), as enumerated in the abstract, are available.

\vskip 0.5cm
We can classify them also into two categories :
%\patrick{ propose d'ajouter: which are explained in the following sub-sections}:

\vskip 1cm
\subsection{Dynamical results at $m_Q=\infty$} 
\
In the infinite quark mass limit, hadronic form factors that describe 
$\overline{B} \to D^{**}$, are determined by the two Isgur-Wise functions: $\tau_{3/2}(w)$ and  $\tau_{1/2}(w)$. Their values at $w=1$ and their $w$ dependence are constrained by theory:

\vskip 0.5cm
a) lattice QCD for $w=1$ at  $N_F=0, 2$ \cite{ref:lqcd-tau}; at  $N_F= 2$ : 

\begin{equation}
\tau_{3/2} = 0.526 \pm 0.023,\,\tau_{1/2}= 0.296\pm0.026\label{eq:latticesBenoit}
%\quad\quad \mbox{\green{ METTRE LES REFERENCES}}
\end{equation}
showing  a striking difference between them, in contrast with a naive  non-relativistic (NR) expectation according to which these two quantities are equal.

These are trustable results  which
{\bf cannot be disregarded}. In  the most recent simulation the lattice spacing is  reasonably small, $a=0.085 fm$, and the volume reasonably large: $24^3 \times 48$, although certain systematic errors are not estimated. Of course, they should be improved.  One notes that, at $N_F = 2$, the inequality between $j = 1/2$ and $ j = 3/2$  values is appreciably  reinforced  with respect to the older $N_F =0$ ones.%\footnote{ A first evaluation of finite quark mass effects can be found in \cite{Bailas}:
%$\tau_{3/2}(1)=0.45 \pm 0.07$ and $\tau_{1/2}(1)=0.39 \pm 0.06$. Quoted uncertainties are such that one can consider that the two functions have similar values or that they are compatible with the results quoted in Eq. (\ref{eq:latticesBenoit}).}
%\vskip 0.5cm

b) quark models, which, although purely phenomenological, have the advantage of providing results for $w \neq 1$ as well. 

Of course, there is a very large variety of quark models.  
We consider a  class of models for current matrix elements where the calculation is decomposed into  two steps : 1) the determination of the wave functions at rest and, 2) a procedure to derive from it the state in motion. Then there is  first  a variety  of spectroscopic  models, describing the spectrum and the wave functions at rest, i.e internal quark motion ; secondly, another variety comes from the way one describes the hadron motion.

{\it As to spectroscopic models} Among many, there is an outstanding one by Godfrey and Isgur (GI) \cite{godfreyj}, which we therefore decide to use.
This model is unique in its covering of a very large number of hadronic states, both with light quarks and with heavy ones. One must underline that most predictions have been confirmed by later experiments. Although it is a complicated model with many parameters, these can be determined thanks to this covering of a very extended spectrum. This model has  a relativistic kinetic energy and its success confirms the necessity of a relativistic treatment of quark internal motion inside hadrons, which is already implied by the fact that excitation energies are of the order of the reduced mass.

{\it As to the description of hadron motion} Here we shall use a specific framework, the one of Bakamjian and Thomas, to describe states in relativistic motion at $m_Q=\infty$. It has several advantages: 
\begin{itemize}
\item[$\alpha$)] it uses the standard three-dimensional wave functions at rest provided by spectroscopic models ;
\item[$\beta$)]it is relativistic as to hadronic motion and even covariant; 
\item[$\gamma$)] it satisfies the standard set of HQET sum rules, like Bjorken \cite{Taron} and curvature \cite{curvature} sum rules, Uraltsev sum rule \cite{Uraltsev}... 
\end{itemize}

The last two points  are important advantages as compared to  non relativistic treatment of quark motion, even if we were adopting the GI spectroscopic model.
Note   that  a NR treatment of hadron motion is not satisfactory for the full range of $w$ since the 3-velocity at $w_{max}$ is large even in the "Equal Velocity Frame" which minimizes velocities  : $v=\sqrt{(w-1)/2} \simeq 0.4$.

Moreover these differences lead to quite different quantitative results. In the NR quark models with non-relativistic treatment of both internal quark velocities and hadron motion,  a general statement \footnote{ In Ref.\cite{IsgurWise}, the calculation is done for   harmonic oscillator wave functions, but this does not alter the generality of the conclusion. Note also that a ``relativisation'' factor $\kappa$ has to be disregarded in their Eq. (44). Possible additional factors expressed as powers of $(w+1)/2$ encountered in the literature  must also be disregarded anyway since  such factors give  a  1/2 contribution  to  the slope  while the dominant NR approximation gives a slope that goes $ \to \infty $ as $m^2 R^2$ (see  Eq. (44) of \cite{IsgurWise})} would be  :
\begin{eqnarray} \label{NR}
\tau_{1/2}(w) = \tau_{3/2}(w) 
\end{eqnarray}
in clear contradiction with the above lattice QCD results. The equality (\ref{NR}) assumes that there is no sizable spin-orbit force surviving at $m_Q = \infty$, which is not a theorem but seems reasonable from the spectroscopic models studies, whence identical rest  frame wave functions.

In the BT framework, using the well known spectroscopic model of Godfrey and Isgur \cite{godfreyj}, one finds very different values for $j = 1/2$ and $j = 3/2$ \cite{ref:morenas}:
\begin{eqnarray}
\tau_{3/2}(1) \simeq 0.5,\tau_{1/2}(1)\simeq 0.25  \label{BTm}
\end{eqnarray}
in full agreement with QCD. This strikingly large difference is a general feature of the BT approach, which provides an intuitive explanation: it is due to the typically relativistic effect of Wigner rotations of spin, included in the BT approach, which acts differently on $j=3/2$ and $j=1/2$ states in motion and which is completely independent of the presence of possible spin-dependent forces in the potential. 
Note that, more generally, the inequality of the $\tau(1)$'s is suggested by the Uraltsev sum rule        :
\begin{eqnarray}
\Sigma_n \left( |\tau^{(n)}_{3/2}(1)|^2 -|\tau^{(n)}_{1/2}(1)|^2\right)=1/4
\end{eqnarray}
which contradicts the equality (\ref{NR}) under the assumption of dominance of $n=0$.

In summary, one should consider the marked difference between $\tau_{3/2}(1)$ and $\tau_{1/2}(1)$ as a solidly settled result (see the extensive discussions by Bigi et al.\cite{bigi_et_al}, see also \cite{DBecirevic} and the work on  zero-recoil sum rules of \cite{BigShif} and \cite{Gambino}. In the latter reference the point of view of the authors on this matter is expressed at the end of subsection 6.3.2). Combined with the appropriate kinematical factors, and converted into  branching ratios, this leads to the still more striking inequality :
\begin{eqnarray} \label{BRll}
{\cal B}(\Bob \to D_{1/2}) \ll {\cal B}(\Bob \to D_{3/2})
\end{eqnarray}
by around one order of magnitude, and {\bf we claim to provide a model which both satisfies this strong inequality  and  fits well the data}.

The quantitative agreement of the BT prediction with the lattice QCD one gives encouragement to trust the predicted shapes at $w \neq 1$, which are another crucial theoretical input.

The full shape is well approximated by a relativistic quark-model inspired
description \cite{ref:bt_model}:%\red{R\'ef\'erence bizarre !!!}:

%\patrick{ notation simplifiee dans la formule}
\begin{eqnarray}\label{les_taus}
\tau_{3/2\,(1/2)}(w) =\tau_{3/2\,(1/2)}(1) \times\left({{2} \over {w+1}} \right)^{2\, \sigma^2_{3/2\,(1/2)}}
\end{eqnarray}

%\brun{One can summarize the predictions by means of  the slopes at $w=1$\\
%$(\tau_{3/2}^\prime(1), \tau^\prime_{1/2}(1))\,=(-\sigma^2_{3/2} \tau_{3/2}(1),-\sigma^2_{1/2} \tau_{3/2}(1))$}

\vspace{1mm}
Numerically :
\begin{eqnarray}
\sigma^2_{3/2} \simeq 1.5 ,\quad\quad \sigma^2_{1/2} \simeq 0.8
\end{eqnarray}

These values are in contrast with what would be given in a fully NR treatment namely  a common and much lower value, of order 0.4 (obtained from \cite{IsgurWise} and skipping the relativisation factor $\kappa$.)

Other analyses use instead a linear approximation for the $\tau$-functions : $\tau_i(w) = \tau_i(1)\times \left ( 1 - \sigma_{i,lin.}^2 \, (w-1)\right )$. It can be checked that  $\sigma_{i,lin.}^2 = \sigma_{i}^2$ when $w=1$. In practice the two descriptions are rather similar because the $w$ variation range, between 1 and $w_{max}=(m_B^2+m_{D^{**}}^2)/(2\,m_B\,m_{D^{**}})$ is limited anyway. Meanwhile, significant differences are expected, from the two parameterizations, when comparing semi-leptonic decays with a light
or an heavy lepton, as evaluated in Table \ref{tab:brsl_dsstar_el}.

\vskip 0.5cm

{\it Why start from $m_Q=\infty$?} 

The visible difficulty is that one has presently trustable quantitative statements only for $m_Q=\infty$. This derives from different reasons :
\vskip 0.5cm
 1) as to lattice QCD, it is the fact that it is very difficult to treat directly the finite masses $m_b, m_c,$  all the more for transitions to excitations. Thus one has been led presently to use a $m_Q=\infty$ framework. Then, to treat finite $w \neq 1$, one would require infinite momenta, whence one is restricted to  $w=1$.
\vskip 0.5cm
 2) quark models have in general no such limitations. However the BT~framework which we want to use is satisfactory only at $m_Q=\infty$. Indeed, it has been shown that the finite mass corrections cannot be trusted : they break covariance as well as certain important sum rules, in contrast with the $m_Q=\infty$ limit. One has not been able to cure this defect.
\vskip 0.5cm
The consequence is that these dynamical results must be complemented by other statements, of a more general nature, which provide constraints on the physical transitions (of course at finite mass) or, equivalently, on finite mass corrections to the above dynamical results.  Those questions are considered in points c) and d) below.

\subsection{Use of general statements or  relations}
\label{genrel}

\vskip 0.2cm
c) First of all, we rely on the validity of the factorization, as does
in principle $\sdcep$. This phenomenon has been firmly established from the theoretical point of view (specially in the thorough demonstration and discussions by BBNS \cite{BBNS}). 
Moreover the order of magnitude of the departure from asymptotic BBNS factorization ($a_1 = 1+{\cal O}(10 \%))$ as observed from the phenomenological
side, is such that it does not seem possible to doubt about this validity for the present decays. Of course, BBNS gives only an asymptotic statement, and a departure from this limit is quite expected (presence of various $1/m_Q$ corrections, $m_c$ not very large, etc...). Factorization implies the well known Bjorken-Neubert relation \cite{ref:bjorken} between the NL and SL decays at some $q^2$ (for example $m_\pi^2$ for decay by pion emission), which is a very strong and useful constraint for the otherwise ill known SL differential $q^2$ distributions.   

Practically, since we have no theoretical quantitative estimates of the departure of 
 $a_1$ from the asymptotic BBNS result, we will apply factorisation  in the following way.  We fit  $a_{1,\,eff.}$   in $\overline{B} \to D_{3/2}$ decays and find  a value compatible with those obtained by analysing $\overline{B} \to D^{(*)}$ transitions (see Sec. \ref{sec:d32}). Then, in our final analysis, we impose a constraint in the fits  :  $a_{1,\,eff.} = 0.93  
\pm 0.07$.    This $a_{1,\,eff.}$ will also be used for cases where experimental data are lacking  or not precise enough. This procedure is, in principle, valid for  Class I decays with emission of a light meson. But we shall extend it to predict $NL$ processes  where a charmed strange meson is emitted like $\overline{B} \to D^{**} D_s^{(*)}$; although in such a case the asymptotic theorem of BBNS does not apply, it is a fact that, once more,  one obtains  a value of $a_{1,\,eff.}$ close to 1 from the measured processes.

\vskip 0.2cm
d) HQET and $1/m_Q$ expansion,
\vskip 0.2cm

Let us comment  about d). The constraints from HQET expansion as performed in \cite{ref:leibo} and further work, including order ${\cal O}(1/m_Q)$  corrections, are  crucial too as a necessary complement to the $m_Q=\infty$ predictions.
 
 \vskip 0.2cm
 {\it On the determination of the ${\cal O}(1/m_Q)$ subdominant functions.}
 \vskip 0.2cm
 
 One must stress that the constraints d) do not lead to quantitative statements  on the finite mass corrections; they  rather yield a parameterization of these  finite corrections. They leave us with a large number of  unknown parameters and functions of 
 $w$.
 Some of those parameters, like $\bar{\Lambda}$, (quoted as Set 1 in the following) can be estimated otherwise, as explained in Section \ref{sec:HQET_param}. One could think of fixing the remaining unknown ones (Set 2) by fits to the experimental data.
But there remains at least  a host of unknown functions of $w$. For instance, for $j=3/2$, one needs in principle the ten functions~: $\tau_1(w), \tau_2(w),\eta^{b,c}_{ke}(w) ,\eta^{b,c}_{1,2,3}(w)$ in $LLSW$ notation\footnote{beware not to confuse the subindices to $\tau$'s (1,2) with the previous subindices $j =1/2, 3/2$ concerning the dominant Isgur-Wise functions.}, which is of course too much to be fitted at present.  This number is reduced thanks to the fact that the $\eta_i^b$'s and $\eta^b_{ke}$ are present only in one combination, $\eta^b = \eta^b_{ke}+ 6 \eta_1^b -2 \eta_2^b(w-1)   + \eta_3^b$ but the problem remains.

Many further assumptions must then be added :

\vskip 0.2cm
 
e) an additional but reasonable assumption of rough proportionality
of the latter functions to the dominant Isgur-Wise functions reduces them to numerical parameters, denoted with "hats", 
 but one remains however with    a host of unknown numbers.  
 For $D_{3/2}$ mesons, present data are sensitive to the values of $\hat{\eta}_{1,\,3}$ and $\hat{\tau}_{1}$, whose values we have determined, while some guesses have to be used to define a possible range of variation for the other two quantities:
$\hat{\eta}_{2}$ and $\hat{\tau}_{2}$. For $D_{1/2}$ mesons, data are much less accurate and fits have the same sensitivity to $\hat{\chi}_{1,\,2}$ or $\hat{\zeta}_{1}$ and one can obtain the value for only one of these quantities. We choose,
{arbitrarily}, to fit $\hat{\chi}_{1}$.

f) of course, one would like, however, to find some reasons for that selection, or to find consistency checks for the physical soundness of the values of the parameters found in this way.
 
-$LLSW$  \cite{LLSW} proposed an additional expansion in $w-1$ (the latter too being  small, of magnitude at most $\simeq 0.3 \simeq \epsilon_c=1/2 m_c $) in the narrow width approximation ($w_{max} \simeq 1.3, m_c \simeq1.5$ $GeV$). Were this valid, it would allow to skip $\tau_{1,\,2}$ terms as being of higher order and to be left with the $\eta$'s. However, of course, this supposes that the $\tau_{1,\,2}$ are not large,  a point for which there is no guarantee (in fact one often retains the reverse, i.e. dropping the $\eta$'s in favor of  the $\tau_{1,\,2}$).
%\patrick{Je propose d ajouter: 
In the present analysis we have not used any 
expansion in $w-1$ and we take the full $1/m_Q$ expansions of the Lorentz invariant form factors given in \cite{ref:leibo}. The validity of this framework requires that unknown quantities that enter in the expansion are of order $\overline{\Lambda}$. This is what we {do} verify for fitted quantities, therefore we have assumed that quantities, not determined by the fit, are also of that order to evaluate their contribution to systematic uncertainties.

-inspiration from quark models can help to check the soundness of fit results, at least qualitatively. As explained above, one cannot trust the BT approach for finite mass corrections, and this is the main obstacle to get physical results in this quark model approach. Then one may return to the NR model of center of mass motion, but only to get a qualitative understanding. An example is the $\eta$ functions, which correspond in naive terms to the modification of the wave functions induced  by the change of $m_Q$ from $\infty$ to its real value in the  Schr\"odinger equation.

Knowing theoretically the "true" (infinite mass) $\tau_{3/2}(w=1)$, the magnitude  of the corrections at $w = 1$  which appear in the combination   $\frac{\eta^b }{2 m_b}+ \frac{\eta^c_{ke}}{2 m_c}$ (corresponding to our  $\hat{\epsilon}_{3/2}$ below ), can be rather well determined with a stable value in the various fits to the data, and it is found to be neatly {\bf negative}.

A fully NR calculation of the effect of ${\cal O}_{ke}(v=(1,0,0,0))$ (naively interpreted as the effect of the change of the kinetic energy on the wave functions in the rest frame) suggests a negative $\eta^c_{ke}(w=1)$ corresponding to the effect on the final state.  $\eta^b_{ke}$  on the contrary should be positive, but the combination $\eta^b$ corresponds to the total effect on the  state  vectors, including ${\cal O}_{mag}(v=(1,0,0,0))$, and, naively interpreted, includes  the large spin-spin force present at finite mass. In the GI model, it is found to be neatly negative by numerical calculation.

Then it is encouraging to find consistency between the theoretically expected sign and the finding of the fits. One could hope to estimate similarly the signs and order of magnitude of the remaining $\eta^c_i$'s, but this requires interpreting  them separately in the quark model (see Appendix {B}).

g) on the whole, at present, there is no theoretical estimate for the values of the different parameters that enter in the $LLSW$ expansion, except, perhaps, a qualitative one of the $\eta$'s if one follows the arguments above (see also Appendix B).

%\green{   A Ecrire ou \`a supprimer  }             {\it A NR calculation of $\eta^c_{kin}$}

%{\color{blue} Introduire les corrections HQET, d\'esign\'ees dans la suite par ``Set 1'' . Proposition \'a modifier.}

{\subsection{Evaluation of HQET parameters}
\label{sec:HQET_param}
In the framework of HQET, masses of charm and beauty mesons are used to evaluate values of the parameters (named Set 1 in the following) that enter in some of the $1/m_Q$ corrections \cite{ref:leibo}, the relation being:

%Masses of heavy mesons are expressed in terms of the parameters that enter in HQET:

\begin{equation}
m_{H_{\pm}}\, = \, m_Q\, + \overline{\Lambda}^H \, -\frac{\lambda_1^H}{2 \, m_Q} \, \pm \frac{n_{\mp}\lambda_2^H}{2\, m_Q}\, +\, ...
\label{eq:massh}
\end{equation}

The total spin ($J_{\pm}$) of the resonance ($H_{\pm}$) is expressed in terms of
the total spin ($s_l$) of the light hadronic system as: $J_{\pm}=s_l \pm 1/2$. while $n_{\pm}=2\,J_{\pm}+1$ is the number of spin states.  Values of these different quantities and those of meson masses, adopted in our analysis, are indicated in Table \ref{tab:heavym}. We have not considered I-spin averaged masses and used only charged $(c\bar{d})$ states for charm and neutral states for beauty $(b\bar{d})$.

\begin{table}[!htb]
\begin{center}
  \begin{tabular}{|c|c|c|c|c|c|}
    \hline
 meson & $J^P$ & $s_l^{\pi}$ & $n_{\pm}$ &  charm mass  &  beauty mass  \\
\hline
 $D^+/\Bob$ & $0^- $ & $1/2^-$ & $n_-=1$ & $1869.65 \pm 0.05$ & $5279.65 \pm 0.12$\\
 $D^{*\,+}/\bar{B}^{*\, 0}$ & $1^- $ & $1/2^-$ & $n+=3$ & $2010.26 \pm 0.05$ & $5324.70 \pm 0.21$\\
\hline
 $D_1^+/ B_1^0$ & $1^+ $ & $3/2^+$ & $n_-=3$ & $2423.8 \pm 1.1$ & $5726.1 \pm 1.4$\\
 $D_2^{*\,+}/ B_2^{*\,0}$ & $1^+ $ & $3/2^+$ & $n_+=5$ & $2465.4 \pm 1.3$ & $5739.5 \pm 0.7$\\
\hline
 $\Dostarp$ & $0^+ $ & $1/2^+$ & $n_-=1$ & $2330 \pm 20$ & $(?)$\\
 $\Dunstarp$ & $1^+ $ & $1/2^+$ & $n_+=3$ & $2452 \pm (?)$ & $(?)$\\
\hline
\hline
  \end{tabular}
  \caption[]{\it {The three doublets of heavy mesons having, respectively, a total spin and parity of their light component, $s_l^{\pi},$ equal to $1/2^-$, $3/2^+$ and $1/2^+$. The masses are given in $MeV$.}
  \label{tab:heavym}}
\end{center}
\end{table}

We use, for $\overline{\Lambda}^H$, the notations 
$\overline{\Lambda}$, $\overline{\Lambda}_{3/2}$ and $\overline{\Lambda}_{1/2}$
for the $s_l^{\pi}$ doublets $1/2^-$, $3/2^+$ and $1/2^+$ respectively. 

Considering only the first order expansion in $1/m_Q$ of Eq. (\ref{eq:massh}), the
ratio ($r_Q$) between charm and beauty quark masses is equal to:
\begin{equation}
r_Q=\frac{m_c}{m_b} = \frac{m_{B^*}-m_B}{m_{D^*}-m_D} = 0.3215 \pm 0.0015.
\end{equation}

The difference between the $\overline{\Lambda}^H$ values for the three doublets
can be expressed in terms of the ratio between heavy quark masses and of the
values of spin averaged masses ($\overline{m}$) \cite{ref:leibo}.
 We have obtained:
\begin{equation} 
\overline{\Lambda}_{3/2} - \overline{\Lambda}= (396.0 \pm 1.5 )\, MeV.
\end{equation}
For the $1/2^+$ doublet the corresponding estimate is more uncertain because the associated $B$-meson states are still un-measured. From the masses of charm states we estimate: $\overline{\Lambda}_{1/2} - \overline{\Lambda} \sim 366 \, MeV$.

These values for $\overline{\Lambda}_{3/2} - \overline{\Lambda}$ and
$\overline{\Lambda}_{1/2} - \overline{\Lambda}$ agree with previous determinations.

To evaluate absolute values for heavy quark masses and $\overline{\Lambda}$ one needs a value for $\lambda_1$. In previous analyses, the value $\lambda_1=-0.2\,GeV^2$ was used; in a recent report from the HFLAV collaboration \cite{ref:hfag2016}, the value $\lambda_1=(-0.362\pm0.067)\,GeV^2$ was obtained from an analysis of $B$-meson semi-leptonic decays. In Table \ref{tab:constants} a summary is given for the
values of the different parameters entering in the analysis ({first}  line, with $\lambda_1 =-0.362\,(GeV)^2$).

\begin{table}[!htb]
\begin{center}
  \begin{tabular}{|c|c|c|c|c|c|}
    \hline
 $\lambda_1$& $\overline{\Lambda}$ & $\overline{\Lambda}_{3/2}$ & $\overline{\Lambda}_{1/2}$ &  $m_c$  & $m_b$ \\
\hline
 % $-0.2$ & $0.310$ & $0.706$ & $0.676$ & $1.602$ & $4.983$\\
  $ -0.362 $ & $ 0.245 $ & $ 0.641 $ & $ 0.611 $ & $ 1.618 $ & $ 5.032 $\\
\hline
 $-0.2$ & $0.4$ & $0.8$ & $0.76$ & $1.4$ & $4.8$\\
\hline
  \end{tabular}
  \caption[]{\it {Values of HQET parameters used in the evaluation of $1/m_Q$ corrections ({first line}). The last line gives the values used in previous analyses. The various quantities are expressed in $GeV$, except for  $\lambda_1\,(GeV^2)$}
  \label{tab:constants}}
\end{center}
\end{table}

}
%\patrick{Je propose de mettre un tableau resumant ce que l on fait pour \'evaluer les param\`etres:}

\subsection{Evaluation of parameters entering in the $LLSW$ parameterization} \label{sec:llsw_parameters}

 In the infinite quark mass limit, hadronic form factors that describe $\overline{B}\to D^{**}$, are determined by the two Isgur-Wise functions: $\tau_{3/2}(w)$ and $\tau_{1/2}(w)$ respectively, which have been introduced in Eq. (\ref{les_taus}).
It has been noted \cite{ref:leibo} that one can define useful effective functions including finite mass corrections to replace the IW functions: namely, particularising at $w=1$, one writes:
%\patrick{IW functions considered at the infinite and finite mass quark values are related by the following expressions:}
\begin{eqnarray}
\tau_{3/2}^{eff.}(1) &=& \tau_{3/2}(1) + \frac{\eta_{ke}(1)}{2\,m_c} + \frac{\eta_{b}(1)}{2\,m_b} = \tau_{3/2}(1) \times | 1 + \hat{\epsilon}_{3/2}|  \nonumber\\  & &\phantom{\tau_{1/2}^{eff.}(1) \tau_{1/2}(1) + \frac{\chi_{ke}(1)}{2\,m_c} + \frac{\chi_{b}(1)}{2\,m_b} }\\
\tau_{1/2}^{eff.}(1) &=& \tau_{1/2}(1) + \frac{\chi_{ke}(1)}{2\,m_c} + \frac{\chi_{b}(1)}{2\,m_b} = \tau_{1/2}(1) \times | 1 + \hat{\epsilon}_{1/2}|  \nonumber\label{eq:normtau}
\end{eqnarray}

In Table \ref{tab:llsw_parameters} are reminded the parameters that enter
in the expressions of Lorentz-invariant form factors for the $LLSW$ parameterization of $\overline{B} \to D^{**}$ transitions.

\begin{center}\begin{table}[!htb]
 \begin{tabular}{|c|c|c|c|}
    \hline
  & parameters  & evaluation &constraints \\
 & list & & from theory\\\hline
  Set 1& $m_{b,\,c},\,\overline{\Lambda},$ & using HQET &see Section \ref{sec:HQET_param}\\
 & $\overline{\Lambda}_{3/2},\,\overline{\Lambda}_{1/2}$ &  &\\
\hline
  & $\tau_{3/2}(1)$ & fitted & $0.53 \pm 0.03$ \\
   Set 2 for                   & $\hat{\epsilon}_{3/2}$ &fitted & \\
  $D_{3/2}$                                 & $\sigma_{3/2}^2$ &fitted & $ 1.5 \pm 0.5$ \\
    mesons                    & $\hat{\eta}_{1,\,3},\,\hat{\tau}_{1}$ & fitted & \\
                        & $\hat{\eta}_{2},\,\hat{\tau}_{2}$ & set to zero& $\pm 0.5\,GeV$ for syst. \\
\hline
  & $\tau_{1/2}^{eff.}(1)$ & fitted & $0.20 \pm 0.06$\\
  Set 2 for                            & $\sigma_{1/2}^2$ & fitted  &$\sigma_{3/2}^2-\sigma_{1/2}^2=0.7\pm0.5$\\
  $D_{1/2}$                      & $\hat{\chi}_{1}$ & fitted & \\
      mesons                  & $\hat{\chi}_{2},\,\hat{\zeta}_{1}$ & set to zero & $\pm 0.5\,GeV$ for syst. \\
\hline
  \end{tabular}
  \caption[]{\it {List of the parameters that enter in $LLSW$ formalism. The two last columns indicate how values of these quantities are obtained in the present analysis. When no information is available on a parameter, it is set to zero and a range of $\pm 0.5\,GeV \sim \pm \overline{\Lambda}$ is used to evaluate a plausible contribution to systematic uncertainties.}
  \label{tab:llsw_parameters}}
\end{table}\end{center}

\section{$\Bob \to D^{**}$ experimental results used as constraints}
Input data, in this analysis, are obtained by averaging branching fraction measurements of non-leptonic Class I, $\Bob \to D^{**,\,+} \pi^- (K^-)$, and semi-leptonic decay $\overline{B} \to D^{**}\ell^- \overline{\nu}_{\ell}$ channels.  These last values, obtained separately for charged and neutral $B$-mesons, are combined, assuming the equality of corresponding partial decay widths for charged and neutral $B$-mesons,  and  {averaged values} are expressed in terms of the $\Bob$.  We have taken into account possible 
correlations between the different uncertainties, corrected intermediate branching fractions \cite{ref:pdg2019}, and used the hypotheses explained in Section
\ref{sec:ddec} to evaluate $D^{**}$  absolute decay branching fractions. Values obtained in this way are compared with those used in a previous analysis \cite{ref:ligeti1}
in Table \ref{tab:constraint}. The first lines are relative to the production of narrow ($D_{3/2}$) states, then they correspond to broad ($D_{1/2}$) states and the last line reports a ratio between the production of narrow and broad states.
Therefore, apart for the last constraint, it is possible to investigate the production of narrow and broad states, independently. 
\begin{widetext}
\begin{center}\begin{table}[!htb]%\begin{center}
  \begin{tabular}{|c|c|c|c|}
    \hline
 decay channel & this analysis & analysis \cite{ref:ligeti1} &PDG(2020)\\
  & &  (2016)&or HFAG $(^3)$\\
\hline
${\cal B}(\Bob \to D_2^{*+} \pim)\times 10^{4}$& $5.85 \pm 0.42$& $5.9 \pm 1.3$&$5.85 \pm 0.43$\\
${\cal B}(\Bob \to D_2^{*+} K^-)\times 10^{5}$& $4.7 \pm 0.8$& not used&$5.0 \pm 0.9$ \\
${\cal B}(\Bob \to D_2^{*+} \ell^- \bar{\nu}_{\ell})\times 10^{3}$ &$3.09 \pm 0.32$& $2.8 \pm 0.4$&$3.18 \pm 0.26$\\
\hline
${\cal B}(\Bob \to D_1^{+} \pim)\times 10^{4}$& $7.12 \pm 1.13$& $7.5 \pm 1.6$&$6.6 \pm 2.0$\\
${\cal B}(\Bob \to D_1^{+} \ell^- \bar{\nu}_{\ell})\times 10^{3}$ &$6.40 \pm 0.44$&$6.2 \pm 0.5$&$6.24 \pm 0.54$\\
\hline \hline
${\cal B}(\Bob \to D_0(2300)^{+} \pim)\times 10^{4}$& $1.19 \pm 0.12$& not used&$1.14 \pm 0.12$\\
${\cal B}(\Bob \to D_0(2300)^{+} \ell^- \bar{\nu}_{\ell})\times 10^{3}$ &$2.2 \pm 1.2$&$4.1 \pm 0.7$&$3.9 \pm 0.7\,(^4)$\\
\hline
${\cal B}(\Bob \to D_1(2430)^{+} \pim)\times 10^{4}$& $0.21 \pm 0.27\,(^1)$& not used&not quoted\\
${\cal B}(\Bob \to D_1(2430)^{+} \ell^- \bar{\nu}_{\ell})\times 10^{3}$ &$1.4 \pm 1.3$&$1.9 \pm 0.5$&$1.8 \pm 1.5\,(^4)$\\
\hline \hline
{\tiny ${\cal B}(\Bob \to D_0(2300)^{+} K^-)/{\cal B}(\Bob \to D_2^{*+} K^-)$}$\,(^2)$& $0.84 \pm 0.36$& not used&$$\\
\hline
  \end{tabular}
  \caption[]{\it {Measured branching fractions used as constraints in the present analysis. In the two last columns are indicated the values used in a previous analysis and those quoted in PDG \cite{ref:pdg2019} or obtained by HFAG \cite{ref:hfag2016}. As compared with our numbers, these values differ mainly in the estimate of the production of broad $D^{**}$ states in $B$-meson semi-leptonic decays. Analysis \cite{ref:ligeti1} differs also by the fact that non-leptonic measurements of $D_{1/2}$ states are not included. $(^1)$: we have not used the measured value for ${\cal B}(\Bob \to D_1(2430)^{+} \pim)$ \cite{ref:belle_d12430pi} as constraint in our nominal fit because this is still a preliminary result and we have instead compared this value to our expectation. $(^2)$: $D^{**,\,+}$ mesons are reconstructed in the $D^0 \pi^+$ final state. $(^3)$: values from PDG and HFAG have been modified using results quoted in Table \ref{tab:br_abs}. $(^4)$: values from HFAG \cite{ref:hfag2016}, the uncertainty on ${\cal B}(\Bob \to D_1(2430)^{+} \ell^- \bar{\nu}_{\ell})$ is multiplied by a factor three to account for the poor compatibility between the three measurements used to obtain the average value.
  \label{tab:constraint}}}%\end{center}
\end{table}\end{center}
\end{widetext}

Production of $D_{1/2}$ mesons in semi-leptonic decays, reported in the two last columns of Table \ref{tab:constraint}, are derived from HFAG \cite{ref:hfag2016}
whereas those used in our analysis are obtained using a different approach, as explained in Section \ref{sec:subtrdv}.  As illustrated from the values given in Table \ref{tab:constraint} and from the results of the fits quoted in Tables \ref{tab:brsl_dsstar_el} and \ref{tab:brsl_dsstar_el_sc2}, estimates for $\Dunstar$
are not accurate, with about 100$\%$ uncertainty, once taking into account the fact that existing measurements are rather incompatible, with $\chi^2/NDF=18/2$, and if we scale the uncertainty, obtained on the average value, using the "PDG recipe" that corresponds to a factor three. Such a scaling factor is not included in the  \cite{ref:ligeti1} analysis and the present value from PDG ($\sim (4 \pm 1)\times 10^{-3}$), not quoted in Table \ref{tab:constraint},  is based on the measurement from BaBar alone, not including the other two results that enter in the HFAG average, and which are rather different, in particular the one from Belle which does not see a signal and quotes a stringent limit.

In addition, the Belle collaboration \cite{ref:belle_dsstarl}
has measured the semi-leptonic decay branching fraction $\Bob \to D_2^{*+} \ell^- \bar{\nu}_{\ell}$ in four bins of the variable $w=v_D \cdot v_B=(m_B^2+m^2_{D_2^*}-q^2)/2\,m_B\,m_{D_2^*};\,q^2=m^2(\ell^- \bar{\nu}_{\ell})$. The sum of these
fractions is normalized to unity therefore, in the following, because we do not have the full error matrix on these measurements available, we use the first three results and assume that they are independent.

\begin{table}[!htb]
\begin{center}
  \begin{tabular}{|c|c|}
    \hline
 $w$ bin & fraction $(\%)$\\
\hline
 $[1.00,\,1.08]$ & $6.0 \pm 2.3 $\\
 $[1.08,\,1.16]$ & $30.0 \pm 5.4 $\\
 $[1.16,\,1.24]$ & $37.5 \pm 6.2 $\\
 $[1.24,\,1.32]$ & $26.5 \pm 6.2 $\\
\hline
  \end{tabular}
  \caption[]{\it {Measured fractions of the $\Bob \to D_2^{*+} \ell^- \bar{\nu}_{\ell}$ decay width in several $w$ bins.}
  \label{tab:fracsl}}
\end{center}
\end{table}

Values quoted in the third column of Table \ref{tab:constraint} are used in Appendix C to demonstrate that our code is able to reproduce the values obtained in \cite{ref:ligeti1} when using similar input values
and hypotheses. Those
given in the last column allow to {perform} a comparison between experimental values retained in our analysis  and those quoted in {"official"  compilations.} 
%considered analyses.
%used obtain expectations from a model (quoted as No-$D_V^{(*)}$), in which the factorization hypothesis is not used in case of $D_{1/2}$ production.}

\subsection{Evaluation of absolute $D^{**,\,+}$ decay branching fractions}
\label{sec:ddec}
At present, absolute $D^{**}$ decay branching fractions are obtained using
hypotheses on the contribution of the various decay channels. For the $\Dun$, it is assumed that it decays only into $D^* \pi$ and $D \pi \pi$ (through $\Dostar \pi$). The $\Ddestar$ is expected to decay only into $D \pi$ and $D^* \pi$. For broad states, we consider that the $\Dostar$ and the $\Dunstar$ decay only into
$D\pi$ and $D^* \pi$, respectively. Expected branching fractions are given in Table \ref{tab:br_abs}; they are similar to those used in \cite{ref:ligeti1}.

\begin{table}[!htb]
\begin{center}
{
  \begin{tabular}{|c|c|}
\hline
${\cal B}(\Ddestar \to D \pi)$ & $0.61 \pm 0.02$\\
\hline
${\cal B}(\Dunp \to D^{*0} \pi^+)$ & $0.45 \pm 0.03$\\
${\cal B}(\Dunp \to D^{+} \pi^+\pi^-)$ & $0.15 \pm 0.02$\\
\hline
${\cal B}(\Dunstar \to D^* \pi)$ & $1$\\
\hline
${\cal B}(\Dostar \to D \pi)$ & $1$\\
\hline
  \end{tabular}}
  \caption[]{\it {Expected absolute $D^{**}$ mesons decay branching fractions.}
  \label{tab:br_abs}}
\end{center}
\end{table}

\subsection{Estimates for ${\cal B}(\Bob \to D_{1/2}^{+} \ell^- \bar{\nu}_{\ell})$}
\label{sec:subtrdv}

Values for $\Dostar$ and $\Dunstar$ production in $\overline{B}$ hadron semi-leptonic decays, quoted in PDG or in HFAG, are obtained by fitting expected $D^{(*)}\pi$ mass distributions on measured $\overline{B} \to D^{(*)}\pi \ell^- \bar{\nu}_{\ell}$ events. This approach is reliable for $D_{3/2}$ mesons which appear as relatively narrow mass peaks. The determination of broad $D_{1/2}$ production is more difficult  because one expects also contributions from $D_V^{(*)}\to D^{(*)}\pi$ and it is not clear how these components have been included in the analyses. For these reasons we have adopted another approach.

To evaluate  ${\cal B}(\Bob \to D_{1/2}^{+} \ell^- \bar{\nu}_{\ell})$ we use ${\cal B}(\bar{B} \to D^{(*)}\pi \ell^- \bar{\nu}_{\ell})$ exclusive measurements from BaBar \cite{ref:babardpi} and Belle \cite{ref:belledpi}, from which we subtract the expected contributions from $D_{3/2}$ and $D^{(*)}_V$ decays.

It has to be noted that, at present, in all these  evaluations, contributions from higher mass resonances, which can decay into $D^{(*)}\pi$, are not specifically evaluated and therefore are fully or partially included in the rates estimated
for $D_{1/2}$ mesons, depending on the approach.

$D^{(*)}_V$ components are normalized in an absolute way using values for   ${\cal B}(\Bob \to D^{(*)+} \ell^- \bar{\nu}_{\ell})$ in which $D^{(*)}$ are on-mass shell, albeit with rather large uncertainties related to their mass dependence \cite{roudeau1}.  

In the $D\pi$ channel, only the $D^*_V$ component contributes. It is expected to decrease with the $D\pi$ mass value. For the $D^{*0}_V$ channel, there is a natural threshold in the decay to $D^+\pi^-$ because $m_{D^{*0}}<m_{D^+}+m_{\pi^-}$. For other charge combinations, the separation between so called $D^*$  and $D^*_V$ components is arbitrary, and therefore the absolute value of the $D^*_V$ component depends on the considered threshold. In Belle, they use the following mass range $m_{D^{(*)}\pi}\in [2.05,\,3]\,GeV/c^2$ whereas, in BaBar they require $m_{D^0\pi^+}-m_{D^0}>0.18\,GeV/c^2$. These cuts are rather similar, being typically $40\,MeV/c^2$ above the nominal $D^*$mass.  

To evaluate the uncertainty on these estimates we assume that the $D^{(*)}_V$ follows  a relativistic Breit-Wigner mass distribution, modified by a Blatt-Weisskopf damping term with parameter $r_{BW}=1.85\,(GeV/c)^{-1}$ which is varied in the range $[1.0,\,3.0]\,(GeV/c)^{-1}$.

\subsubsection{${\cal B}(\Bob \to \Dostarp \ell^- \bar{\nu}_{\ell})$}
\label{sec:dv_dostar}

 Averaging experimental measurements one obtains: 
${\cal B}(\Bob \to D \pi \ell^- \bar{\nu}_{\ell})=(6.14 \pm 0.53)\times 10^{-3}$.
We estimate the $D^*_V \to D\pi$ contribution, with $m_{D\pi}> 2.05\,GeV$, to be
equal to: $(2.0 \pm 0.6)\times 10^{-3}$, in which the uncertainty corresponds to the variation range used for $r_{BW}$.
The $\Ddestar$ contribution is obtained from Tables \ref{tab:constraint}  and \ref{tab:br_abs}; it amounts to $(1.90 \pm 0.18)\times 10^{-3}$.

This gives:
\begin{equation}
{\cal B}(\Bob \to \Dostarp \ell^- \bar{\nu}_{\ell}) = (2.2 \pm 1.2)\times 10^{-3}
\end{equation}
as reported, after rounding,  in Table \ref{tab:constraint}. The uncertainty on the $D^*_V$ estimate is added linearly to the total uncertainty evaluated for the other sources.

\subsubsection{${\cal B}(\Bob \to \Dunstarp \ell^- \bar{\nu}_{\ell})$}
 \label{sec:dv_d1star}

We use the same approach for the $D^* \pi$ hadronic final state. 
Averaging experimental measurements, one obtains: 
${\cal B}(\Bob \to D^{*} \pi \ell^- \bar{\nu}_{\ell})=(8.39 \pm 0.54)\times 10^{-3}$.
We estimate the $D^{*+}_V$ contribution, using the coupling $g_{D^*D^*\pi}= \sqrt{2\frac{m_{D^*}}{m_D}}g_{D^*D\pi} $ which corresponds to a fictitious $\Gamma(D^* \to D^* \pi)=2\, \Gamma(\Dstar\to D\pi ) $, and  we have added the expected (small) contribution
from $D_V \to D^* \pi$. This gives: $(1.4 \pm 0.6) \times 10^{-3}$. $D_V$ and $D^*_V$ contributions are added incoherently because of helicity conservation and of the vanishing of $A-V$ interferences for the zero-helicity after the  angular integration.

Subtracting the $\Dunp$ ($(4.32 \pm 0.41) \times 10^{-3}$)and $\Ddestarp$ ($(1.21 \pm 0.14) \times 10^{-3}$) contributions,  obtained from Tables \ref{tab:constraint}  and \ref{tab:br_abs}, this gives:
\begin{equation}
{\cal B}(\Bob \to\Dunstarp \ell^- \bar{\nu}_{\ell}) = (1.4 \pm 1.3)\times 10^{-3}
\end{equation}
as reported in Table \ref{tab:constraint}.  The uncertainty on the $D^{(*)}_V$ estimate is added linearly.

\section{Parameterization of semi-leptonic and non-leptonic decay widths}

%{\color{orange} Section modifi\'ee selon les indications d Alain. Cependant les modifs propos\'ees pour la partie sur les corrections en $1/m_Q$ ont \'et\'e d\'eplac\'ees dans la section 3.4. Ce n est sans doute toujours pas satisfaisant.}
We indicate here the expressions we use  to compute  semi-leptonic and non-leptonic {Class I} decay branching fractions  and explain how we have taken into account effects from contributions of virtual states.
\subsection{Semi-leptonic transitions to real, discrete charmed states}
Differential semi-leptonic $\overline{B} \to D_X \ell^- \bar{\nu}_{\ell}$ decay widths, with $D_X = D^{(*,\,**)}$, can be written as \cite{ref:ligeti1}:
\begin{eqnarray}
\frac{d\Gamma}{dq^2}&=& C |\vec{p}| q^2 \left ( 1-\frac{m_{\ell}^2}{q^2}\right )^2\nonumber\\
& \times & \left [ (H_+^2 + H_-^2 + H_0^2) \left (  1+\frac{m_{\ell}^2}{q^2}\right )+ \frac{3 m_{\ell}^2}{2 q^2}H_t^2 \right ] \label{eq:dgdq2heli}
\end{eqnarray}
with $C=G_F^2 \Vcb^2 \eta_{EW}^2/(96\,\pi^3\,m_B^2)$. $\vec{p}$ is the  three-momentum of the $D_X$ in the $B$ rest frame.
\begin{equation}
|\vec{p}| = m_{D_X} \, \sqrt{w_{D_X}^2-1}
\end{equation}
in which $w_{D_X}=v_B \cdot v_{D_{X}}$ is the product of the 4-velocities of the two mesons.

$H_{\pm,0,t}$ are helicity form factors which are expressed in terms of, $q^2$ dependent, Lorentz invariant form factors ($FF(q^2)$) and depend on the considered meson $D_X$.  
Accurate parameterizations of $FF(q^2)$  are obtained for $D$ (we use \cite{ref:ff_dlnu}) and $D^*$
mesons (we use \cite{ref:cln,ref:fajfer}). For $D^{**}$ mesons, expressions are taken from \cite{ref:leibo}. They correspond to expansions at first order in $1/m_{c,b}$ and $\alpha_s.$ 
%(see Section \ref{sec:oneoverm}).
For $D_{3/2}$ mesons, expressions that contain $1/m_{c,b} \times \alpha_s$ corrections are also available and have been included.

\subsection{Semi-leptonic transitions to virtual charmed states}
Let us now consider the physical case where the charmed state terminates on a two-body continuum like $D^{(*)} \pi$. As a useful intermediate step, one now considers fictitious weak transitions to intermediate virtual $D^{**}$ or $D^{(*)}$, which represent the weak vertex part of the overall process, with the charmed leg bearing a momentum squared $p^2=m^2\left (=m^2_{D^{(*)}\pi}\right )$, different from the nominal squared mass of the state $m^2_{D_X}$ (of course, in the overall process, leading for instance to a $D\pi$ final state, one has to introduce a $D^*$ propagator relating the weak vertex to the strong vertex which couples the virtual state to $D\pi$). If one considers production of virtual $D^{**}$ mesons (that will get a Breit-Wigner distribution for $m$) or of {virtual $D^{(*)},$  noted usually $D^{(*)}_V,$ mesons,    with $m$ higher than their nominal mass value $m_{D^{(*)}},$ the} invariant form factors at the weak vertex are expected to be also dependent on $m \neq m_{D_X}$.

However, in the absence of a theoretical knowledge of this dependence we assume simply:
\begin{equation}
FF(m,q^2) = FF(m_{D_X},q^2)
\end{equation}
 in the expressions of helicity form factors, while keeping $m \neq m_{D_X}$ dependence
of the kinetic quantities, such as momenta, in the additional factors entering in  these expressions. This procedure is illustrated in the following example of the production of a virtual $D_V$ meson of mass $m\neq m_D$. There are two invariant form factors noted : $f_{+,0}(m,q^2)$. Helicity form factors are expressed as:
\begin{eqnarray}
q \, H_0 & = & 2 \, m_B\, |\vec{p}| \,f_+(m_D,q^2)\nonumber\\
q\, H_t & = & (m_B^2-m^2)\, f_0(m_D,q^2)\label{eq:hd}.
\end{eqnarray}
The two other form factors, $H_{\pm}$,  vanish because of helicity conservation. In these expressions and in Eq. (\ref{eq:dgdq2heli}), the decay
momentum is evaluated at the virtual mass $m$: $|\vec{p}| = m \, \sqrt{w^2-1}$ with $w=(m_B^2+m^2-q^2)/(2\,m\,m_B)$.
Expressions relating helicity and invariant form factors when a $D^*$ meson is emitted can be found in \cite{ref:fajfer} and in \cite{ref:ligeti1} in case of $D^{**}$ mesons. {\color{black}Since these formula are devised for real, discrete charmed states, one must modify the factors affecting the form factors when considering virtual charmed states.}

\subsection{Non-leptonic decays}
Thanks to factorization, non-leptonic Class  I decay widths
$\Gamma(\Bob \to D_X^+ P^-)$ are related to corresponding differential semi-leptonic decay widths, through the $H_t^{D_X}$ helicity form factor:

\begin{equation}\begin{split}
\Gamma&(\Bob \to D_X^+ P^-)\\&= \frac{\left | a_{1,eff.}^{D_X\pi} \right |^2 f_P^2\,G_F^2\,|V_{ij}|^2 |V_{cb}|^2|\vec{p}| }{16 \pi\,m_B^2}  \left . q^2 (H_t^{D_X})^2 \right |_{w_P}.
\label{eq:nltosl}
\end{split}\end{equation}

In this expression, $q^2=m_P^2$, $f_P$ is the leptonic decay constant of the  emitted charged meson  $P$, and $V_{ij}$ is the corresponding CKM matrix element. 
\begin{equation}
w_P=\frac{m_B^2+m_{D_X}^2-m_P^2}{2 \,m_B\,m_{D_X} }.
\end{equation}
If the virtual ``mass'' $m$ of the $D_X$ meson doesn't have the nominal value $m_{D_X}$,  we will still use values of invariant form factors evaluated at $q^2=m_P^2$ as above while taking the running mass to compute the other terms that enter in $H_t^{D_X}$ and in $\vec{p}$.

We term the factor $a_{1,\,eff.}^{D_X\pi}$  "effective" because the non-leptonic decay width, that enters in Eq. (\ref{eq:nltosl}), corresponds to the sum of the Class I diagram amplitude and of subdominant terms which correspond to exchange or penguin mechanisms, {while  the remaining factors in the right hand are those provided by the analytic expression for a pure Class I process}.

Expressions for $\Bob \to D^{**,\,+} P^-$ partial decay widths, with
$P\,=\,\pi^-$ or $\Dsm$,  and for $\Bob \to D^{**,\,+} D_s^{*-}$  are given in Appendix \ref{app:nl}.
Corresponding quantities, obtained in the infinite quark mass limit, for non-leptonic - $\Gamma(\overline{B}^0 \to D^{**} \pi^-)$ - and for the differential semi-leptonic - $d\Gamma(\overline{B}^0 \to D^{**,\,+} \ell^- \bar{\nu}_{\ell})/dw$ -  can be found, for example, in \cite{ref:jugeau}.

\subsection{Finite width effects}

To include effects from the mass distribution of resonances we express the differential decay width for a process in which a $D^{**}$ is reconstructed in a given final state ($i$) as \cite{roudeau1}:
\begin{equation}
\frac{d\Gamma_i}{ds} = \frac{1}{\pi} \frac{\Gamma_0(s) \,\sqrt{s}\, \Gamma_{i,\,D^{**}}(s)}{(s-m_{D^{**}}^2)^2+s\, \Gamma^2_{D^{**}}(s)}
\label{eq:breit}.
\end{equation}
$\Gamma_0(s)$ is the decay width for the process computed in the hypothesis of a virtual $D^{**}$ of mass equal to $\sqrt{s}$. $\Gamma_{i,\,D^{**}}(s)$ is the partial decay width for the $D^{**}$, of mass $\sqrt{s}$, reconstructed in the $i$ observed channel. $\Gamma_{D^{**}}(s)$ is the total $D^{**}$ decay width at the mass
$\sqrt{s}$.

When the current mass ($\sqrt{s}$) is higher than the threshold for the $i$ decay channel:
{\small \begin{equation}
\Gamma_{i,\,D^{**}}(s) = \Gamma_{i,\,D^{**}}(m_{D^{**}}^2)\left(\frac{p_i}{p_i^0} \right)^{2\,L+1} \left(\frac{m_{D^{**}}}{\sqrt{s}} \right)^2 \left(\frac {F_{i,\,L}(p_i)}{F_{i,\,L}(p_i^0)}\right)^2.
\end{equation}}
$p_i$ and $p_i^0$ are the break-up momenta for the $D^{**}$ decaying into the $i$-channel at masses equal to $\sqrt{s}$ and $m_{D^{**}}$, respectively.  $F_{i,\,L}(p_i)$ is a Blatt-Weisskopf damping factor.
For decay channels with a threshold above the resonance mass, $p_i^0$ becomes imaginary and $F_i(p_i^0)$ cannot be evaluated. In these conditions  we take:
\begin{equation}
\Gamma_{i,\,D^{**}}(s) = \frac{g_i^2}{24\,\pi}\frac{p_i^{2L+1}}{s} F_i^2(p_i)
\end{equation}
in which $g_i$ is the coupling of the $D^{**}$ to the $i$-channel. %\patrick{\'a supprimer:, can be usually evaluated.} 
This expression is used, in the following, to evaluate contributions from virtual $D$ or $D^*$ mesons to $D^{(*)}\pi$ final states.

%\patrick{Expression qui suit modifi\'ee)}
$\Gamma_{D^{**}}(s)$ is the sum of all partial decay widths $\Gamma_{i,\,D^{**}}(s)$, opened at the mass $\sqrt{s}$.

In non-leptonic decays, the expression in Eq. (\ref{eq:breit}) is multiplied by another damping factor (noted usually $(F_{B,\,i}(p^{\prime}_i)/F_{B,\,i}(p^{\prime,\,0}_i))^2$) to account for strong interaction effects due to the hadron emitted with the $D^{**}$. $p^{\prime}_i$ is the momentum of the $B$ decay products, evaluated in the $B$ rest-frame \footnote{In some analyses, $p^{\prime}$ is evaluated in the resonance rest-frame. We consider that our choice is more physical. Effects of changing the convention to compute $p^{\prime}$ are given in Table \ref{tab:brsl_dsstar_el}.}.

\section{Constraints used in our analysis}
\label{sec:constraint-th}

The measurements used in our analysis are listed in Table \ref{tab:constraint}, second column.

As to the theoretical constraints, there are three of them (see Table \ref{tab:llsw_parameters}):

1) factorization. The validity of the factorization  is checked, using $D_{3/2}$ events (see Section \ref{sec:d32}), and then it is used as a constraint in the final fits with
$a_{1,\,eff.}^{D^{**}\pi}=0.93 \pm 0.07$ for all $D^{**}$ states.

{2) at $w=1$ we use the constraint : $ \tau_{3/2}(1) = 0.53 \pm 0.03$ as expected from LQCD \cite{ref:lqcd-tau}. Because $ \tau_{3/2}$ is defined in the infinite quark mass limit it is necessary to introduce mass corrections and one parameter characterizing them $\hat{\epsilon}_{3/2}$, (see Eq. (\ref{eq:normtau})) which is a useful combination of the basic ones introduced by $LLSW$. $\hat{\epsilon}_{3/2}$ is fitted, as well as  a part of  the other parameters.  The data on $D_{1/2}$ mesons are less accurate and it is not  possible to fit $\hat{\epsilon}_{1/2}$.

3) at $w \ne 1$, LQCD does not provide any information on the variation with respect to $w$ of the two IW functions  for which we use quark models,  namely the BT calculations explained in the second section. Unfortunately, quark models cannot provide errors. Therefore, to use them in a fit, we have considered  a rather large range of values of the slopes around the predicted one.

\subsection{Comparison with the analysis of \cite{ref:ligeti2}.}
\label{sec:ligeti_comp}

To validate our code, we check that, using the same input data (given in the third column of Table \ref{tab:constraint})
and the same hypotheses, we reproduce the results published in  \cite{ref:ligeti2}(see Appendix \ref{app:bl}).

\section{Production of $D_{3/2}$ mesons : a check of factorization}
\label{sec:d32}

In a first step, the analysis is restricted to the production of $D_{3/2}$ mesons to quantify the importance of $1/m_Q$ corrections and to check the {applicability }%validity
 of the factorization property. We have required that $\tau_{3/2}(1)=0.53 \pm 0.03$ and fitted the $\hat{\epsilon}_{3/2}$ correction. This is essentially equivalent to fitting directly $\tau^{eff.}_{3/2}$. No constraint is used on  $a_{1,\,eff.}^{D_{3/2}\pi}$ and $\sigma_{3/2}^2$.

Data are first analyzed without any $1/m_Q$ correction. The fit probability is below $10^{-13}$. This is  mainly due to the fact that, in the infinite quark mass limit, theory predicts a production higher for the $D_2^*$ as compared to the $D_1$ whereas it is measured two times lower.

Adding the Set 1 of $1/m_Q$ corrections improves the situation but
the fit probability is still below $10^{-4}$. 

Therefore it is necessary to fit additional parameters which control the other $1/m_Q$ corrections.  Fitting successively one among all other parameters,
we end up with the results given in Table \ref{tab:lq-3corr}. %
 All fit probabilities are higher than $10\%$.
 
\begin{widetext}
{ \begin{center}\begin{table}[!htb]
{
  \begin{tabular}{|c|c|c|c|c|c|}
    \hline
 X param. & $a_{1,\,eff.}^{D_{3/2}\pi}$ & $\sigma_{3/2}^2$ & $\hat{\epsilon}_{3/2}$  & X $(GeV)$ & $\chi^2/NDF$  \\
\hline
 $\hat{\eta}_1$ & $0.90 \pm 0.05$  & $-0.4\pm 0.7$& $-0.50 \pm 0.09$& $-0.40 \pm 0.11$ & $3.4/4$ \\
 $\hat{\eta}_3$ & $0.82 \pm 0.05$  & $-2.1 \pm 0.7$ & $-0.77 \pm 0.07$& $3.2\pm 1.1$ & $7.0/4$ \\
 $\hat{\tau}_1$ & $0.81 \pm 0.07$ & $1.0 \pm 0.7$ & $-0.34 \pm 0.12$& $0.75\pm 0.24$ & $6.4/4$\\
 $\hat{\tau}_2$ & $0.89 \pm 0.06$ & $2.1 \pm 0.5$ & $-0.19 \pm 0.11$& $2.9\pm 0.8$ & $2.5/4$ \\
 $\hat{\eta}_2$ & $0.86 \pm 0.06$  & $1.2 \pm 0.6$ & $-0.30 \pm 0.11$& $-1.63\pm 0.44$ & $3.8/4$\\
\hline
  \end{tabular}}
  \caption[]{\it {Results obtained using the constraint from theory on  $\tau_{3/2}(1)$. Set 1 $1/m_Q$ corrections are used and one additional parameter from Set 2 is fitted, in addition to $\hat{\epsilon}_{3/2}$.} %
  \label{tab:lq-3corr}}
\end{table}\end{center}}\end{widetext}

It can be noted that, depending on the choice of the additional fitted parameter, values of the slope ($\sigma_{3/2}^2$) and of the correction ($\hat{\epsilon}_{3/2}$ ) to the normalization of the IW function fluctuate. This comes from the fact that fitted individual parameters are also changing the $w$ dependence of form factors and there are not enough measurements of the differential decay branching fractions versus $w$ to constrain these variations. It results that fitted values for the IW slope can be highly correlated with the value fitted for some of the additional parameters. In the following we therefore use the constraint expected from QM :  $\sigma_{3/2}^2 = 1.5 \pm 0.5$ so that the $w$-dependence of the IW function  verifies expectations from theory while we have no constraints on the precise values of all parameters entering in $1/m_Q$ corrections. One has to check, a posteriori, that such quantities are not too large so that the model we are using remains valid.

Results obtained for all possible fitted pairs of parameters are given in Table \ref{tab:lq-4corr}. The fitted correction $\hat{\epsilon}_{3/2}$ is now rather independent on the choice of the fitted pair.

\begin{widetext}\strut\hspace{-1.cm}\begin{center}\begin{table}[!htb]
{\small
  \begin{tabular}{|c|c|c|c|c|c|c|}
    \hline
 X-Y  & $a_{1,\,eff.}^{D_{3/2}\pi}$ & $\sigma_{3/2}^2$ & $\hat{\epsilon}_{3/2}$  & X  & Y & $\chi^2/NDF$ \\
 param. &  &  &   & $(GeV)$ & $(GeV)$ &\\\hline
 $\hat{\eta}_1-\hat{\eta}_3$ & $1.06 \pm 0.08$ & $1.45 \pm 0.45$ & $-0.19 \pm 0.11$& $-0.45 \pm 0.08$ & $-0.78 \pm 0.25$ & $1.4/4$\\
 $\hat{\eta}_1-\hat{\tau}_1$ & $0.83 \pm 0.11$ & $1.34 \pm 0.57$ & $-0.28 \pm 0.10$ & $0.05 \pm 0.26$& $0.68\pm 0.73$ & $6.7/4$ \\
 $\hat{\eta}_1-\hat{\tau}_2$ & $0.88 \pm 0.06$  & $1.55 \pm 0.50$ & $-0.27 \pm 0.09$ & $-0.07 \pm 0.10$ & $2.34\pm 1.24$ & $2.7/4$ \\
 $\hat{\eta}_1-\hat{\eta}_2$ & $0.86 \pm 0.06$  & $1.55 \pm 0.50$ & $-0.24\pm 0.10$ & $0.09 \pm 0.17$ & $-2.1\pm 1.1$ & $3.7/4$ \\
\hline
 $\hat{\eta}_3-\hat{\tau}_1$ & $0.82 \pm 0.05$  & $1.75 \pm 0.50$ & $-0.18 \pm 0.12$ & $-0.52 \pm 0.33$& $1.04 \pm 0.31$ & $4.6/4$ \\
 $\hat{\eta}_3-\hat{\tau}_2$ & $0.86 \pm 0.05$  & $1.66 \pm 0.50$ & $-0.27 \pm 0.11$ & $0.10 \pm 0.23$ & $2.9 \pm 0.9$ & $3.0/4$\\
 $\hat{\eta}_3-\hat{\eta}_2$ & $0.89 \pm 0.05$ & $1.67 \pm 0.48$ & $-0.19 \pm 0.11$& $-0.30 \pm 0.26$ & $-2.0 \pm 0.5$ & $2.8/4$ \\
\hline
 $\hat{\tau}_1-\hat{\tau}_2$ & $0.86 \pm 0.06$  & $1.75 \pm 0.46$& $-0.25 \pm 0.09$ & $0.07 \pm 0.33$& $2.8 \pm 1.5$ & $3.2/4$ \\
 $\hat{\tau}_1-\hat{\eta}_2$ & $0.98 \pm 0.10$ & $1.60 \pm 0.41$& $-0.20 \pm 0.10$ & $-1.1 \pm 0.7$& $-3.9 \pm 1.6$ & $1.9/4$ \\
\hline
 $\hat{\tau}_2-\hat{\eta}_2$ & $0.87 \pm 0.05$ & $1.65 \pm 0.48$& $-0.25 \pm 0.09$ & $2.1 \pm 2.2$& $-0.5 \pm 1.1$ & $3.0/4$ \\
\hline
  \end{tabular}}
  \caption[]{\it {Results obtained using the two constraints from theory on  $\tau_{3/2}(1)$ and $\sigma_{3/2}^2$. Set 1 $1/m_Q$ corrections are used and two additional parameters from set 2 are fitted, in addition to $\hat{\epsilon}_{3/2}$.}
  \label{tab:lq-4corr}}
\end{table}\end{center}\end{widetext}

The value of $a_{1,\,eff.}^{D_{3/2}\pi}$, fitted in each model, (see Tables \ref{tab:lq-3corr} and \ref{tab:lq-4corr}), varies between $0.81$ and $1.06$ with uncertainties between $0.05$ and $0.11$. This is compatible with estimates 
of  $a_{1,\,eff.}^{D \pi} = 0.880 \pm 0.024$ and  $a_{1,\,eff.}^{D^* \pi} = 0.981 \pm 0.025$ we obtain by analysing  corresponding decay channels. { These values differ %
  somewhat from the one given by BBNS, but this is not unexpected since we are far from the asymptotic situation considered by those authors.}

Therefore, in our final results we add, as a constraint, that $a_{1,\, eff.}^{D^{**} \pi}= 0.93 \pm 0.07$ obtained from the measurements when either a $D$ or a $D^*$ is produced. This constraint is important, when evaluating systematic uncertainties, to avoid that effects of a variation of a given parameter induce a large variation on $a_1$ and therefore correspond to effects that are outside the fact that the present analysis is done in the framework of  factorization. This constraint is softer than the one used for this same property in previous analyses of these channels.

\subsection{Fitted model parameters for $D_{3/2}$ production}

As observed in Tables \ref{tab:lq-3corr} and \ref{tab:lq-4corr}, the $\chi^2$ obtained when fitting  present data is mainly sensitive to the parameters $\hat{\eta}_1$, $\hat{\eta}_3$, and $\hat{\tau}_1$. There are enough measurements and constraints to determine these quantities with some accuracy. Values of the two other parameters, $\hat{\tau}_2$ and $\hat{\eta}_2$ cannot be determined from present data. We have evaluated effects from this indetermination by changing the values of these two quantities by $\pm 0.5\,GeV$ and by redoing the fit of the other parameters. The variation range we consider is obtained by noting that these quantities
can have, at most, values similar to $\overline{\Lambda}$ so that the model remains valid.
 
\subsection{Summary on $D_{3/2}$ production}
\label{sec:32alone}
In summary, production of $D_{3/2}$ in non-leptonic Class I $B$-meson decays is compatible with  factorization. 

The analysis can be done using the value expected from LQCD for $\tau_{3/2}(1)=0.53\pm0.03$ but this differs from the $\tau_{3/2}^{eff.}(1)$ introduced in equation (\ref{eq:normtau}) %
by the quantity $\hat{\epsilon}_{3/2}=-0.2\pm0.1$,  
whose value has been fitted. % 
We have verified, {%\color{orange} 
in Appendix \ref{app:epsneg}}, that the minus sign of this correction agrees with theory. 
Combining these values we obtain: 
\begin{equation}
\tau_{3/2}^{eff.}(1)=0.42\pm0.06,
\label{eq:tau32eff}
\end{equation}
in agreement with previous analyses.

To have reasonable agreement between data and expectations we find that it is necessary to fit at least one, among the five parameters that control $1/m_Q$ corrections. In this case there remain four parameters that are unknown and it is difficult to obtain model uncertainties. Hopefully, present measurements and constraints from theory allow us to evaluate values of the three most important parameters that control the model. In this way we estimate to have a better control of systematic uncertainties that come from estimates of the values of quantities that are not fitted on data. We obtain:
\begin{eqnarray}
\hat{\eta}_1 \,(GeV)& = & -0.32 \pm 0.13 \pm 0.03\nonumber\\ 
\hat{\eta}_3 \,(GeV)& = & -0.77 \pm 0.28 \pm 0.21\nonumber\\ 
\hat{\tau}_1 \,(GeV)& = & 0.36 \pm 0.35 \pm 0.35. 
\end{eqnarray}
in which the second uncertainty corresponds to the largest variations induced by changing the values of $\hat{\tau}_2$ and $\hat{\eta}_2$ by $\pm0.5\, GeV$. We consider that this model uncertainty has to be added linearly to the one coming from the fit because we cannot favour any value for $\hat{\tau}_2$ and $\hat{\eta}_2$, within their considered variation range.

One cannot directly compare values obtained for  $\hat{\eta}_{1,\,3}$ and $\hat{\tau}_1$ with  previous determinations in which one or at most two of these parameters have been fitted on data.  {It can be noted that fitted values do not preclude the assumed validity of the $1/m_Q$ expansion because these quantities are of order $\overline{\Lambda}$. In Table \ref{tab:r32} we compare %, instead, 
expected values for the ratios ${\cal R}_{D_{3/2}}={\cal B}(\Bob \to D_{3/2}^+ \tau^-\bar{\nu}_{\tau})/{\cal B}(\Bob \to D_{3/2}^+ \ell^-\bar{\nu}_{\ell})$  with  previous determinations. }

\begin{table}[!htb]
\begin{center}
{
  \begin{tabular}{|c|c|c|c|}
    \hline
${\cal R}_{D_{3/2}}\,(\%)$  & our analysis & \cite{ref:ligeti1}(2016) & \cite{ref:ligeti2}(2017) \\
\hline
$ {\cal R}_{D_2^*}$ & $6.1 \pm 0.5 \pm 0.2$ & $7 \pm 1 $ & $7 \pm 1$\\
$ {\cal R}_{D_1}$ & $9.9 \pm 0.7 \pm 0.1$ & $10 \pm 1 $ & $10 \pm 2$\\
\hline
  \end{tabular}}
  \caption[]{\it {Comparison between expected values for ${\cal R}_{D_{3/2}}$.}
  \label{tab:r32}}
\end{center}
\end{table}

\section{Production of $D_{1/2}$ mesons: $\tau_{1/2}^{eff.}(1) \ll \tau_{3/2}^{eff.}(1)$}
\label{sec:d12}

%{\color{blue} texte modifi\'e: 
Data on $B$-meson semi-leptonic decays into $D_{1/2}$ states are rather uncertain. In agreement with previous analyses (see Table \ref{tab:ligeti12} in Appendix \ref{app:bl}), we find that present data do not allow to determine the slope $\sigma_{1/2}^2$ of the corresponding IW function. In addition, the effet of $1/m_Q$ parameters is to modify the observed $w$ dependence of form factors, therefore it is important, as explained in the previous section, to ensure that the variation of the IW function remains physical. Following relativistic quark model  (RQM)  expectations, we use $\sigma_{1/2}^2 = 0.8$  with a conventional error $\pm 0.5$.

For the same reasons it is not possible to fit 
the parameter $\hat{\epsilon}_{1/2}$.

Using measurements, relative to $D_{1/2}$ production in Table \ref{tab:constraint}, apart for ${\cal B}(\Bob \to \Dunstarp \pi^-)$ which is not published, and factorization with $a_1=1$, we have fitted $\tau_{1/2}^{eff.}(1)$ (see Table \ref{tab:d12}). 

Without fitting any additional $1/m_Q$ parameter, we obtain:
\begin{equation}
\tau_{1/2}^{eff.}(1)= 0.147 \pm 0.025,
\label{eq:tau12eff}
\end{equation}
with  a $24\,\%$  fit probability.  
This value is much smaller
than $\tau_{3/2}^{eff.}(1)$ given in Eq. (\ref{eq:tau32eff}), in agreement with the theoretical  expectations and in contradiction  with $\sdcep$.

Let us now take into account the $1/m_Q$ corrections. The $\hat{\chi}_{1,\,2}$ and $\hat{\zeta}_{1}$ correction parameters provide enough flexibility in decay rate expressions to accommodate essentially any measured values, with an acceptable $\chi^2$ probability. It is therefore needed to use additional constraints from theory.

LQCD expectation gives $\tau_{1/2}(1)= 0.296 \pm 0.026 $. We have no predicted value for the quantity $\hat{\epsilon}_{1/2}$ which corresponds to $1/m_Q$ corrections on
$\tau_{1/2}(1)$. If we assume that  $\hat{\epsilon}_{1/2}=\hat{\epsilon}_{3/2}$, we expect $\tau_{1/2}^{eff.}(1)=0.24 \pm 0.02$, which is higher than the measurement in Eq. (\ref{eq:tau12eff}). This can be due to either the fact that $\hat{\epsilon}_{1/2}$ differs from $\hat{\epsilon}_{3/2}$ or that other $1/m_Q$ corrections, not considered for the evaluation in Eq. (\ref{eq:tau12eff}) can have some effect. 
For these reasons, in the following, we  use, as a constraint, the value:
$\tau_{1/2}^{eff.}(1)=0.20 \pm 0.06$ where the uncertainty is large enough to cover the two previous estimates.

Using these two constraints (on $\tau_{1/2}^{eff.}(1)$ and $\sigma_{1/2}^2$ ), and fitting one additional parameter, we obtain the values given in Table \ref{tab:d12}.

\begin{table}[!htb]
{%\color{magenta}
\begin{center}
{
  \begin{tabular}{|c|c|c|c|c|}
    \hline
 X param. & $\tau_{1/2}^{eff.}(1)$ & $\sigma_{1/2}^2$ &  X & $\chi^2/NDF$ \\
\hline
no X  & $0.155 \pm 0.025$ & $1.0 \pm 0.5$ & no value & $3.5/3$ \\
\hline
$\chi_1$ & $0.21 \pm 0.06$ & $0.80 \pm 0.50$ & $-0.27 \pm 0.22$ & $2.6/2$\\
$\chi_2$ & $0.21 \pm 0.06$ & $0.78 \pm 0.50$ & $0.37 \pm 0.27$ & $2.5/2$\\
$\zeta_1$ & $0.22 \pm 0.06$ & $0.76 \pm 0.50$ & $0.75 \pm 0.44$ & $2.3/2$\\
\hline
  \end{tabular}}
  \caption[]{\it {Fitted values of one of the parameters entering in $1/m_Q$ corrections. $\tau_{1/2}^{eff.}(1)$ and $\sigma_{1/2}^2$ are constrained as explained in the text. }
  \label{tab:d12}}
\end{center}
}
\end{table}
{%\color{magenta}
 Fit probabilities are close to $30\,\%$ and values for the different parameters are reasonable, of the order of $\overline{\Lambda}$,} but comparing with $D_{3/2}$ production one cannot identify a parameter to which the analysis is most sensitive and we are not able to fit more than one of these quantities, with reasonable accuracy, using present data. In the following we adjust $\hat{\chi}_1$ and evaluate model systematic uncertainties changing $\hat{\chi}_2$ and $\hat{\zeta}_1$ by $\pm 0.5\,GeV$.

\section{Production of $D_{3/2}$ and  $D_{1/2}$  mesons: combined analysis and systematic uncertainties}
\label{sec:d32d12}

We include in the analysis the data given in Table \ref{tab:constraint}
excluding the unpublished measurement of ${\cal B}(\Bob \to D_1(2430)^{+} \pi^-)$. {\color{black} Constraints from theory on the parameters of the IW functions  and the list of fitted quantities are given in Table \ref{tab:llsw_parameters}, Section \ref{sec:llsw_parameters}}.

The analysis is done taking into account the  validity of factorization.
{\color{black}for all $D^{**}$-meson production} and using: $a_{1,\,eff.}^{D^{**} \pi} = 0.93 \pm 0.07$, see Section \ref{sec:d32}.

{ 

The ratio $\chi^2/NDF=6.3/7$ corresponds to a fit probability of $51\,\%$.

Values of fitted parameters, given in  Table \ref{tab:fitref}, are almost identical with those obtained when considering separate productions of $D_{3/2}$ and $D_{1/2}$ mesons (see Sections \ref{sec:32alone} and \ref{sec:d12}).

\begin{center}\begin{table}[!htb]
{\small \begin{tabular}{|c|c|c|c|c|}
    \hline
 $a_{1,\,eff.}^{D^{**} \pi}$ & $\tau_{3/2}(1)$ & $\sigma_{3/2}^2$ & $\tau_{1/2}^{eff.}(1)$   & $\sigma_{1/2}^2$ \\

$0.944\pm 0.062$ & $0.53 \pm 0.03$ & $1.50 \pm 0.50$ & $0.21 \pm 0.06$ & $0.80 \pm 0.70$ \\
\hline
 $\hat{\epsilon}_{3/2}$ & $\hat{\eta}_1\,(GeV)$ & $\hat{\eta}_3\,(GeV)$ & $\hat{\tau}_1\,(GeV)$ & $\hat{\chi}_1\,(GeV)$ \\

$-0.18\pm 0.11$ & $-0.32 \pm 0.13$ & $-0.77 \pm 0.28$ & $0.36 \pm 0.35$ & $-0.24 \pm 0.26$ \\
\hline
  \end{tabular}}
  \caption[]{\it {Fitted values of the reference model parameters, for $B \to D^{**}$ Lorentz invariant form factors. The first line contains parameters that are constrained by theory or from external measurements ($a_{1,\,eff.}^{D^{**} \pi}$). Quoted uncertainties are obtained from the fit.}
  \label{tab:fitref}}
\end{table}\end{center}

Fitted quantities allow to  obtain values for different branching fractions of a
$\Bob$ meson decaying into $D^{**}\ell^- \bar{\nu}_{\ell}$, with a light or the $\tau$ lepton, as well as for non-leptonic decays. We have also considered $\Bob \to D^{**,\,+}D_s^{(*),-}$ decays, in the {framework} of factorization. 

To evaluate systematic uncertainties, values of not-fitted parameters are      
changed to $\pm 0.5\,GeV$ and the largest variation on a fitted or derived quantity, that depends on fitted values, is used as systematic model uncertainty. For some of these quantities, mainly related to the production of $D_{1/2}$ mesons, these variations are asymmetric, when compared to the value obtained with the reference model (in which the non-fitted parameters are set to zero). In this case we have symmetrized systematic uncertainties and corrected central values accordingly. Model uncertainties are added  linearly to those from the fit
because there is no reason to prefer a value for unfitted quantities, within their variation range.

Other sources of systematic uncertainties are considered as:
\begin{itemize}
\item effects from uncertainties on HQET parameters (Set 1). They are illustrated by using the values adopted in previous analyses (see Table \ref{tab:constants});
\item effects of using  a linear parameterization for the IW functions, versus $w$, in place of the dipole distribution (Eq. (\ref{les_taus}));
\item the effect of changing the parameterization of Blatt-Weisskopf terms in non-leptonic decays when the momentum of the emitted hadron is computed in the frame of the resonance in place of the $B$-meson rest frame (the former has been used by some experimental collaborations).
\end{itemize}}
These observed variations are only indicative and cannot be considered as really representative of corresponding systematic uncertainties values. In most cases, these sources are less important than uncertainties from the fit or from the model.

When evaluating a ratio between two derived quantities, correlations between the different uncertainties are taken into account.  

 {\color{black}In the following we detail our expectations and give comparisons with those obtained in another analysis, {quoted as $\sdce$,} which is close to those done in previous publications \cite{ref:ligeti1,ref:ligeti2}. Differences relative to our approach are listed in Appendix \ref{app:second}.
Numerical values, obtained in this way, are quoted also in appendices whereas corresponding expected distributions are compared with our results on the different Figures that follow.}

\subsection{Comparison between our model and the $\sdce$ analysis for  $D_{1/2}$ production in non-leptonic Class I decays}
\label{sec:sc2_nl}
{%\color{blue}
 In Table \ref{tab:sc2dpi} we illustrate the differences between our model and the $\sdce$ analysis for Class I non-leptonic decays.

\begin{widetext}\begin{center}\begin{table}[!htb]
{
  \begin{tabular}{|c|c|c|c|}
    \hline
 channel & measured & our model & $\sdce$ model\\
\hline
$ {\cal B}(\Bob \to \Dostarp \pi^-)\times 10^{4}$ & $1.19 \pm 0.12$ & $1.21 \pm 0.12 $ & $10.0 \pm2.5$\\
\hline
$  {\cal B}(\Bob \to \Dunstarp \pi^-)\times 10^{4}$ & $0.21 \pm 0.27$ & $0.7 \pm 0.7$ & $3.2 \pm 2.8$\\
\hline
$  {\cal R_K}(\Dostar,\Ddestar)$ & $0.84 \pm 0.36$ & $0.35 \pm 0.04$ & $2.8 \pm 0.7$\\
\hline
  \end{tabular}}
  \caption[]{\it {Comparison between measured and expected
   values for
   non-leptonic $\Bob \to D_{1/2}$  transitions. ${\cal R_K}(\Dostar,\Ddestar)$ is the ratio between the branching fractions ${\cal B}(\Bob \to \Dostarp K^-)$ and ${\cal B}(\Bob \to \Ddestarp K^-)$, with the two $D^{**}$ mesons decaying into $D^0\pi^+$.}
  \label{tab:sc2dpi}}
\end{table}\end{center}\end{widetext}

{The measured values for  $ {\cal B}(\Bob \to \Dostarp \pi^-)$ and $  {\cal R}(\Dostar,\Ddestar)$ enter
our model through the use of factorization. Therefore it is a check of consistency
that the corresponding fitted values are in agreement with the data, as well as with  theory, which predicts indeed that $j=1/2$ transitions should be much smaller than $j=3/2$ ones.}

{On the
other hand, one can see that predictions of $\sdce$ are in disagreement with the
data by more that four standard deviations and too large by around one order of magnitude, which leads to discard the model.
Keeping this in mind, it may nevertheless be useful to apply the same model to
semi-leptonic decays for the sake of comparison with our own results, especially since $LLSW$ is used by experimentalists in the analyses of the background to  decays such as $\Bb \to D^{*} \ell \overline{\nu_\ell}$.
} 
}

\section{$D^{**}$ meson production in $\Bob$ semi-leptonic decays}
 Our results on decay branching fractions of $\Bob$ mesons into the four $D^{**,\,+}$ mesons are explained. In Sections \ref{sec:btodstarpi_el} and \ref{sec:btodpi_el}, expected hadronic mass and $q^2$ distributions are obtained for  the $D^*\pi$ and $D\pi$ hadronic final states, respectively. Spectra are compared with the $\sdce$ analysis 
and, to ease the comparison, total decay rates, expected in the two cases, have been scaled (only for plotting purposes) to the central values measured for the considered final states. {Corresponding uncertainties are scaled also to agree with those obtained on measurements for $ {\cal B}(\Bob \to D^{(*),+}\pi^- \ell^-\bar{\nu}_{\ell})$, with light leptons. Additional uncertainties, from the fit and the model, which appear when considering decays with a $\tau$ lepton or a $D_s^-$ meson, are included.}}

{
\subsection{Expected values for ${\cal B}(\Bob \to D_{i}^{**,\,+} \ell^- \bar{\nu}_{\ell})$ and corresponding $q^2$ distributions.}
Expected $q^2$ distributions in $\Bob$ semi-leptonic decays with a  light or a $\tau$ lepton are given in Figures 
\ref{fig:d2starsl_q2}, \ref{fig:d1sl_q2}, \ref{fig:d0starsl_q2} , and \ref{fig:d1starsl_q2} for $\Ddestarp$, $\Dunp$, $\Dostarp$, and $\Dunstarp$, respectively. Hatched areas correspond to uncertainties from the fit. Curves indicated with dots are the expected systematic uncertainties from the un-measured $\hat{\eta}_2$ and $\hat{\tau}_2$ parameters. Those indicated with triangles are from the un-measured $\hat{\chi}_2$ and $\hat{\zeta}_1$ parameters.
In each figure the left plot is for light leptons and the right one for the $\tau$.

{\begin{figure*}[!htb]
  \begin{center}
  \mbox{\epsfig{file=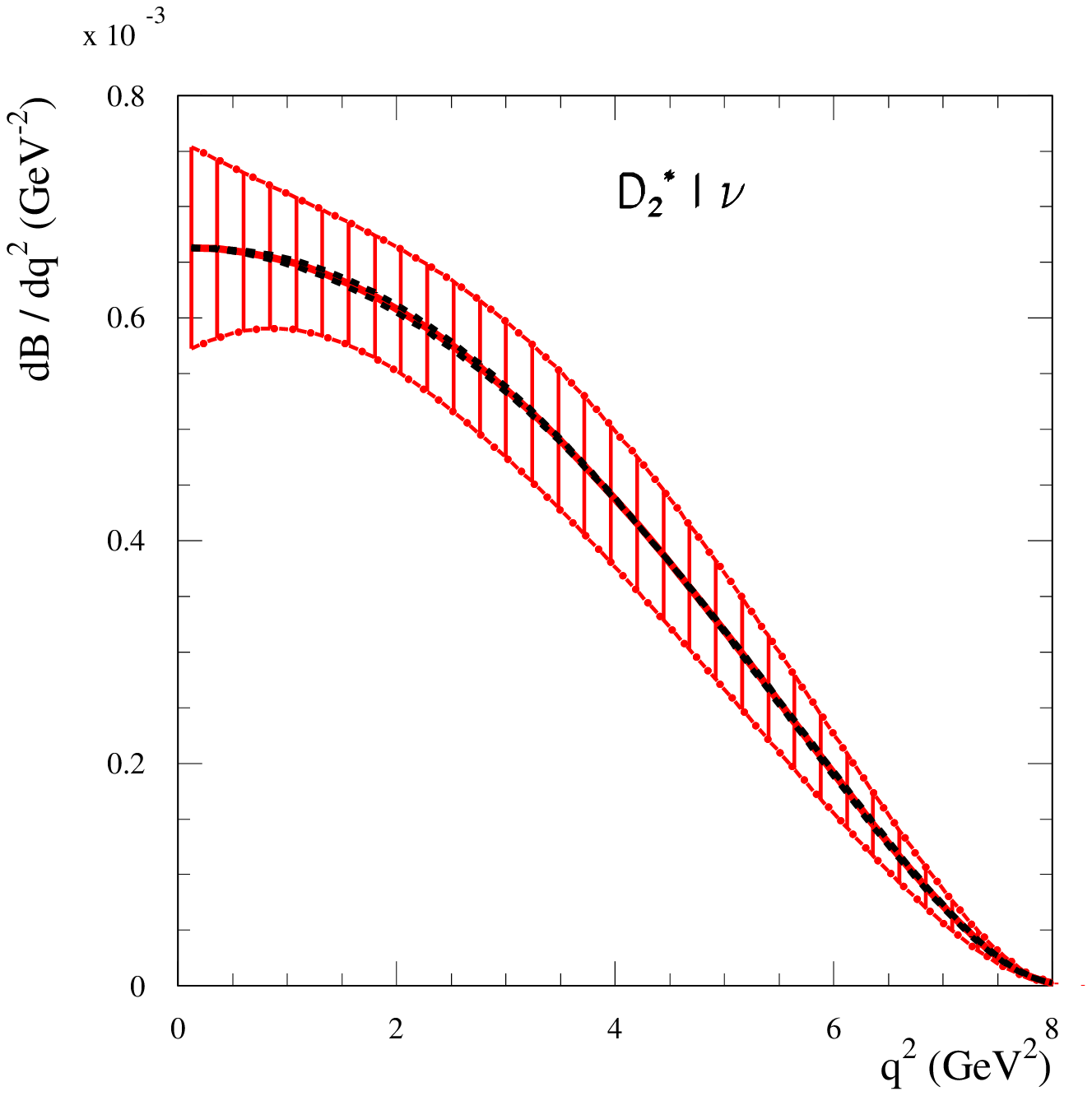,width=0.5\textwidth}
\epsfig{file=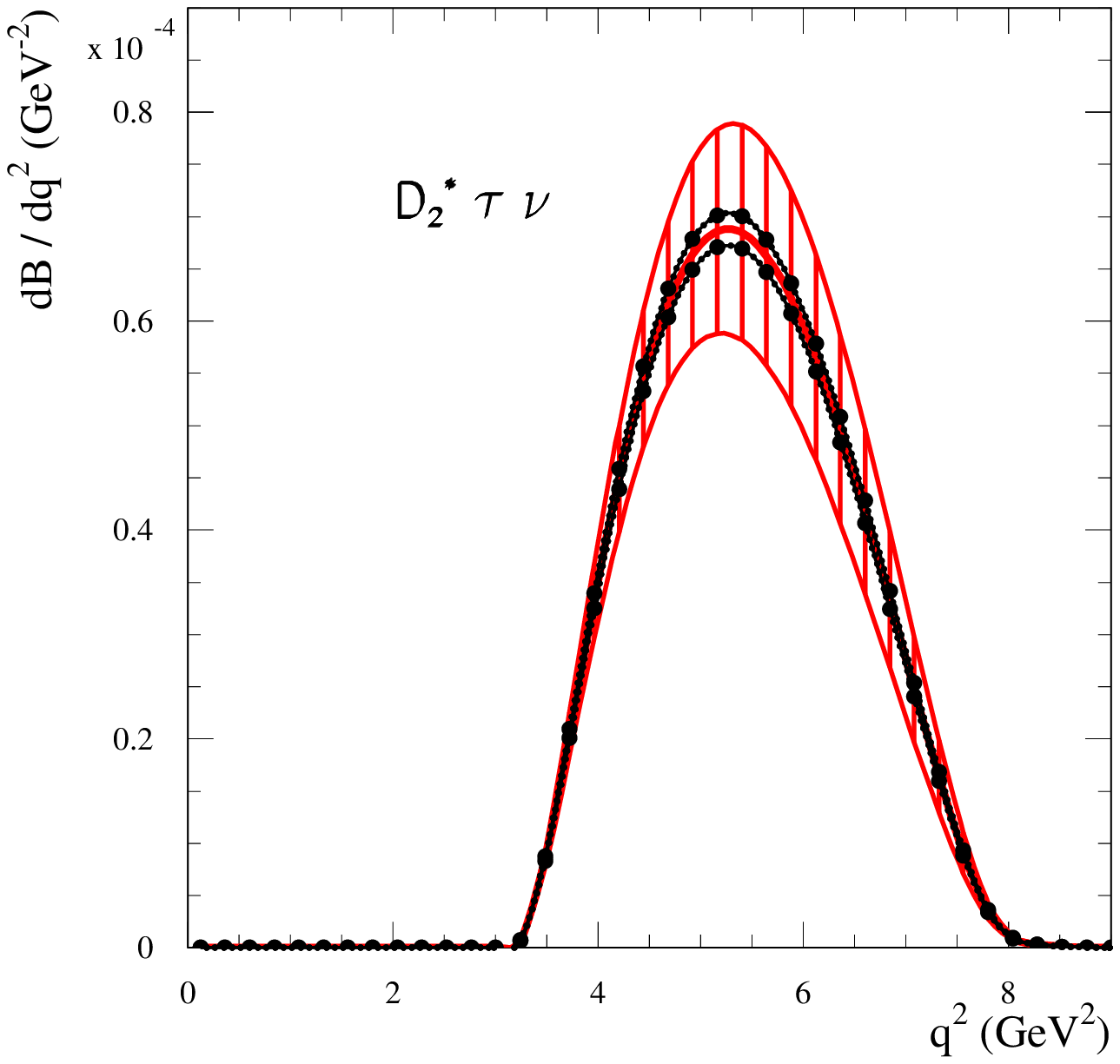,width=0.5\textwidth}}
 \end{center}
  \caption[]{ Expected $q^2$ distributions for $\Bob \to \Ddestarp$ in semi-leptonic decays. Hatched areas correspond to uncertainties from the fit, curves with dots indicate the model uncertainty. Other systematic uncertainties, indicated in Table \ref{tab:brsl_dsstar_el}, are not displayed.}
\label{fig:d2starsl_q2}
\end{figure*}
}

\begin{figure*}[!htb]
  \begin{center}
  \mbox{\epsfig{file=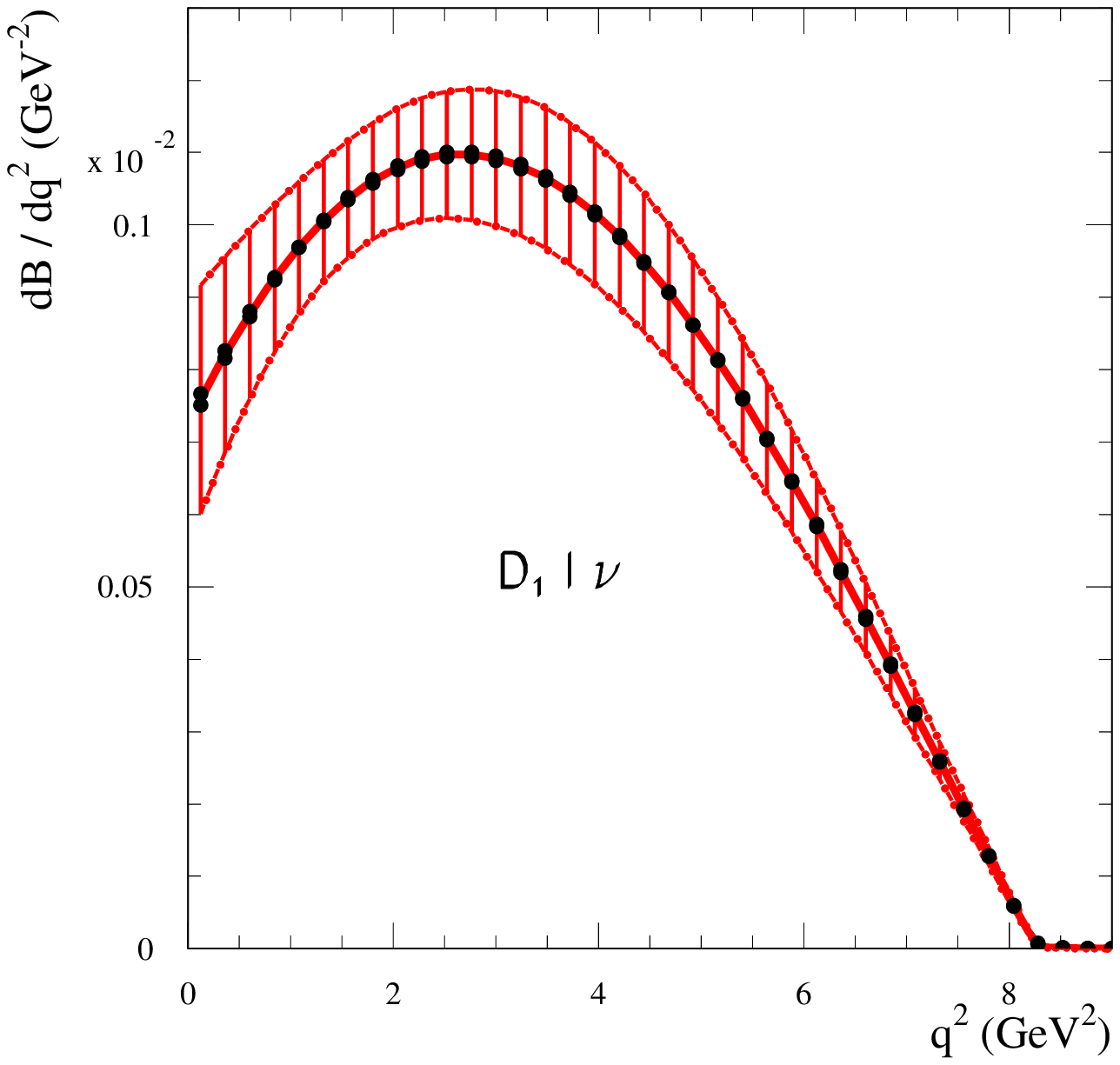,width=0.5\textwidth}
\epsfig{file=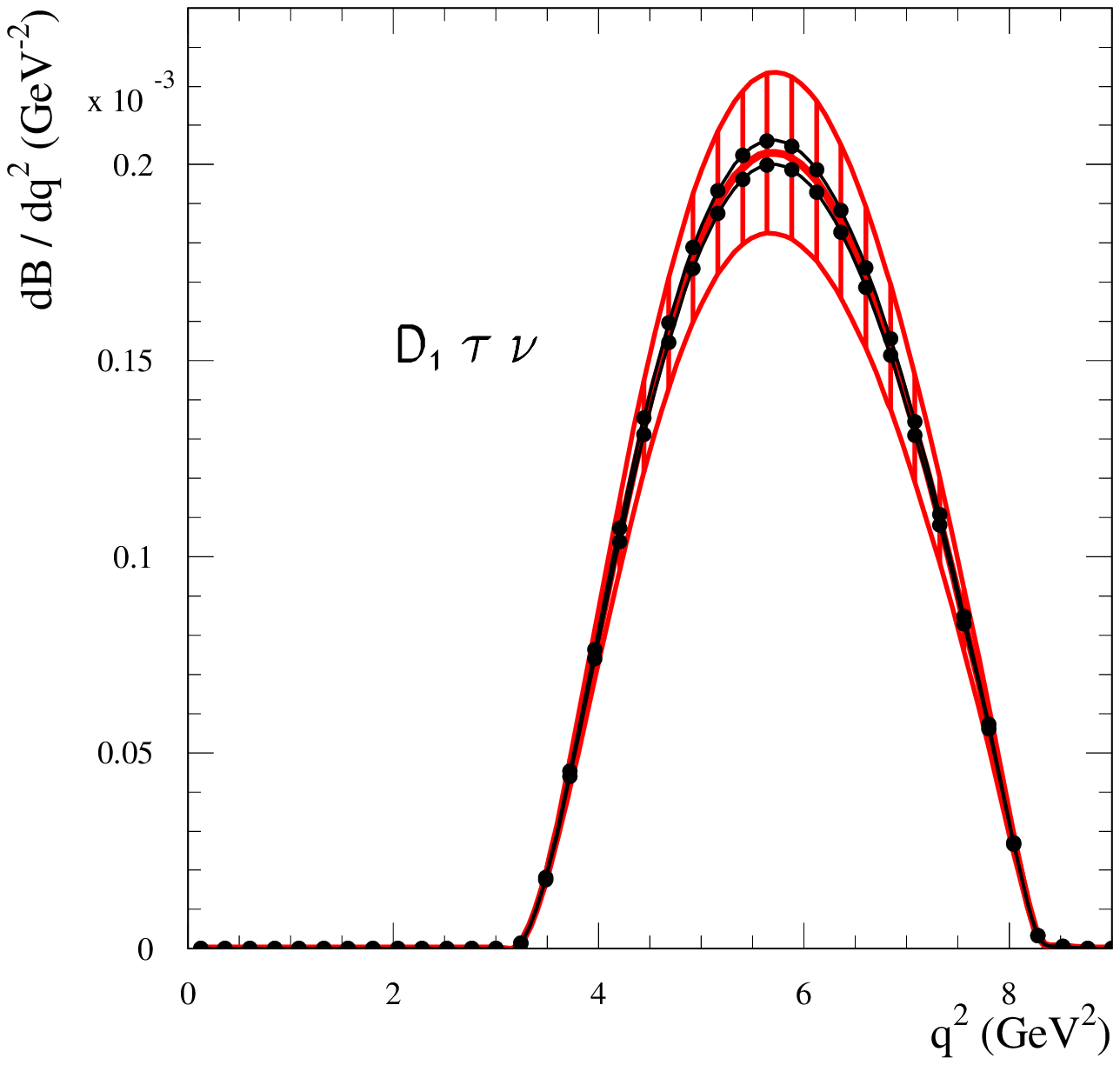,width=0.5\textwidth}}
 \end{center}
  \caption[]{ Expected $q^2$ distributions for $\Bob \to \Dunp$ in semi-leptonic decays. Same conventions are used as in Fig.\ref{fig:d2starsl_q2}.
}
\label{fig:d1sl_q2}
\end{figure*}

\begin{figure*}[!htb]

  \begin{center}
  \mbox{\epsfig{file=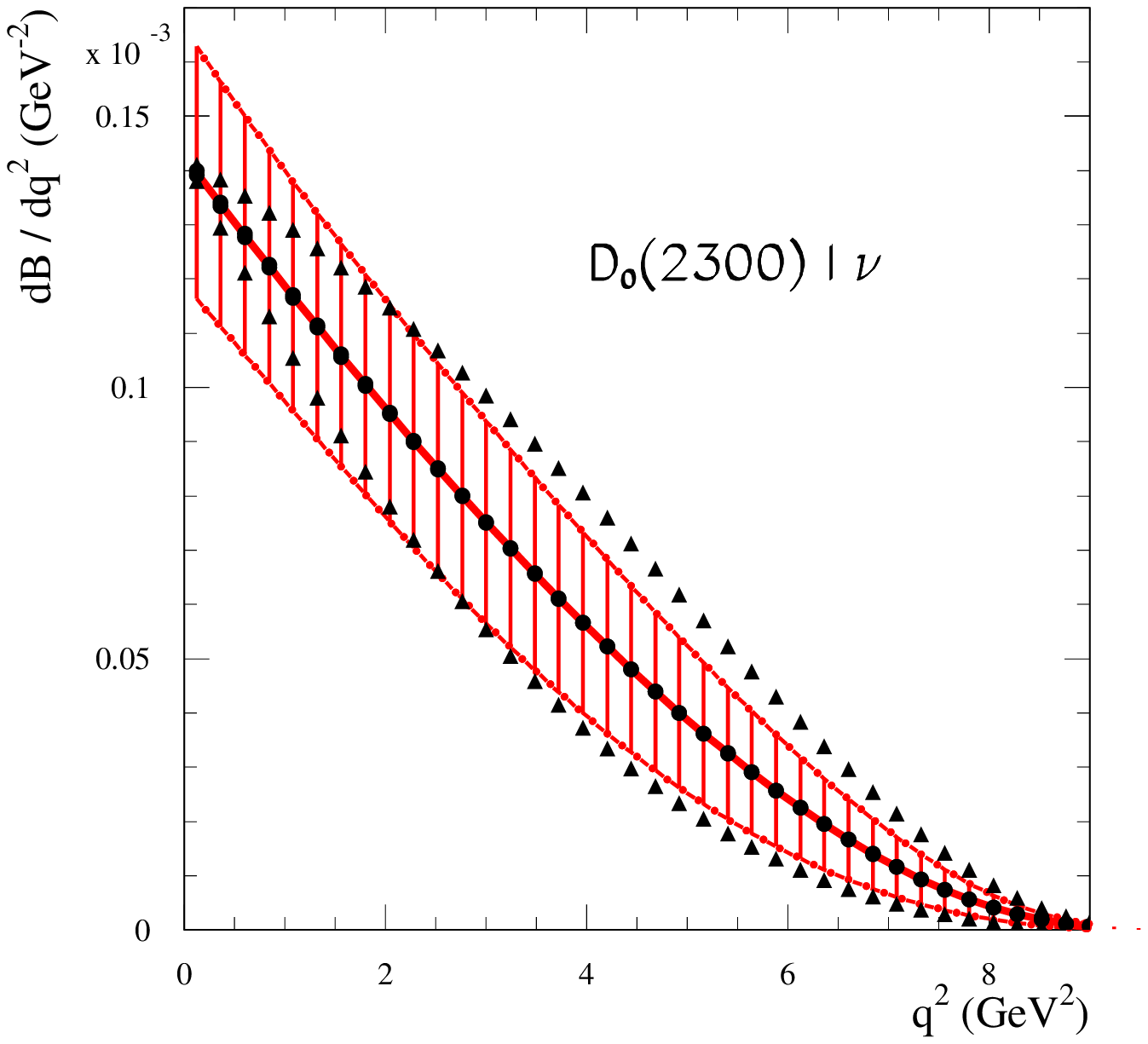,width=0.5\textwidth}
\epsfig{file=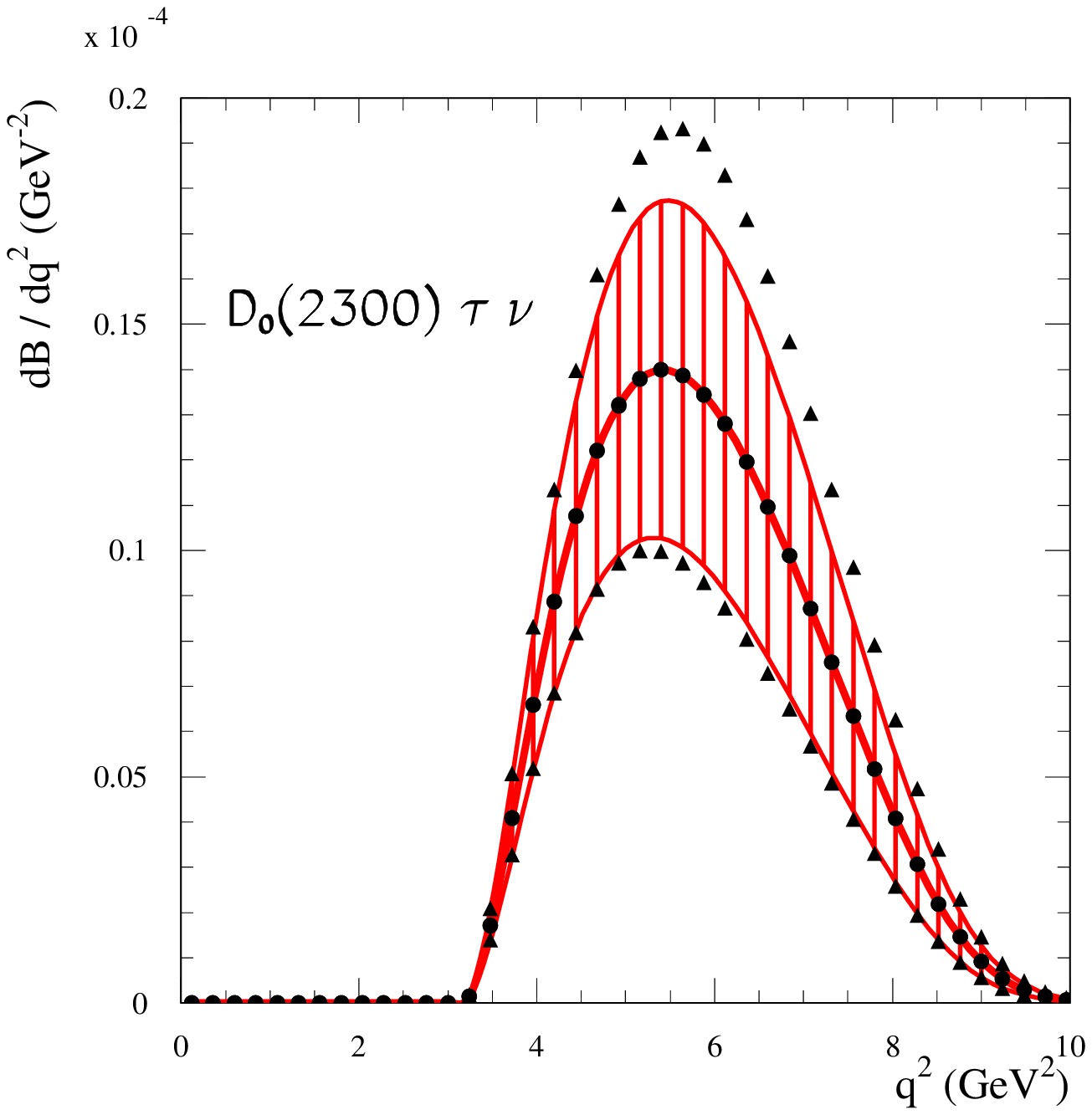,width=0.5\textwidth}}
  \end{center}
  \caption[]{ Expected $q^2$ distributions for $\Bob \to \Dostarp$ in semi-leptonic decays. Hatched areas correspond to uncertainties from the fit, curves with dots indicate the model uncertainty expected from $\hat{\tau}_2$ and $\hat{\eta}_2$ parameters whereas, those with triangles, correspond to the uncertainty from $\hat{\chi}_2$ and $\hat{\zeta}_1$. Other systematic uncertainties, invoked in Table \ref{tab:brsl_dsstar_el}, are not displayed.
}
\label{fig:d0starsl_q2}
\end{figure*}
\begin{figure*}[!htb]
  \begin{center}
  \mbox{\epsfig{file=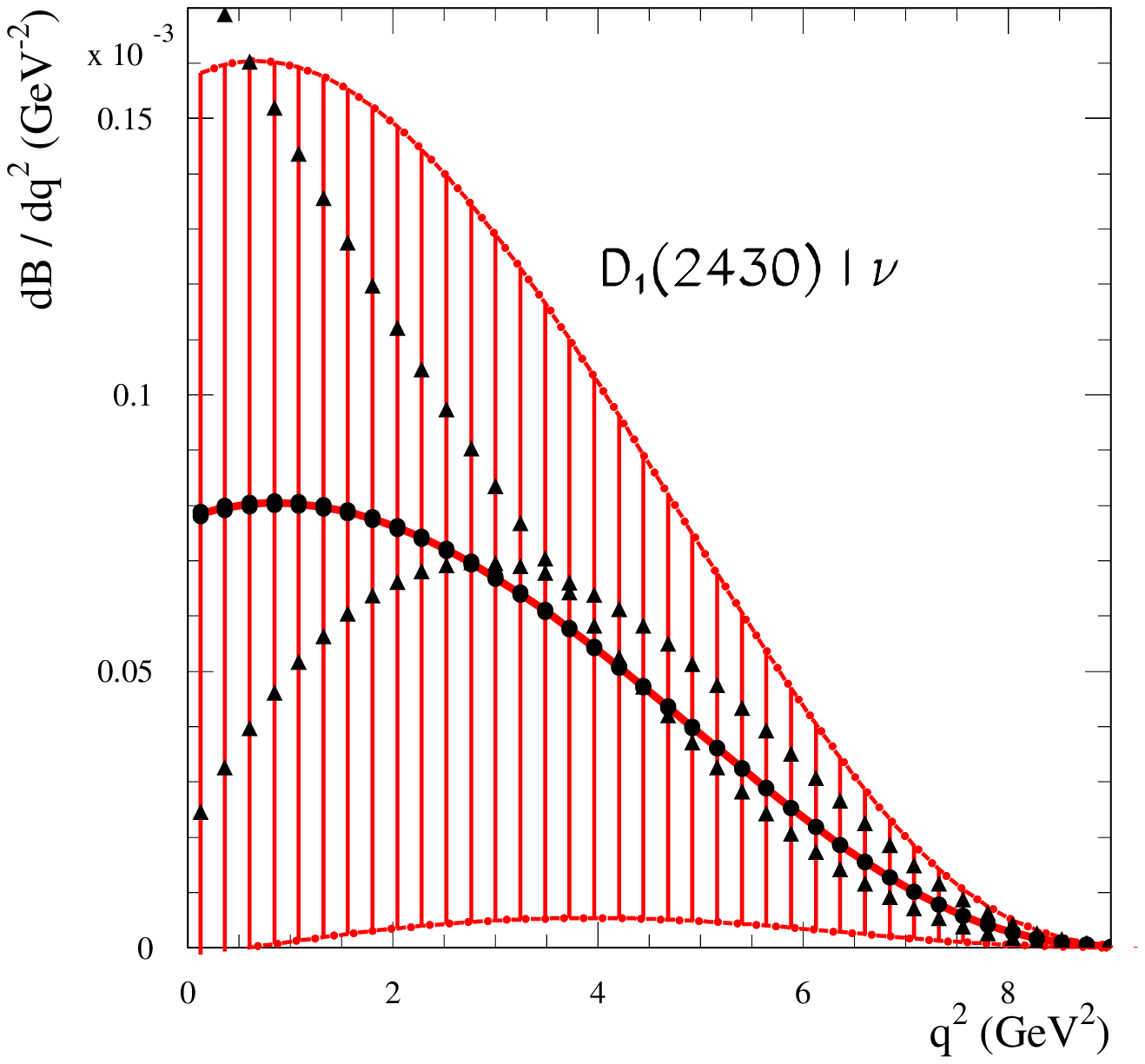,width=0.5\textwidth}
\epsfig{file=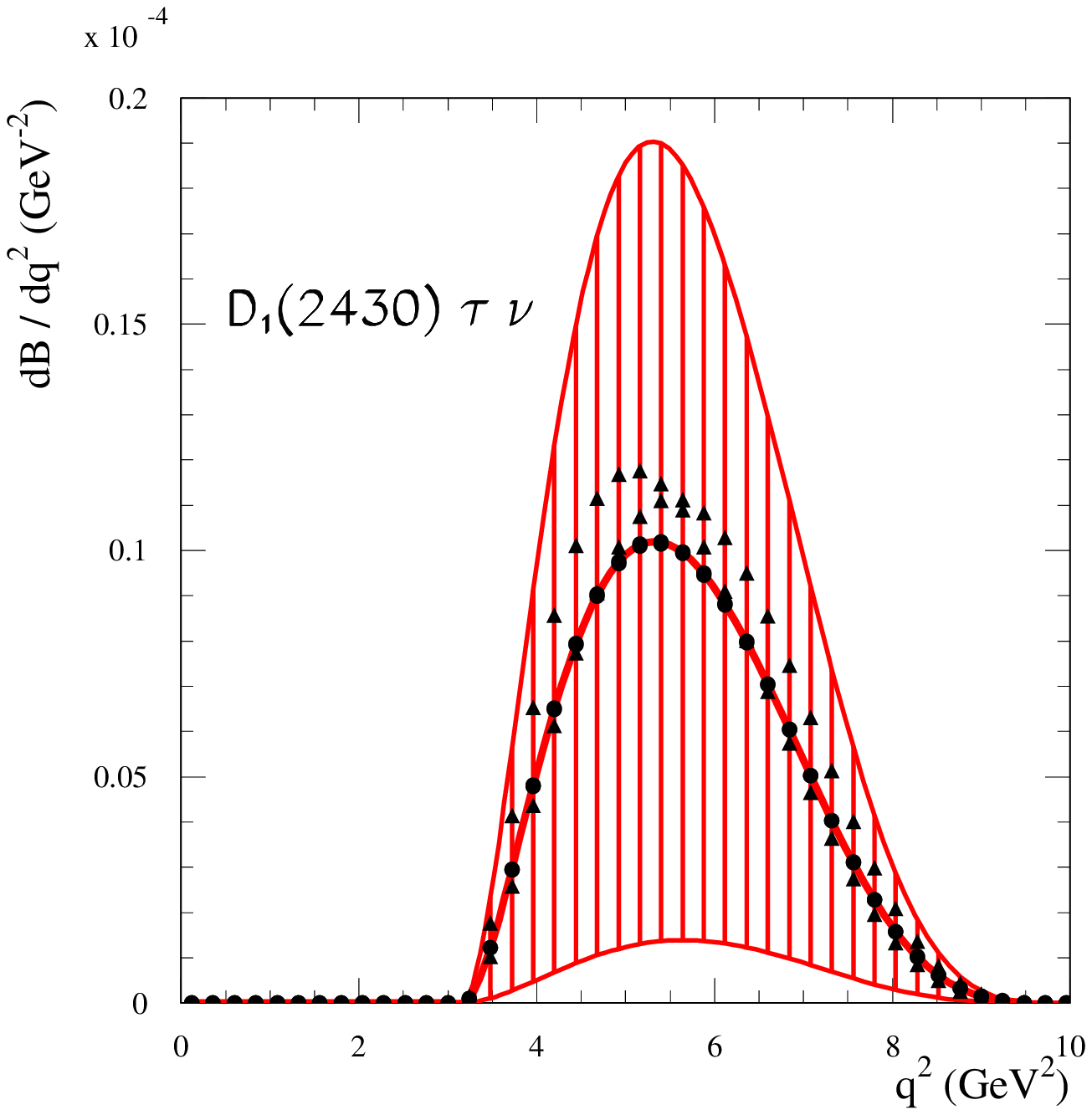,width=0.5\textwidth}}

 \end{center}
  \caption[]{ Expected $q^2$ distributions for $\Bob \to \Dunstarp$ in semi-leptonic decays. Same conventions are used as in Fig.\ref{fig:d0starsl_q2}
}
\label{fig:d1starsl_q2}
\end{figure*}
}

Expected values for semi-leptonic branching fractions, with a light and the $\tau$ lepton, are given in Table \ref{tab:brsl_dsstar_el}. Those obtained in the $\sdce$ analysis are given in Appendix \ref{app:sl_sc2}.  To evaluate uncertainties on quoted values we have added linearly uncertainties from the fit and from the model. Values for other possible sources of uncertainties, quoted in Table \ref{tab:brsl_dsstar_el} are simply indicative and usually are not dominant when compared with the others. It can be noted that ratios of branching fractions with a $\tau$ or a light lepton have a better accuracy because of correlations between the different uncertainties.

%{\color{red} Commenter ce qui est diff\'erent et donner la valeur obtenue pour $ {\cal B}(\Bob \to \Dostarp \ell^-\bar{\nu}_{\ell})$.}

\begin{widetext}
\begin{center}
\begin{table}[!htb]
{
  \begin{tabular}{|c|c|c|c|c|c|}
    \hline
 channel & value $\pm$ fit & model & HQET & IW linear& Blatt-W. \\
\hline
$ {\cal B}(\Bob \to \Ddestarp \ell^-\bar{\nu}_{\ell})\times 10^3$ & $3.15 \pm 0.30$ & $0.$ & $0.$ & $0.01$ & $0.02$\\
$ {\cal B}(\Bob \to \Ddestarp \tau^-\bar{\nu}_{\tau})\times 10^4$ & $1.90 \pm 0.29$ & $0.05$ & $0.$ & $0.09$ & $0.03$\\
${\cal R}_{\Ddestar}\times 10^2$& $6.03 \pm 0.52$ & $0.15$ & $0.02$ & $0.24$ & $0.06$\\
\hline
$ {\cal B}(\Bob \to \Dunp \ell^-\bar{\nu}_{\ell})\times 10^3$ & $6.40 \pm 0.44$ & $0.$ & $0.$ & $0.$ & $0.$\\
$ {\cal B}(\Bob \to \Dunp \tau^-\bar{\nu}_{\tau})\times 10^4$ & $6.30 \pm 0.59$ & $0.10$ & $0.30$ & $0.70$ & $0.07$\\
${\cal R}_{\Dun}\times 10^2$& $9.84 \pm 0.68$ & $0.15$ & $0.47$ & $1.10$ & $0.11$\\
\hline\hline
$ {\cal B}(\Bob \to \Dostarp \ell^-\bar{\nu}_{\ell})\times 10^4$ & $5.1 \pm 1.2$ & $1.2$ & $-0.2$ & $0.4$ & $-0.1$\\
$ {\cal B}(\Bob \to \Dostarp \tau^-\bar{\nu}_{\tau})\times 10^5$ & $5.0 \pm 1.3$ & $1.7$ & $-0.1$ & $0.6$ & $0.1$\\
${\cal R}_{\Dostar}\times 10^2$& $9.9 \pm 1.5$ & $1.0$ & $0.4$ & $0.4$ & $0.1$\\
\hline
$ {\cal B}(\Bob \to \Dunstarp \ell^-\bar{\nu}_{\ell})\times 10^4$ & $4.6 \pm 3.7$ & $0.9$ & $0.3$ & $-0.5$ & $0.$\\
$ {\cal B}(\Bob \to \Dunstarp \tau^-\bar{\nu}_{\tau})\times 10^5$ & $3.4 \pm 2.7$ & $0.6$ & $0.3$ & $-0.3$ & $0.$\\
${\cal R}_{\Dunstar}\times 10^2$& $7.4 \pm 1.2$ & $1.6$ & $0.1$ & $0.4$ & $0.1$\\
\hline
  \end{tabular}}
  \caption[]{\it %
   Our expectations for semi-leptonic branching fractions with a light or a $\tau$ lepton and their ratio for individual $D^{**}$-meson states. 
  \label{tab:brsl_dsstar_el}}
\end{table}\end{center}\end{widetext}

\subsubsection{$D_{3/2}$ production}
Values for semi-leptonic branching fractions with a light lepton and a $D_{3/2}$ meson are essentially identical with input measurements. This is because one has basically no measurement of the $q^2$ dependence of the different decay rates and because the normalisation is fitted through the $\tau_{3/2}$ and $\hat{\epsilon}_{3/2}$ parameters. Expected uncertainties
on the production of $D_1$ and $D_2^*$, with a $\tau$ lepton are of about $20\,\%$. It can be noted that, in $D_1$ production, uncertainties on Set 1 (HQET) parameters and on the $w$ expected dependence of $\tau_{3/2}$ may be not negligible. This is expected because HQET Set 1 corrections change the computed branching fraction with a $D_1$ by a large factor, as compared with the infinite quark mass limit prediction.

\subsubsection{$D_{1/2}$ production}
 The value expected for $ {\cal B}(\Bob \to \Dostarp \ell^-\bar{\nu}_{\ell})=(5.1 \pm 2.4)\times 10^{-4}$ is much smaller than the one usually anticipated from %
 {$\sdce$: $(39.1 \pm 7.2)\times 10^{-4}$}, as given in Appendix \ref{app:sl_sc2}. This result is a direct consequence of the use of { factorization}. Values expected for $\Dunstar$ are similar but being affected by larger uncertainties one cannot draw any conclusion. 
Production of $D_{1/2}$ mesons in $B$-meson semi-leptonic decays is expected to be an order of magnitude smaller than the one of narrow states.  In the production of $D_{1/2}$ mesons, model uncertainties are dominant when compared with the other considered sources of systematic uncertainties.

 Expected branching fractions for $\Dostar$ and $\Dunstar$ production have about $50\,\%$ and $100\,\%$ uncertainty, respectively. These relative uncertainties are even larger when a $\tau$ lepton is emitted.

\subsubsection{Conclusions}
In our analysis, the expected low production rates of broad, relative to narrow, $D^{**}$ mesons comes from theoretical arguments and is {in agreement with the factorization property}. This low value implies that it has to be complemented by another source of events, to explain the measured broad mass distributions in
$D^{(*)} \pi$ hadronic final states. We examine, in the following sections, if the contributions expected from $D^{(*)}_V$ components can fill these gaps. Such components have been ignored in previous analyses which consider that $D_{1/2}$ decays, alone, are enough to explain measurements.

\subsection{Analysis and predictions for the $\Bob \to D^* \pi \ell^- \bar{\nu}_{\ell}$ final state}

\label{sec:btodstarpi_el}

We examine if measured $\Bob \to D^* \pi \ell^- \bar{\nu}_{\ell}$ decays can be explained using a sum of $D^{**}$ and $D^{(*)}_V$ components.
All quoted numbers are relative to the sum of $D^{*0}\pi^+$ and  $D^{*+}\pi^0$ final states and we refer to Section \ref{sec:dv_d1star} for input measurements. 

The measured branching fraction into broad components amounts to:
\begin{equation}
{%\color{magenta}
{\cal B}(\Bob \to \left. D^* \pi\right |_{broad} \ell^- \bar{\nu}_{\ell})=(2.86 \pm 0.69)\times 10^{-3}.
\label{eq:dstarpi_broad}}
\end{equation}

{In the hypothesis that contributions from  higher mass resonances can be neglected, the expected contribution from $\Dunstarp$ decays, equal to $(0.46\pm 0.46)\times 10^{-3}$, %\patrick{ 
has to be complemented by a $D^{(*)}_V$ component equal to: $(2.4 \pm 0.8)\times 10^{-3}$. 
For $r_{BW}$ values varying between $1$ and $3\,GeV^{-1}$ our estimates for this component  are in the range $[2.0,\,0.9]\times 10^{-3}$ (see Section \ref{sec:dv_d1star}). These values are compatible with the needed contribution.}

\begin{figure*}[!htb]
  \begin{center}
    \mbox{\epsfig{file=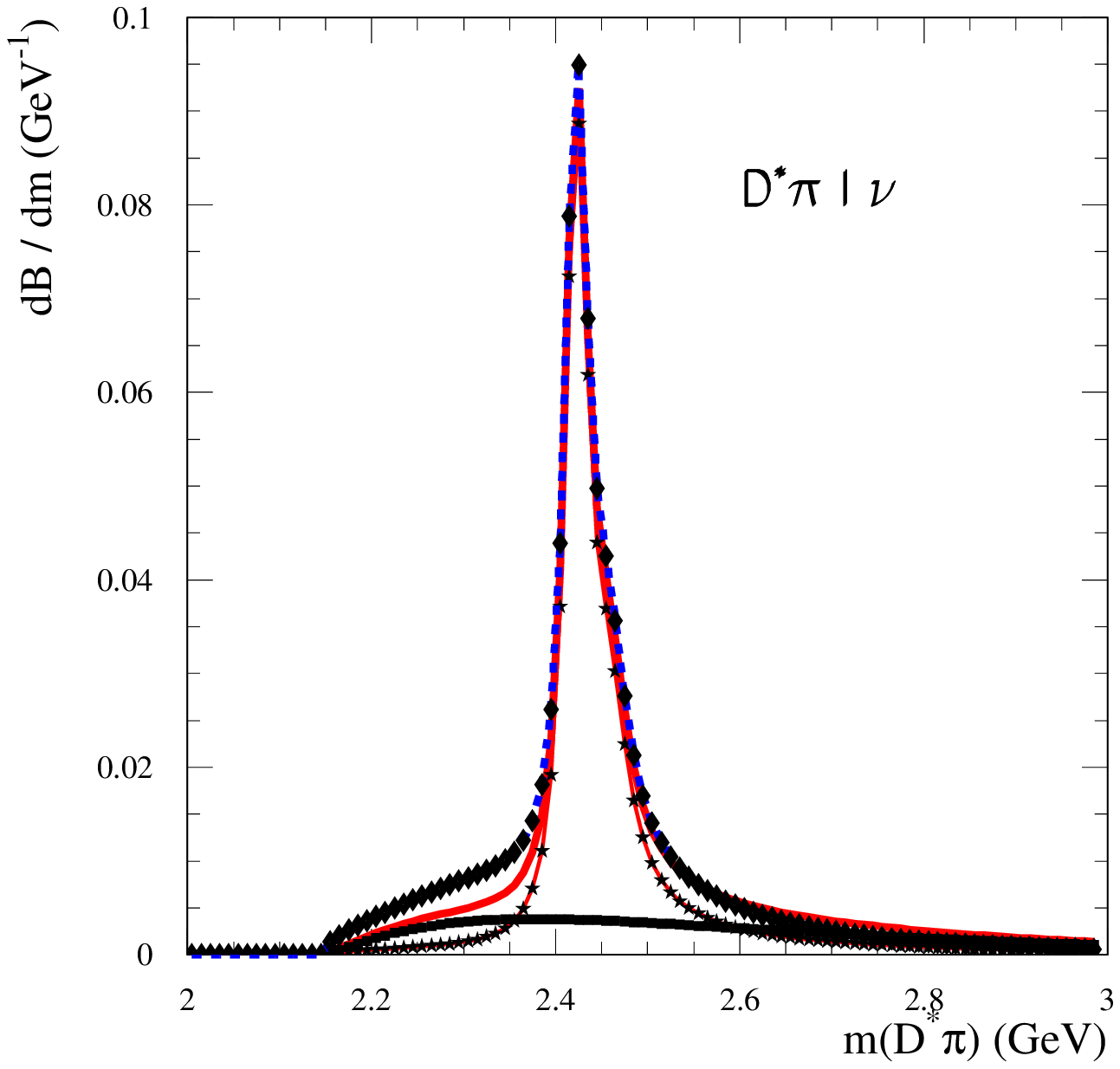,width=0.5\textwidth}
\epsfig{file=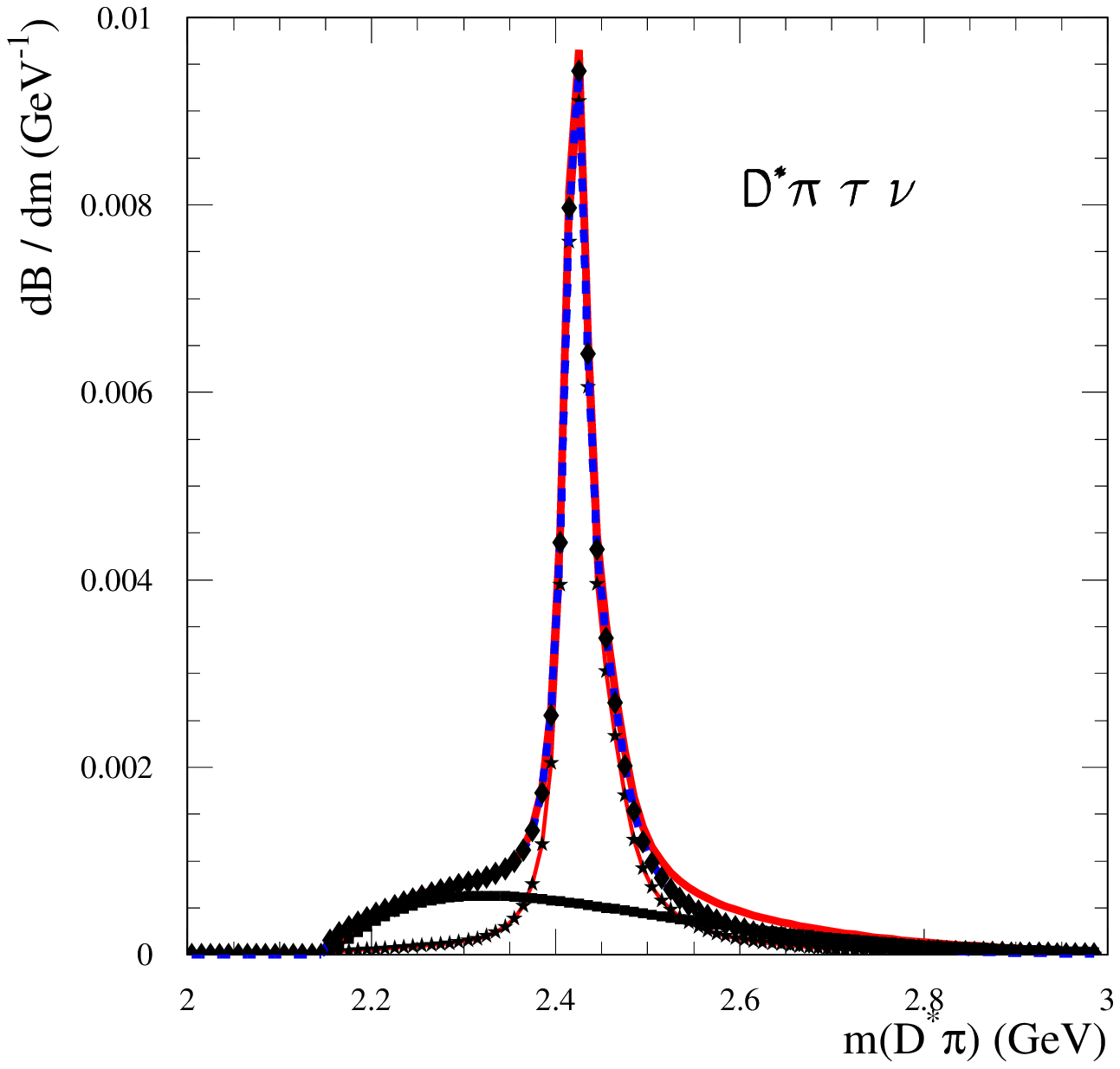,width=0.5\textwidth}}
  \end{center}
  \caption[]{ Expected $D^*\pi$ mass distributions for $\Bob \to D^* \pi$ in semi-leptonic decays. Full lines correspond to our model and the dashed line is for $\sdce$. The line with stars gives the $D^{**}$ contribution and the one with squares is for the $D^{(*)}_V$ component. The red line corresponds to the sum of these two contributions. {Only central values of the distributions are displayed.} }
\label{fig:dstarpisl_all_m}
\end{figure*}

In Figure \ref{fig:dstarpisl_all_m}, expected $D^*\pi$ mass distributions from our model (full line) and from the $\sdce$ analysis (dashed line), are compared. To ease the comparison, central values expected from the two models are scaled to agree with the measured one. To do so, in our model, the $D^{(*)}_V$ component only is scaled while, for other components, expected values are used. In {$\sdce$}, the $\Dunstarp$ component, only, is scaled. Scaling factors are obtained for decays into light leptons and their values are used for the other analyzed transitions which involve a $\tau$ lepton or a $D_s$ meson.

 Spectra are dominated by the contributions from $D_{3/2}$ mesons.

\subsubsection{Expected differences between our model and $\sdce$, for broad mass components}

In the following we compare expected mass and $q^2$ distributions, for the $D^*\pi$ broad mass component, in our model and in $\sdce$, and after having normalized expectations to agree with the measured central value, for light leptons. {% 
Hatched areas correspond, only, to the uncertainty quoted in Eq. (\ref{eq:dstarpi_broad})}. {%
Therefore, they are mainly indicative and do not illustrate the uncertainty
on the shape of the distributions which comes from the model dependence of the two analyses.

\begin{figure*}[!htb]
{
  \begin{center}
    \mbox{\epsfig{file=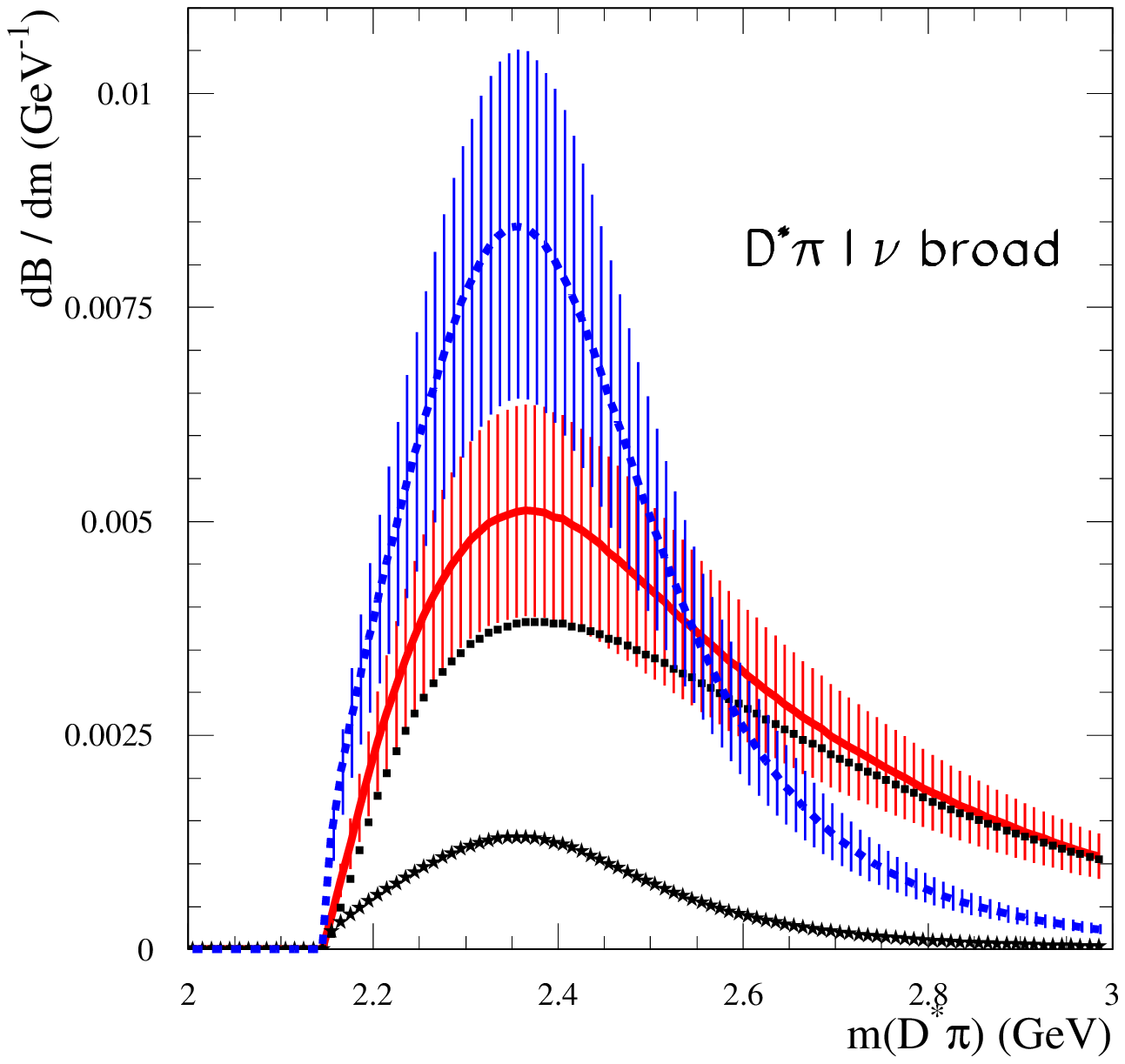,width=0.5\textwidth}
\epsfig{file=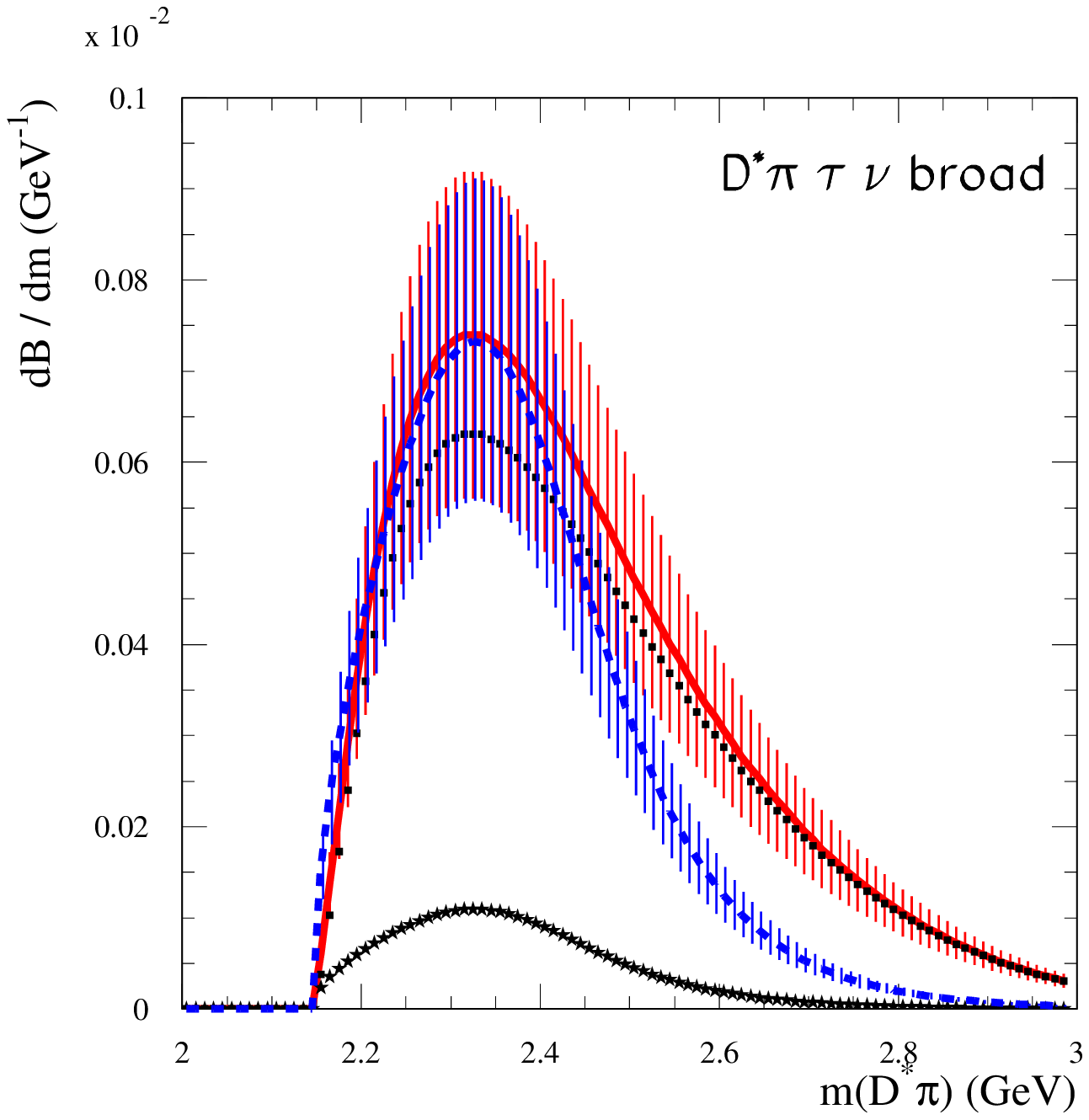,width=0.5\textwidth}}
  \end{center}
  \caption[]{ Expected broad $D^*\pi$ mass distributions for $\Bob \to D^* \pi$ in semi-leptonic decays. The same conventions as in Figure \ref{fig:dstarpisl_all_m} are used. Hatched areas include only the experimental 
  uncertainty given in Eq. (\ref{eq:dstarpi_broad}).}
}
\label{fig:dstarpisl_m}
\end{figure*}

For $\tau$ events one expects 1.3 times more events in our analysis  whereas estimates were equal, by convention, for light leptons. This comes from the different dependence of $D^{(*)}_V$ and $\Dunstar$ components versus $q^2$. The $D^{(*)}_V$ component corresponds to a $D^*\pi$ mass distribution which can mimic a broad
resonance in the absence of an analysis of the angular distribution. 

\begin{figure*}[!htb]
{%\color{magenta}
  \begin{center}
    \mbox{\epsfig{file=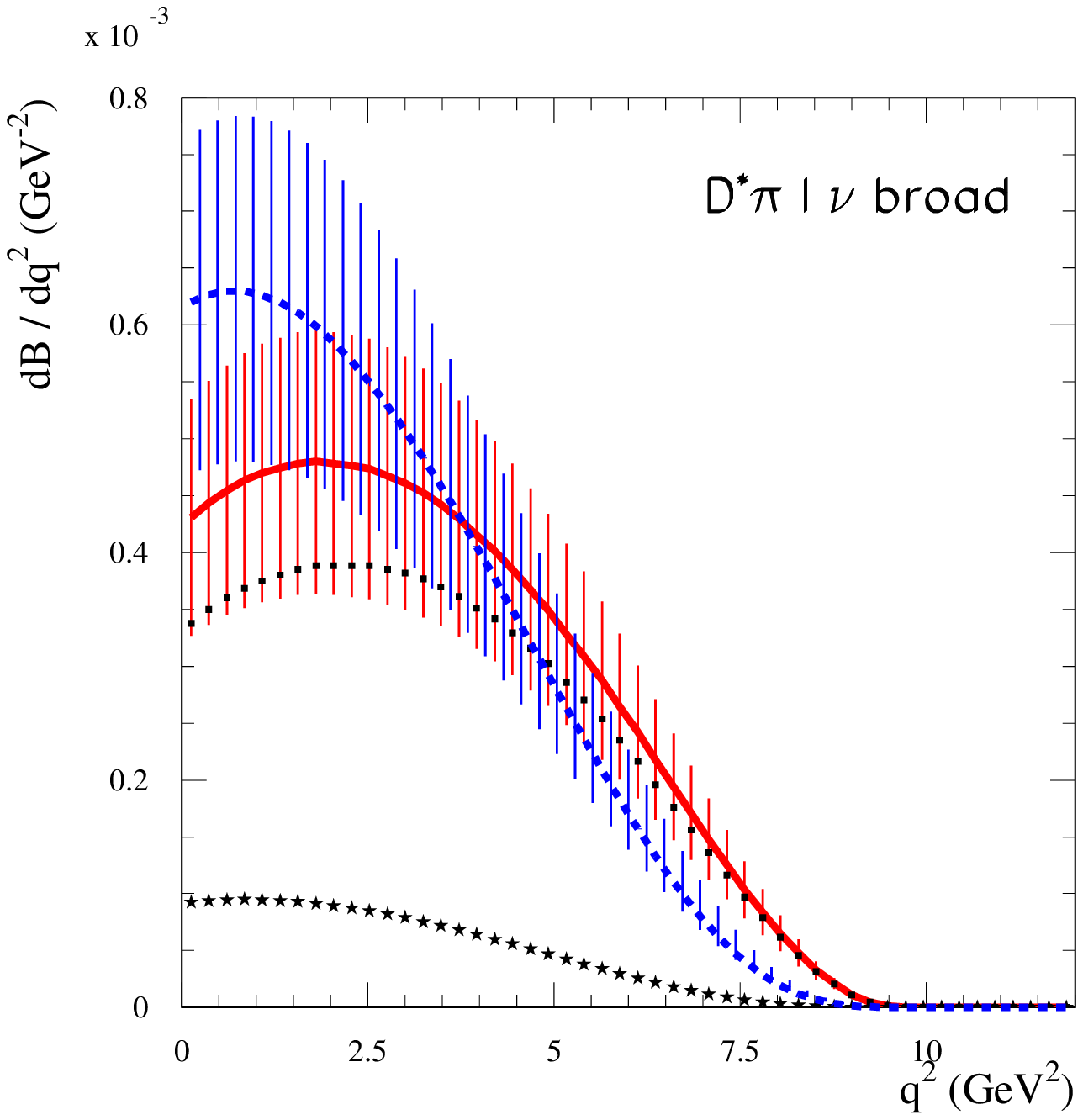,width=0.5\textwidth}
\epsfig{file=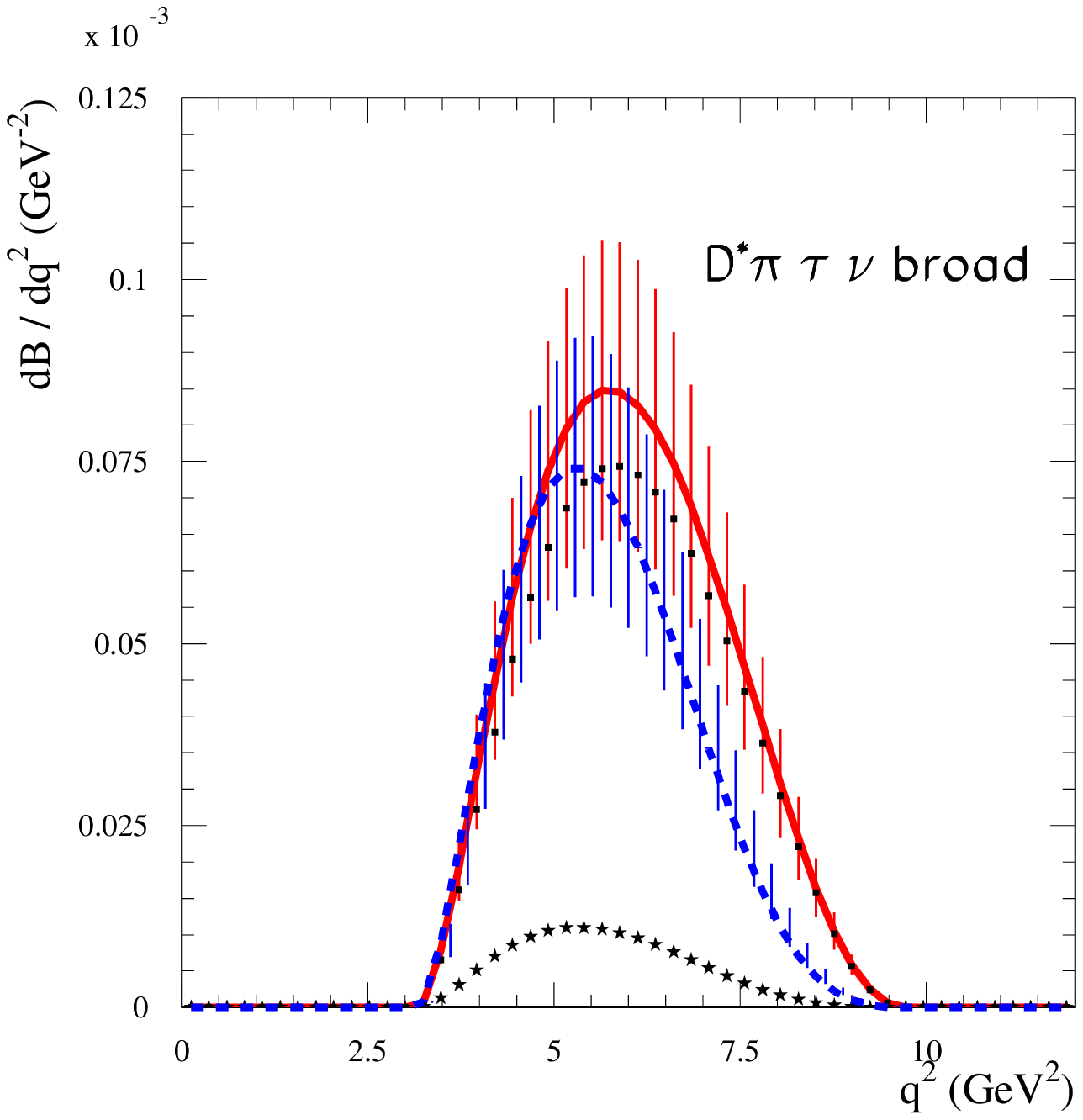,width=0.5\textwidth}}
  \end{center}
  \caption[]{ Expected $q^2$ distributions for $\Bob \to D^* \pi$, broad mass component, in semi-leptonic decays. The same conventions as in Figure \ref{fig:dstarpisl_all_m} are used.}
\label{fig:dstarpisl_q2}
}
\end{figure*}

\subsection{Analysis and predictions for the $\Bob \to D \pi \ell^- \bar{\nu}_{\ell}$ final state}

\label{sec:btodpi_el}
All quoted numbers are relative to the sum of $D^{0}\pi^+$ and  $D^{+}\pi^0$ final states. Input measurements are given in Section \ref{sec:dv_dostar}.

The measured branching fraction into broad components amounts to:
\begin{equation}
{\cal B}(\Bob \to \left. D \pi\right |_{broad} \ell^- \bar{\nu}_{\ell})=(4.24 \pm 0.56)\times 10^{-3}.
\label{eq:dpi_broad}
\end{equation}
In the hypothesis that contributions from  higher mass resonances can be neglected,
the expected contribution from $D_{1/2}$ decays is equal to $(0.51\pm 0.24)\times 10^{-3}$. To account for the observed broad mass $D\pi$ distribution, the $\Dostarp$ contribution has to be complemented by a $D^{*}_V$ component equal to: $(3.7 \pm 0.6)\times 10^{-3}$. For $r_{BW}$ values varying between $1$ and $3\,GeV^{-1}$ our estimates are in the range $[2.6,\,1.5]\times 10^{-3}$. Therefore, taking into account present uncertainties, this scenario is compatible (marginally) with present measurements.

In Figure \ref{fig:dpisl_all_m}, we compare expected $D\pi$ mass distributions from our model (full line) and from the $\sdce$ analysis (dashed line). There is a narrow peak from the $\Ddestarp$ meson located on top of a broad mass distribution which is very different in the two scenarios. As in the $D^* \pi$ channel analysis, broad mass distributions have been scaled so that total decay rates agree with measurements in the case of light leptons.

\begin{figure*}[!htb]
  \begin{center}
    \mbox{\epsfig{file=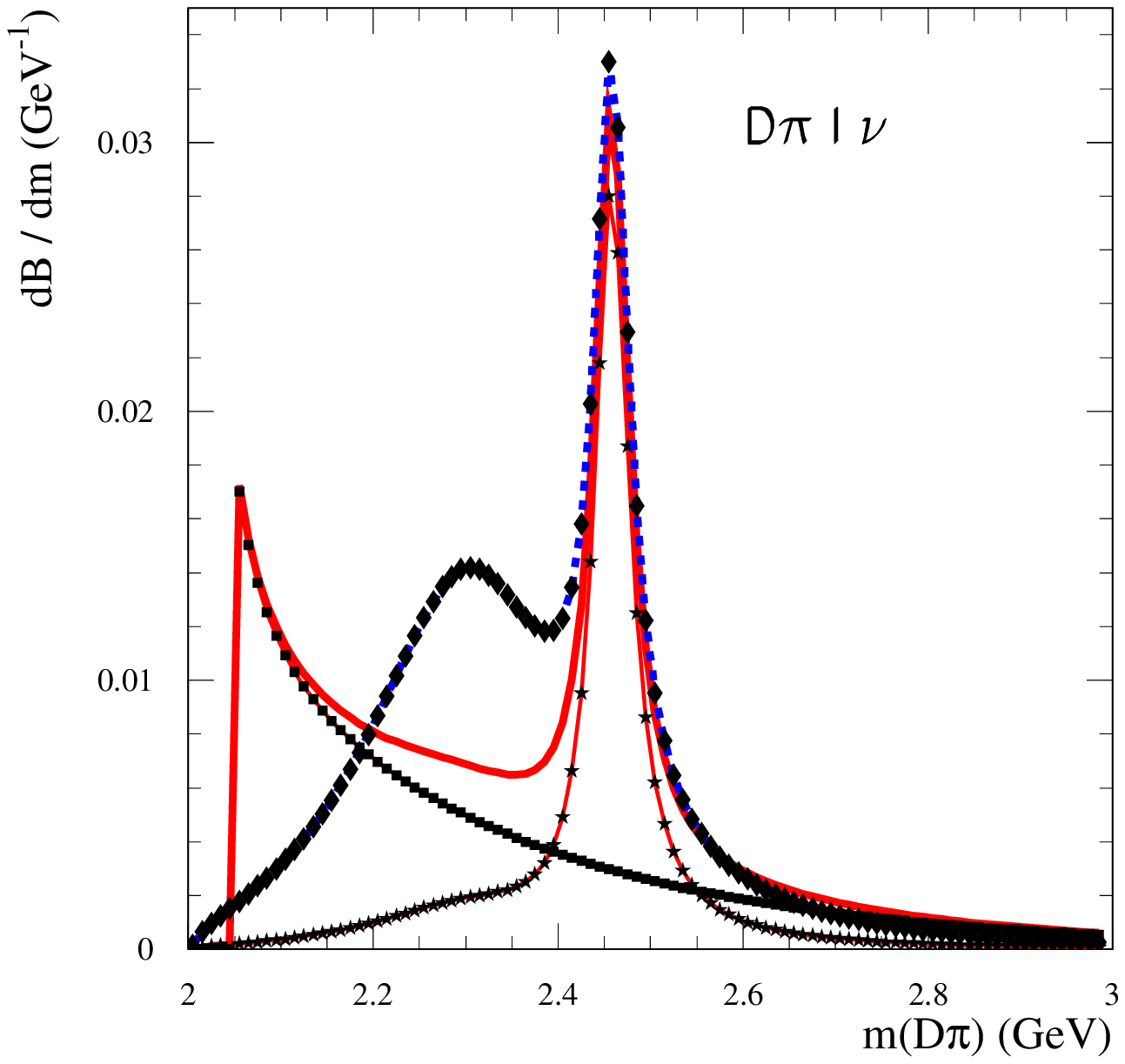,width=0.5\textwidth}
\epsfig{file=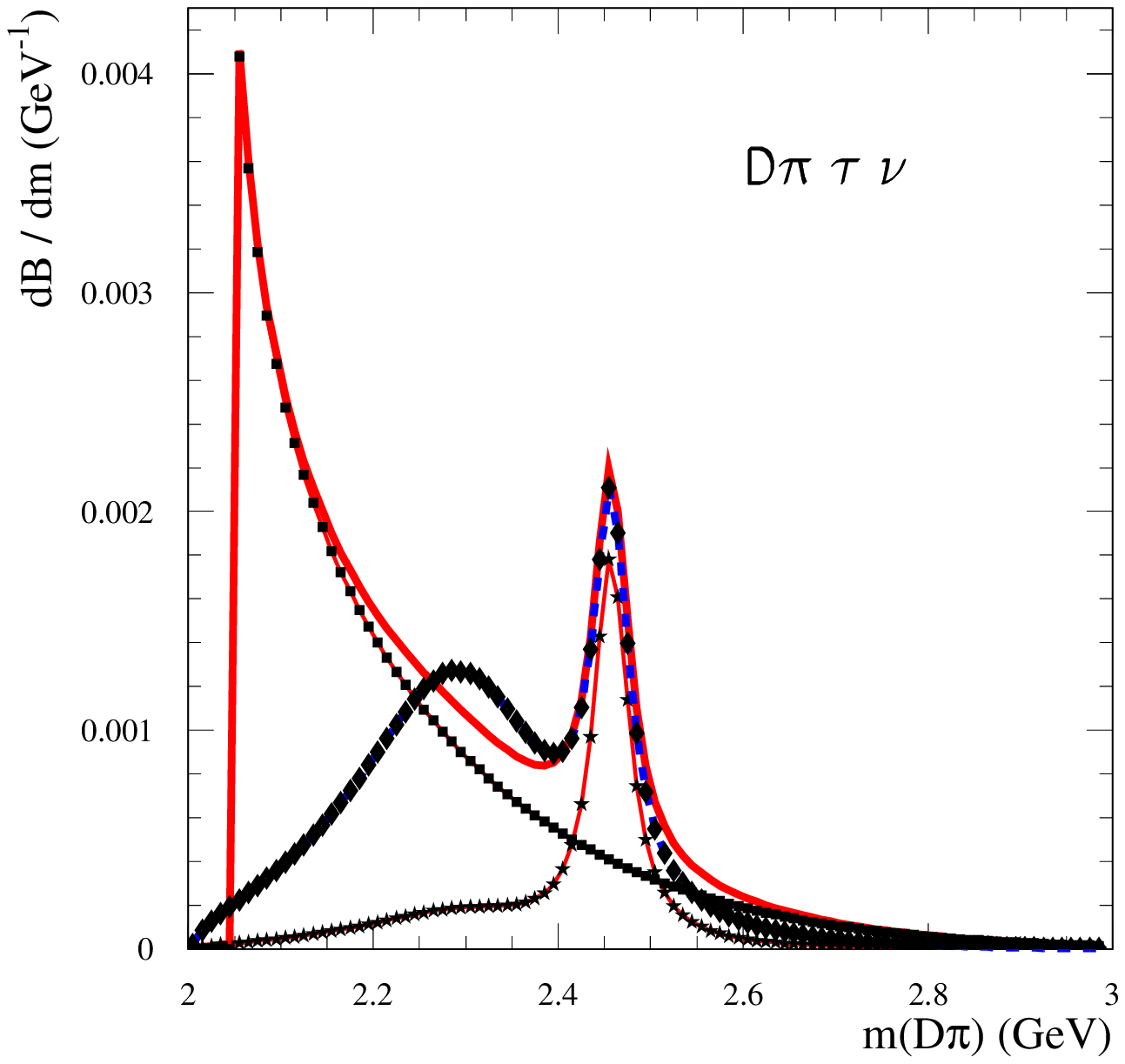,width=0.5\textwidth}}
  \end{center}
  \caption[]{ Expected $D\pi$ mass distributions for $\Bob \to D \pi$ in semi-leptonic decays. Only central values are displayed.
}
\label{fig:dpisl_all_m}
\end{figure*}

\subsubsection{Expected differences between our model and $\sdce$, for broad mass components}

In the following we  compare expected mass and $q^2$ distributions, for the $D\pi$ broad mass component, in our model and $\sdce$, and after having normalized expectations to agree with the measured central value, for light leptons.

\begin{figure*}[!htb]
  \begin{center}
    \mbox{\epsfig{file=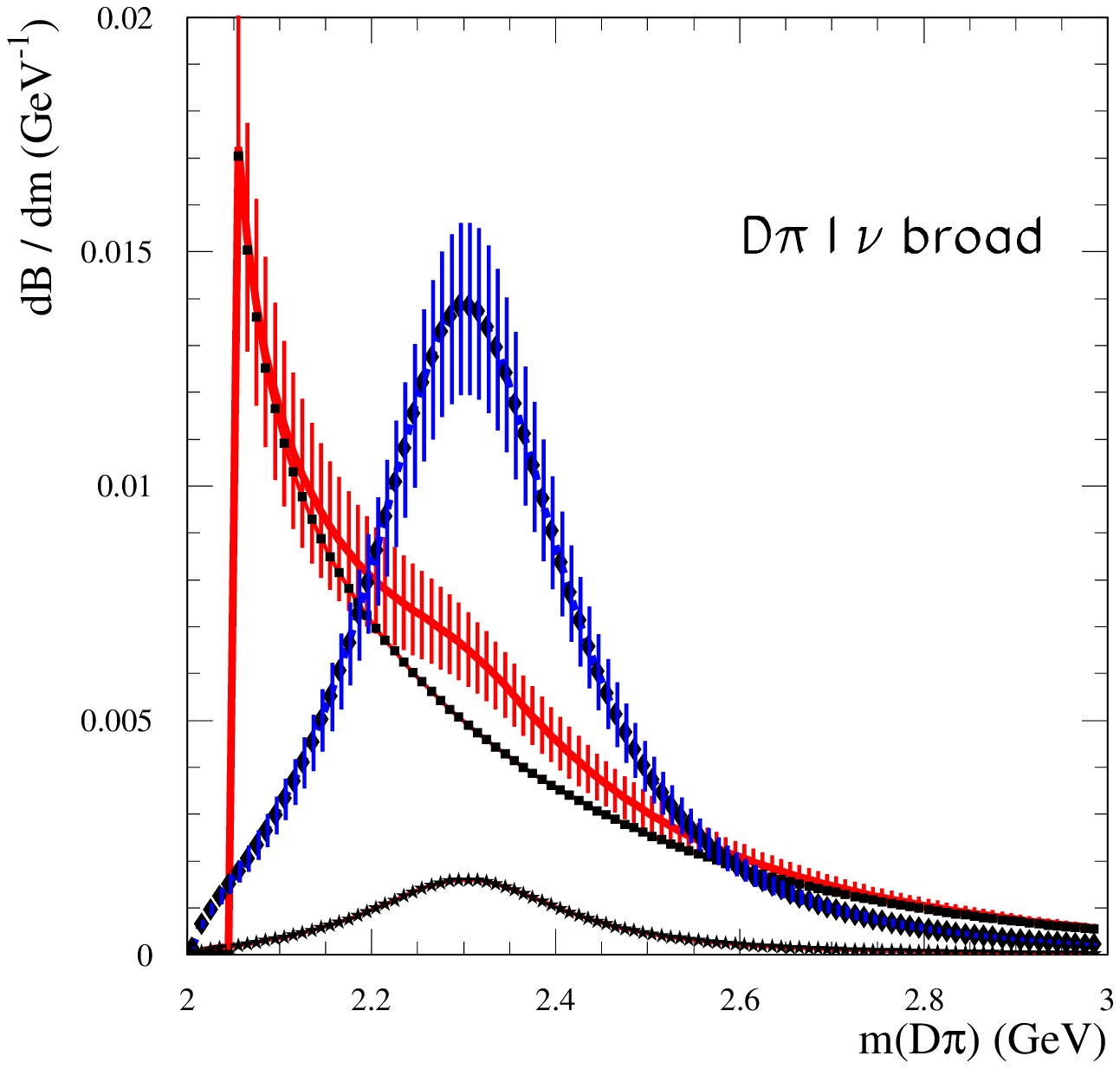,width=0.5\textwidth}
\epsfig{file=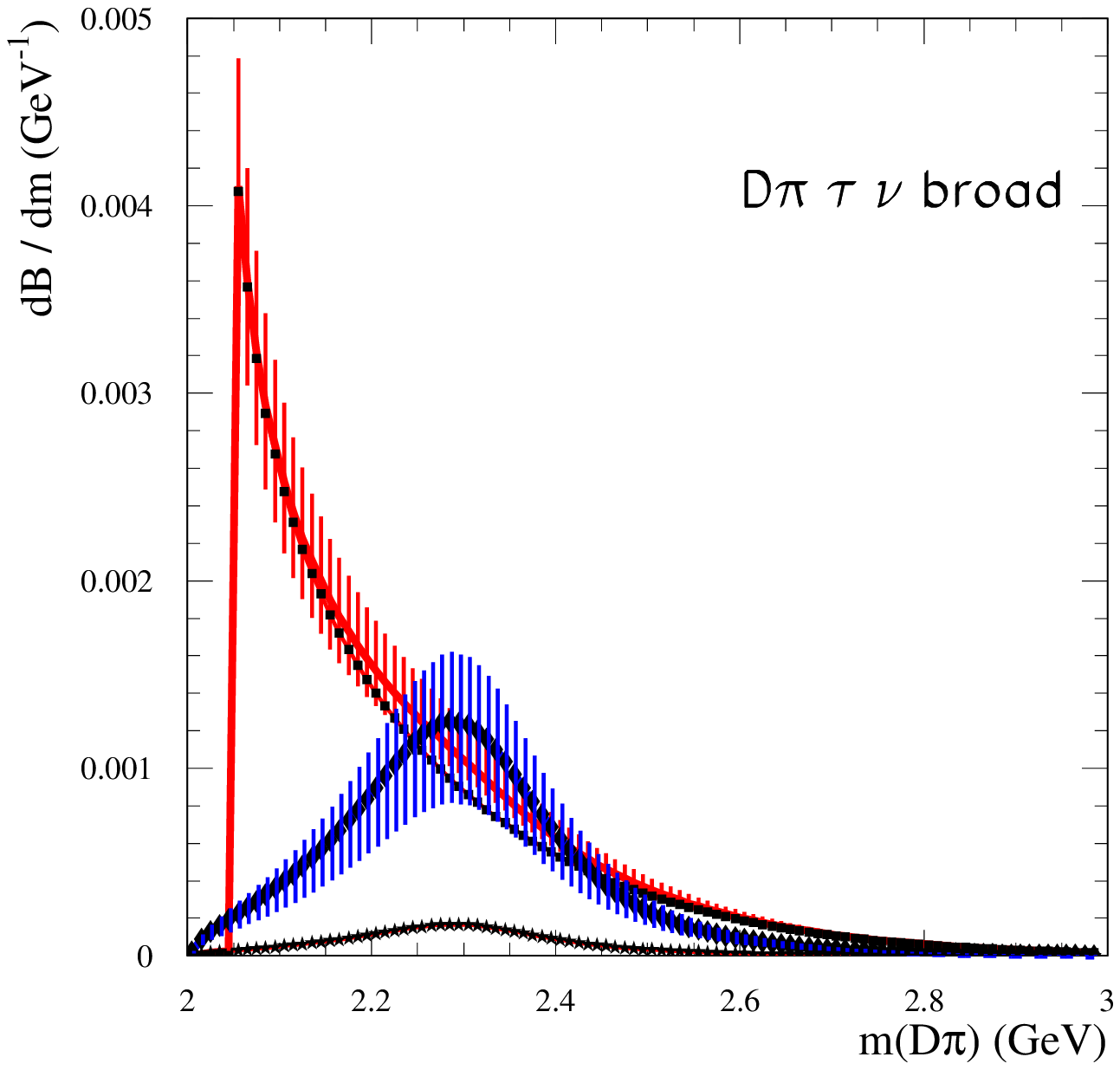,width=0.5\textwidth}}
  \end{center}
  \caption[]{ Expected broad $D\pi$ mass distributions for $\Bob \to D \pi$ in semi-leptonic decays. 
}
\label{fig:dpisl_m}
\end{figure*}

Our model and $\sdce$ show marked differences. In particular, for $\tau$ events one expects two times more candidates in our model and very different mass and $q^2$ distributions. Dashed areas correspond to measured uncertainties of the broad mass component, used for the normalization, as given in Eq. (\ref{eq:dpi_broad}) and do not include model uncertainties which can affect also the shape of the distributions.

\begin{figure*}[!htb]
  \begin{center}
    \mbox{\epsfig{file=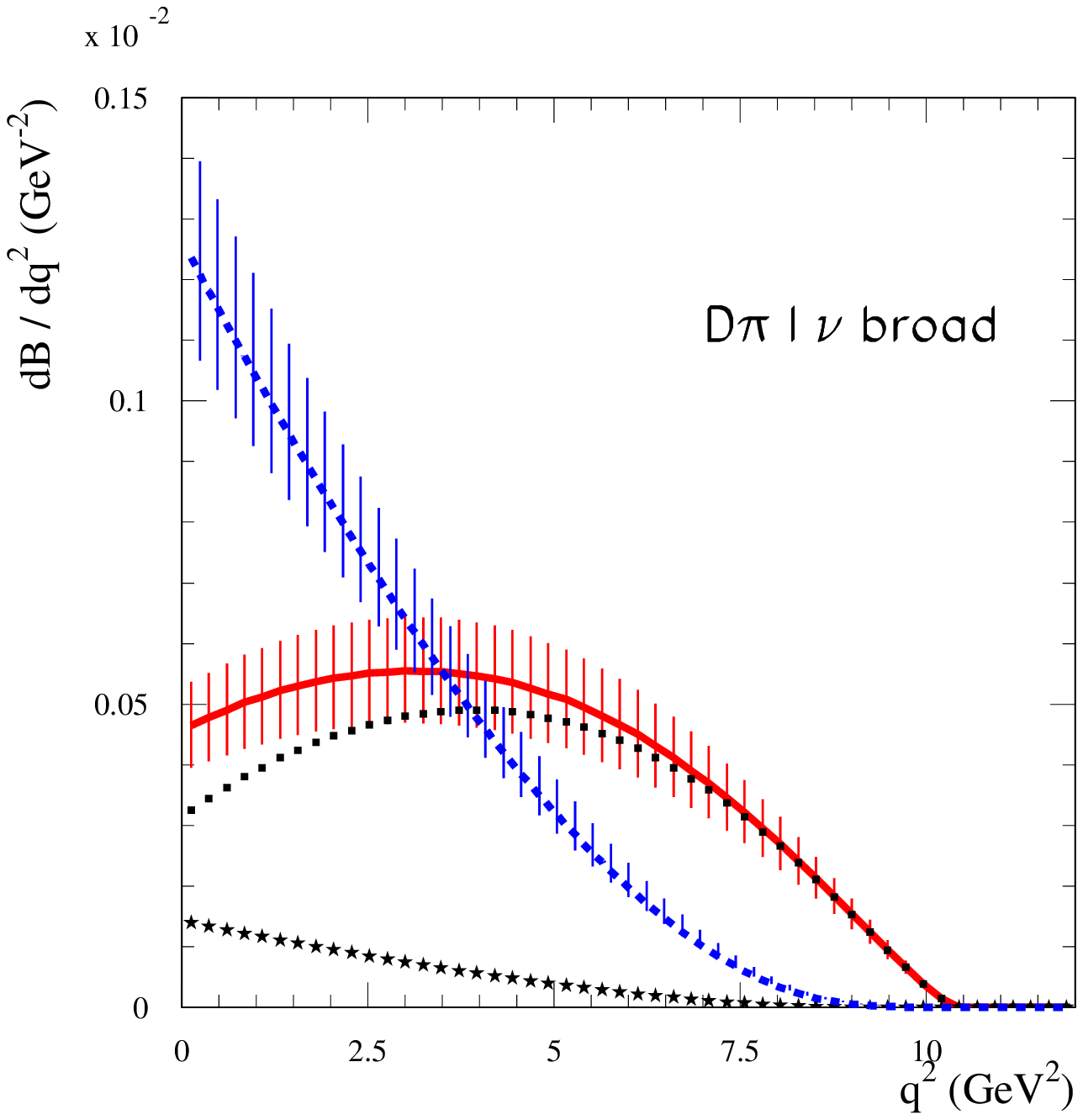,width=0.5\textwidth}
\epsfig{file=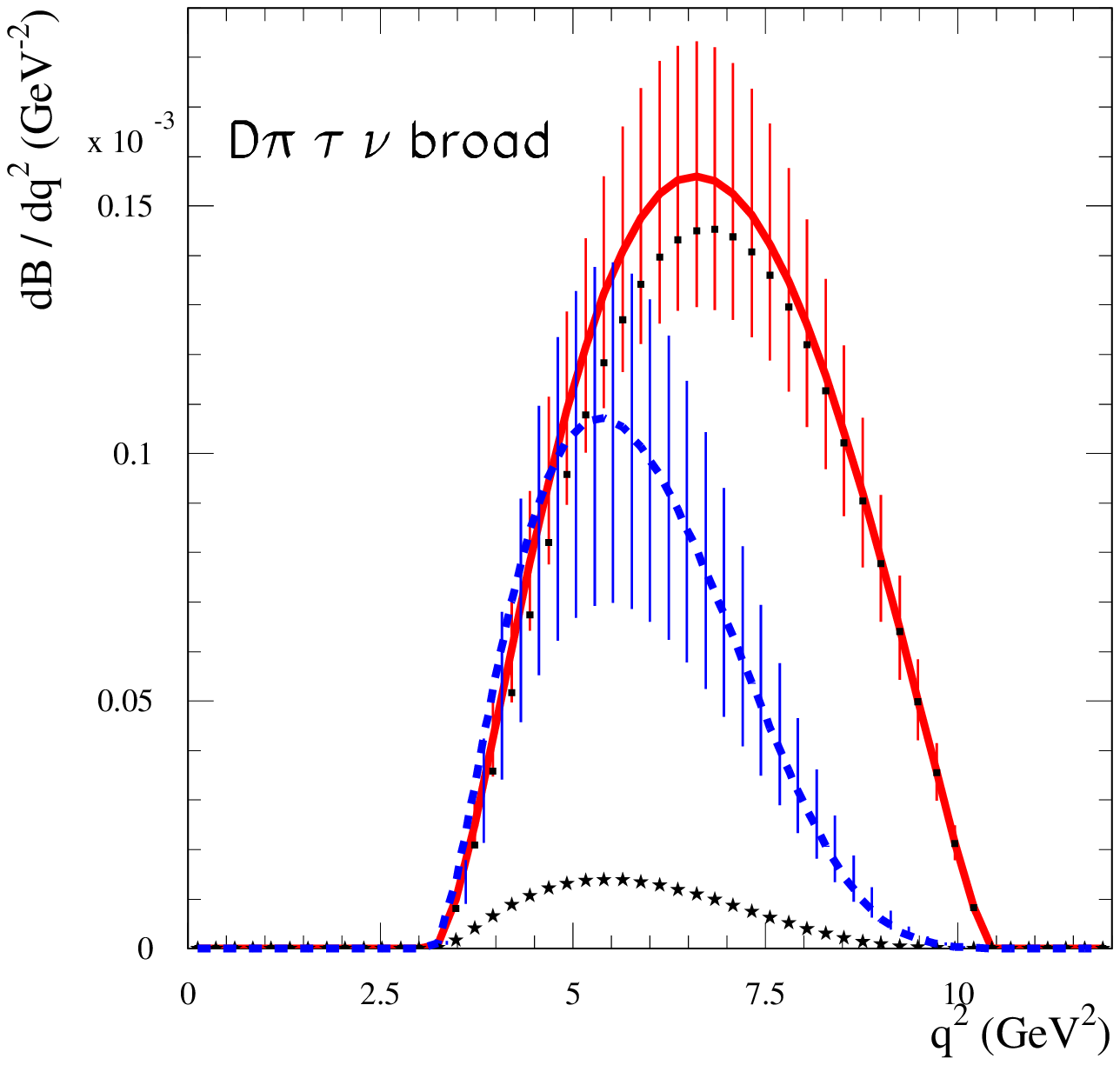,width=0.5\textwidth}}
  \end{center}
  \caption[]{ Expected $q^2$ distributions for $\Bob \to D \pi$, broad mass component, in semi-leptonic decays.
}
\label{fig:dpisl_q2}
\end{figure*}

\section{Predictions for $\Bob \to D^{**,\,+}\Dsm$ decays}

After having described our predictions for decay branching fractions of $\Bob$ mesons into the four $D^{**,\,+}$ accompanied by a $\Dsm$,  we give expectations for the hadronic $D^*\pi$ and $D\pi$ final states.

Because the $D_s$ meson and the $\tau$ lepton have a similar mass, we have also evaluated \cite{ref:GW}  the ratios:
\begin{equation}
{\cal R}^{\tau,\,D_s}_{D^{**}_i}=\frac{{\cal B}(\Bob \to D_{i}^{**,\,+} \tau^- \bar{\nu}_{\tau})}{{\cal B}(\Bob \to D_{i}^{**,\,+} \Dsm)}
 \end{equation}
expecting that, because of correlations between the different sources of uncertainties, they are more accurate than individual measurements of the corresponding decay rates,  as observed already for semi-leptonic channels (see Table \ref{tab:brsl_dsstar_el}).

Results can be used also for corresponding charged $B$-meson decays,
$B^- \to D^{**,\,0}\Dsm$, because they are of Class I (once penguin terms are neglected). Branching fractions have simply to be corrected by the ratio between charged and neutral $B$-meson lifetimes and ratios ${\cal R}^{\tau,\,D_s}_{D^{**}_i}$ are the same.

The validity of  factorization in $B$ meson decays, with emission of a heavy meson has been questionned, see the comment in Sec. \ref{genrel} supra, and the  previous experimental tests had large uncertainties. 
Using ${\cal B}(\bar{B} \to D^{(*)}\Dsm)$ and ${\cal B}(\bar{B} \to D^{(*)}\ell^- \bar{\nu}_{\ell})$ measurements, we have  checked that this property is verified, once the evaluated contribution from penguin terms is taken into account. Therefore we use the same constraint, $a_{1, \, eff.}= 0.93\pm 0.07$ and, in the absence of theoretical evaluations for penguin amplitudes in $\bar{B} \to D^{**}\Dsm$ decays,  we assume that they can be neglected.

 \vspace{5mm}
 
\subsection{Predicted values for ${\cal B}(\Bob \to D_{i}^{**,\,+} \Dsm)$ } 

 Predicted values for $\Bob \to D_i^{**,\,+}\Dsm$ decay branching fractions, in our model, are given in Table \ref{tab:brds_dsstar_el}.

\begin{widetext}\begin{center}\begin{table}[!htb]
{
  \begin{tabular}{|c|c|c|c|c|c|}
    \hline
 channel & value $\pm$ fit & model & HQET & IW linear& Blatt-W. \\
\hline
$ {\cal B}(\Bob \to \Ddestarp \Dsm)\times 10^4$ & $5.8 \pm 0.8$ & $0.6$ & $0.1$ & $1.7$ & $0.1$\\
${\cal R}_{\Ddestar}^{\tau,\,D_s}$& $ 0.33\pm 0.06$ & $0.03$ & $0.$ & $-0.07$ & $0.02$\\
\hline
$ {\cal B}(\Bob \to \Dunp \Dsm)\times 10^4$ & $13.1 \pm 3.5$ & $0.9$ & $-0.1$ & $6.4$ & $0.9$\\
${\cal R}_{\Dun}^{\tau,\,D_s}$& $0.48 \pm 0.12$ & $0.03$ & $0.03$ & $-0.15$ & $-0.03$\\
\hline\hline
$ {\cal B}(\Bob \to \Dostarp \Dsm)\times 10^4$ & $2.3 \pm 0.4$ & $0.2$ & $0.1$ & $0.2$ & $0.0$\\
${\cal R}_{\Dostar}^{\tau,\,D_s}$& $0.22 \pm 0.04$ & $0.05$ & $-0.01$ & $0.08$ & $-0.03$\\
\hline
$ {\cal B}(\Bob \to \Dunstarp \Dsm)\times 10^4$ & $1.4 \pm 1.1$ & $0.9$ & $0.1$ & $-0.1$ & $0.$\\
${\cal R}_{\Dunstar}^{\tau,\,D_s}$& $0.25 \pm 0.05$ & $^{+0.45}_{-0.10}$ & $0.01$ & $0.04$ & $-0.03$\\
\hline
  \end{tabular}}
  \caption[]{\it 
   Our expectations for  ${\cal B}(\Bob \to D^{**,\,+}_i \Dsm)$ branching fractions, and their ratio, to corresponding semi-leptonic decays with a $\tau$ lepton. The first quoted uncertainty corresponds to the error from the fit. The model uncertainty is evaluated by changing the values of parameters, that are fixed to zero, by $\pm0.5\, GeV$. Then, as explained in the text, are indicated variations of fitted parameters which correspond to different hypotheses that enter in the parameterization of fitted expressions. 
  \label{tab:brds_dsstar_el}}
\end{table}\end{center}\end{widetext}

{

Values obtained in the $\sdce$ analysis are given in Table \ref{tab:brds_dsstar_el_sc2} in Appendix \ref{app:ds_sc2}. 

Values for $\Dunstarp$ are very uncertain because of the lack of control of two parameters that enter in $1/m_Q$ corrections ($\hat{\chi}_2$ and $\hat{\zeta}_1$ in the present analysis). These
model uncertainties affect also the ratio ${\cal R}^{\tau,\,D_s}_{\Dunstar}$. Measurements of non-leptonic Class I $B$ decays  with $\Dunstar$ emission are therefore desirable to improve the present situation.

To evaluate branching fractions when the hadronic final state, accompanying the $\Dsm$ meson, is $D^* \pi$ or $D \pi$, we have to multiply values given in Table \ref{tab:brds_dsstar_el} for the production of the different $D^{**}_i$ meson by corresponding branching fractions listed in Table \ref{tab:br_abs}. In our model we have also evaluated the 
$ \Bob \to D_V^{(*),\,+} \Dsm,\, D_V^{(*),\,+}\to D^{(*)} \pi $ expected contributions. 

In practice it is not possible to compute really the expected mass distributions because of strong phase-shifts between the different hadronic amplitudes, that are unknown. One can evaluate only absolute values of these amplitudes. Therefore mass distributions, displayed in the following, are obtained by adding incoherently contributing individual channels.

}

\subsection{Analysis of the $\Bob \to D^* \pi  \Dsm$ final state}
All quoted numbers are relative to the sum of $D^{*0}\pi^+$ and  $D^{*+}\pi^0$ final states. 
{ Expected branching fractions for $D^{**}$ channels, multiplied by ${\cal B}(D^{**} \to D^* \pi)$,  are the following:}
{%\color{red}
\small
\begin{eqnarray} 
 {\cal B}(\Bob \to \Ddestarp \Dsm) &=& (2.3 \pm 0.3 \pm 0.2)\times 10^{-4} \nonumber \\
 {\cal B}(\Bob \to \Dunp \Dsm) &=& (8.8 \pm 2.3 \pm 0.6)\times 10^{-4}\nonumber \\
 {\cal B}(\Bob \to \Dunstarp \Dsm) &=& (1.4 \pm 1.1 \pm 0.9)\times 10^{-4}
\end{eqnarray} 
}

{\color{black}
To evaluate the $D^* \pi$ mass distribution, $D^{**}$ channels are complemented by the $D_V^{(*),\,+}$ contribution of  $(3.0 \pm 1.0)\times 10^{-4}$. This gives a total broad $D^*\pi$ component of $(4.4\pm 3.0 )\times 10^{-4}$ which can be compared with the estimate from $\sdce$: $(8.8\pm 7.4 )\times 10^{-4}$.

It can be noted that the $D^{(*)}_V$ contribution can mimic a broad resonance and an analysis of the alignment distribution is necessary to separate the two possibilities.}

\begin{figure*}[!htb]
  \begin{center}
    \mbox{\epsfig{file=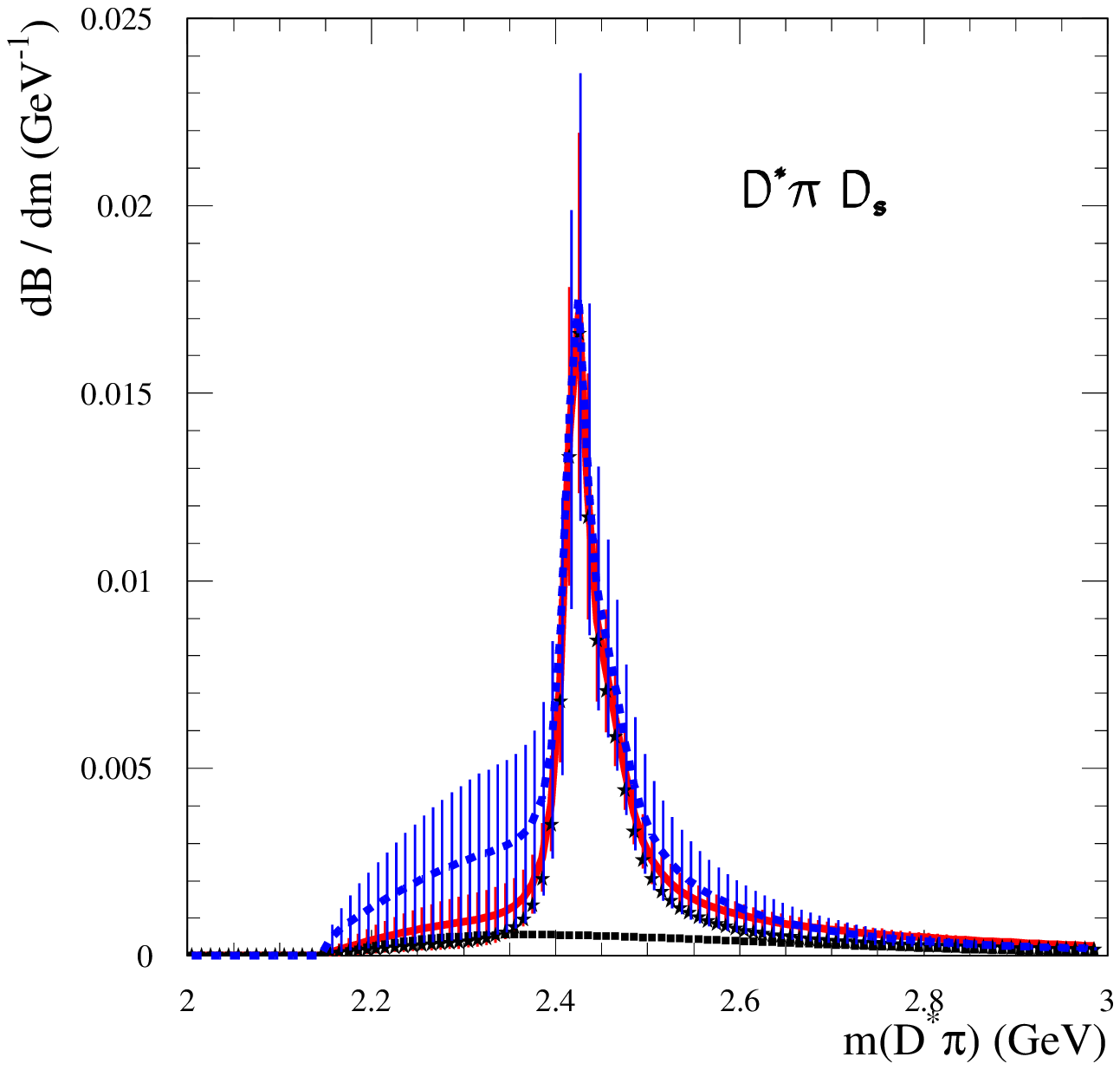,width=0.5\textwidth}
\epsfig{file=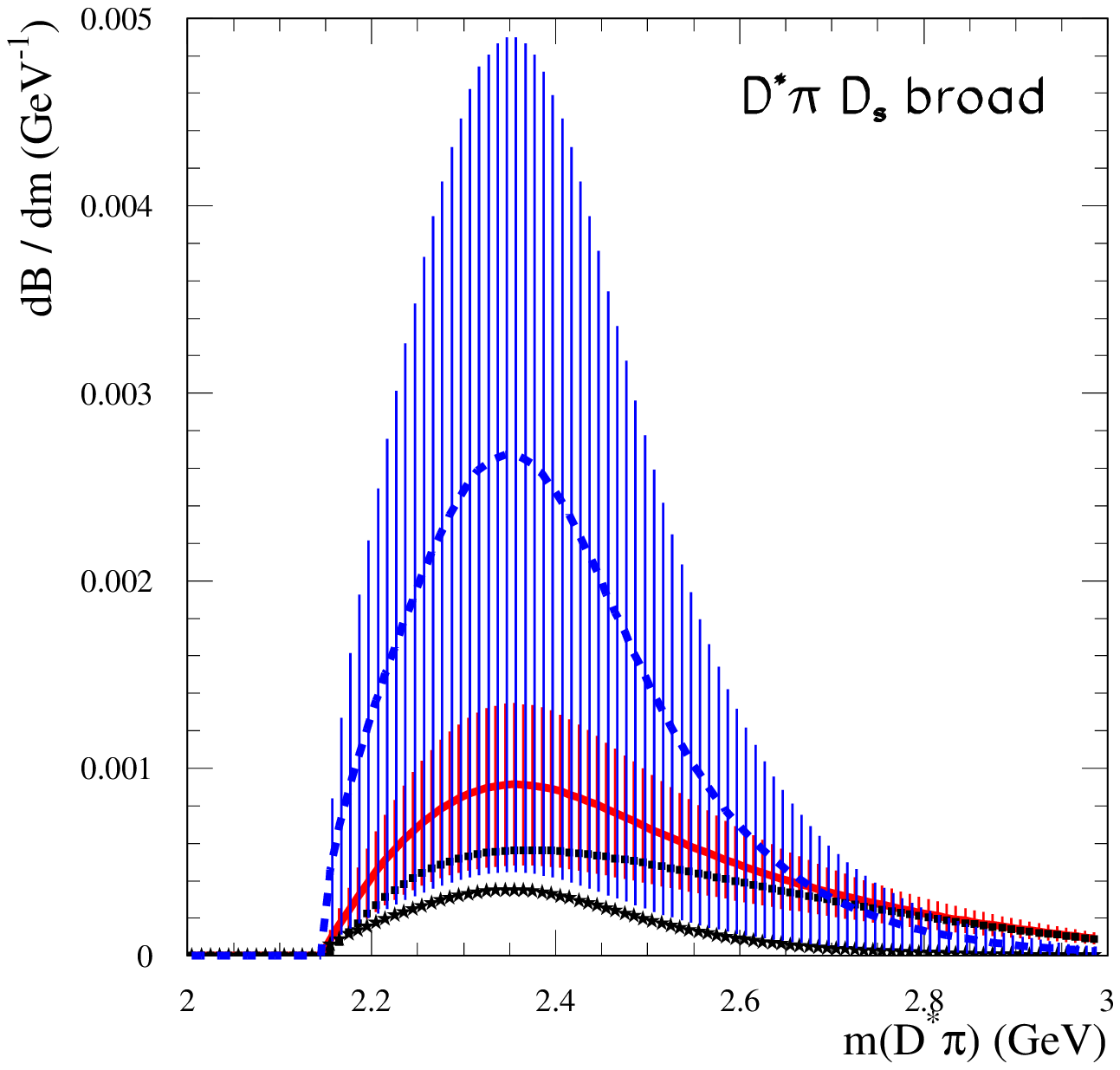,width=0.5\textwidth}}
  \end{center}
  \caption[]{ {\color{black}Expected $D^*\pi$ mass distributions from our analysis (full red curve) and from $\sdce$  (dashed blue curve). Left plots correspond to all expected contributing components whereas the plots on the right are for broad mass components only.} 
}
\label{fig:dstarpids_m}
\end{figure*}

\subsection{Analysis of the $\Bob \to D \pi  \Dsm$ final state}
All quoted numbers are relative to the sum of $D^{0}\pi^+$ and  $D^{+}\pi^0/\gamma$ final states. 
{%\color{red}
Expected branching fractions for channels with a $D^{**}$ meson, multiplied by ${\cal B}(D^{**} \to D \pi)$, are the following:
{\small
\begin{eqnarray} 
 {\cal B}(\Bob \to \Ddestarp \Dsm) &=& (3.48 \pm 0.46 \pm 0.33)\times 10^{-4}\nonumber\\
 {\cal B}(\Bob \to \Dostarp \Dsm) &=& (2.30 \pm 0.43 \pm 0.22)\times 10^{-4}
\end{eqnarray}}

To evaluate the $D \pi$ mass distribution, the $D^{**}$ component is complemented by a $D_V^{*}$ contribution evaluated to be  $(5.5 \pm 0.9)\times 10^{-4}$. This gives a total broad $D\pi$ component of $(7.8\pm 1.5 )\times 10^{-4}$ which can be compared with the estimate from the $\sdce$ analysis: $(17\pm 8 )\times 10^{-4}$.}

\begin{figure*}[!htb]
  \begin{center}
    \mbox{\epsfig{file=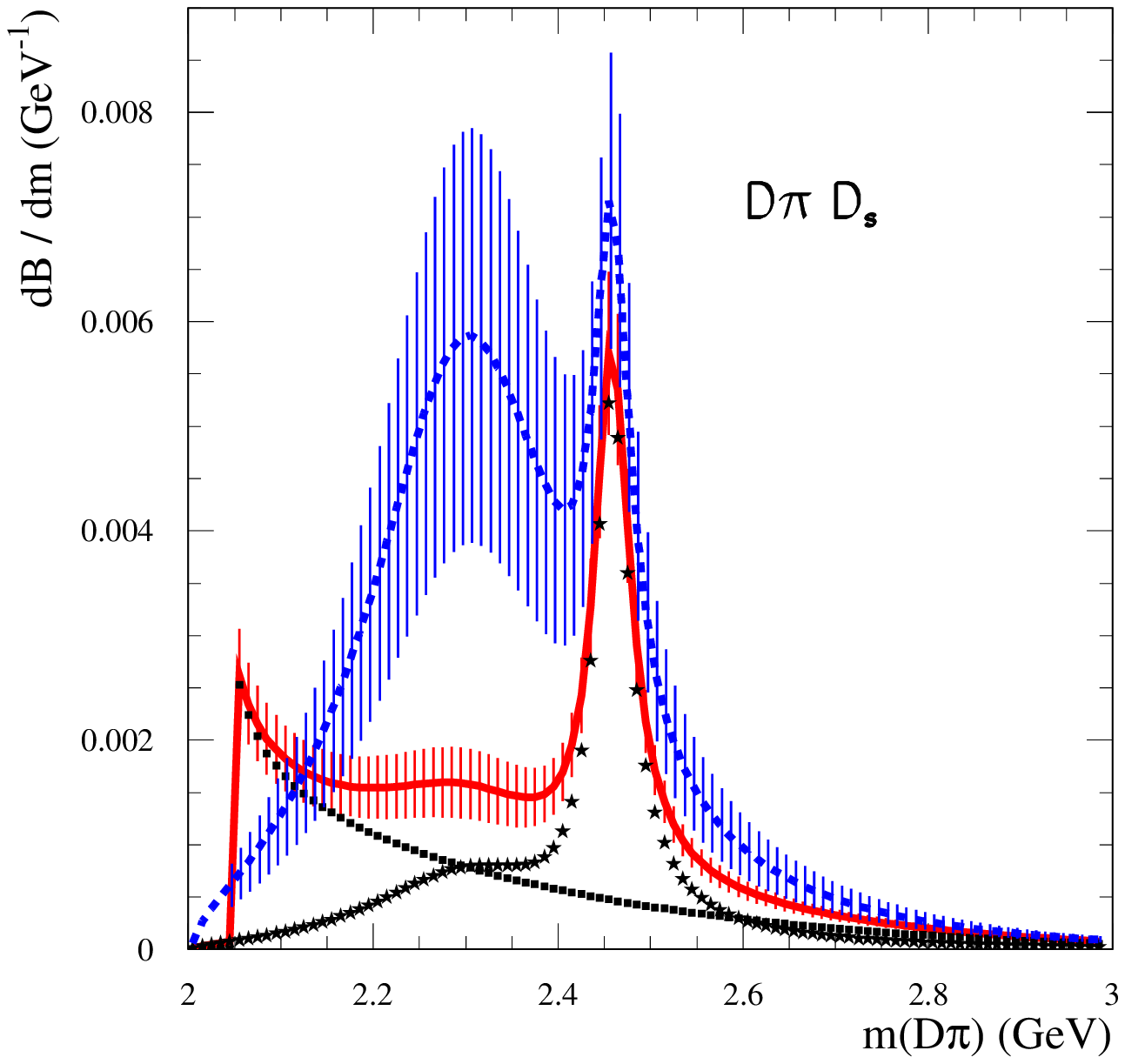,width=0.5\textwidth}
\epsfig{file=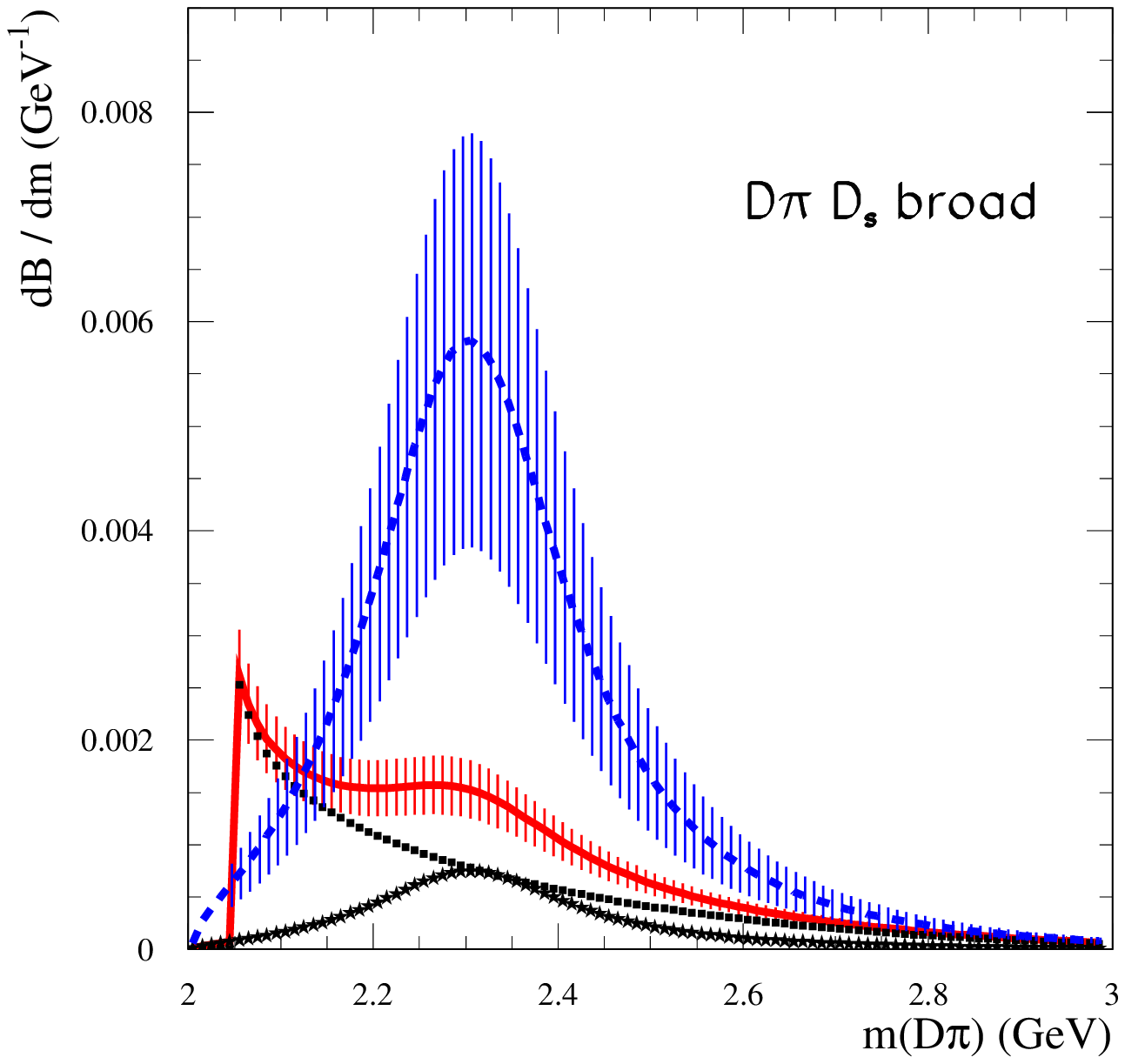,width=0.5\textwidth}}
  \end{center}
  \caption[]{ Expected $D\pi$ mass distributions obtained in our model (red full line) and in $\sdce$ (blue dashed line). Left plots correspond to all expected contributing components whereas, plots in the right are for broad mass components only. 
}
\label{fig:dpids_m}
\end{figure*}

\section{Predictions for $\Bob \to D^{**,\,+}\Dstarsm$ decays}

Expected results on decay branching fractions of $\Bob$ mesons into the four $D^{**,\,+}_i$ accompanied by a $\Dstarsm$ are given in Table \ref{tab:brdstars_dsstar_el}, including different ratios that compare these branching fractions with those expected for $\Bob \to D^{**,\,+}_i\Dsm$ and $\Bob \to D_{i}^{**,\,+} \tau^- \bar{\nu}_{\tau}$ decays. Expressions for ${\cal B}(\Bob \to D_{i}^{**,\,+} \Dstarsm)$ are given in Appendix \ref{app:nl}. They are obtained using  factorization  and considering only Class I decay amplitudes. These expressions are valid for charged or neutral $B$-meson decays.

\begin{equation}
{\cal R}^{\tau,\,D_s^*}_{D^{**}_i}=\frac{{\cal B}(\Bob \to D_{i}^{**,\,+} \tau^- \bar{\nu}_{\tau})}{{\cal B}(\Bob \to D_{i}^{**,\,+} \Dstarsm)}
 \end{equation}
and 
\begin{equation}
{\cal R}^{D_s,\,D_s^*}_{D^{**}_i}=\frac{{\cal B}(\Bob \to D_{i}^{**,\,+} \Dsm)}{{\cal B}(\Bob \to D_{i}^{**,\,+} \Dstarsm)}
 \end{equation}

\begin{table}[!htb]
{%\color{red}
\begin{center}
{
  \begin{tabular}{|c|c|c|}
    \hline
 channel & value $\pm$ fit & model \\
\hline
$ {\cal B}(\Bob \to \Ddestarp \Dstarsm)\times 10^3$ & $1.9 \pm 0.4$ & $0.$ \\
${\cal R}_{\Ddestar}^{\tau,\,D_s^*}$& $ 0.100\pm 0.014$ & $0.$\\
${\cal R}_{\Ddestar}^{D_s,\,D_s^*}$& $ 0.30\pm 0.08$ & $0.03$\\
\hline
$ {\cal B}(\Bob \to \Dunp \Dstarsm)\times 10^3$ & $4.7 \pm 0.9$ & $0.1$ \\
${\cal R}_{\Dun}^{\tau,\,D_s^*}$& $0.134 \pm 0.024$ & $0.$ \\
${\cal R}_{\Dun}^{D_s,\,D_s^*}$& $0.28 \pm 0.10$ & $0.02$ \\
\hline\hline
$ {\cal B}(\Bob \to \Dostarp \Dstarsm)\times 10^4$ & $2.4 \pm 0.7$ & $1.0$ \\
${\cal R}_{\Dostar}^{\tau,\,D_s^*}$& $0.20 \pm 0.04$ & $0.02$\\
${\cal R}_{\Dostar}^{D_s,\,D_s^*}$& $0.96 \pm 0.26$ & $^{+0.44}_{-0.24}$\\
\hline
$ {\cal B}(\Bob \to \Dunstarp \Dstarsm)\times 10^4$ & $2.2 \pm 2.0$ & $^{+0.6}_{-0.1}$\\
${\cal R}_{\Dunstar}^{\tau,\,D_s^*}$& $0.14 \pm 0.02$ & $0.02$ \\
${\cal R}_{\Dunstar}^{D_s,\,D_s^*}$& $0.52 \pm 0.02$ & $^{+0.52}_{-0.36}$\\
\hline
  \end{tabular}}
  \caption[]{\it %\color{orange}
   Our expectations for  ${\cal B}(\Bob \to D^{**,\,+}_i \Dstarsm)$ branching fractions, and their ratio, to corresponding semi-leptonic decays with a $\tau$ lepton and non-leptonic $\Dsm$ production. The first quoted uncertainty corresponds to the error from the fit. The model uncertainty is evaluated by changing the values of parameters, that are fixed to zero, by $\pm0.5\, GeV$.  
  \label{tab:brdstars_dsstar_el}}
\end{center}
}
\end{table}
}

\section{Conclusions}

We have analyzed $\bar{B} \to D^{**}$ decays in semi-leptonic and non-leptonic Class I processes that can be related { using factorization}.

We have verified that {factorization} is well satisfied in $\bar{B} \to D_{3/2}$ decays and also in Class I non-leptonic transitions with $D_s^-$ emission.

Assuming naturally that this factorization is valid also in $\bar{B} \to D_{1/2}$ decays,  one expects a quite small contribution from $D_{1/2}$ relative to $D_{3/2}$ mesons in semi-leptonic decays  as is the case in non-leptonic Class I processes. This  is in contrast with the results of $\sdce$  and in agreement with LQCD computations of the corresponding IW  functions at maximum transfer \cite{ref:lqcd-tau} and with relativistic QM calculations \cite{ref:morenas}.

To evaluate branching fractions for the different decay channels in which a $D^{**}$ meson is produced, we use the expressions  derived in \cite{ref:leibo}. They depend on several parameters that control $1/m_Q$ corrections. Using present experimental measurements and constraints from theory we have determined the most important of these parameters for $D_{3/2}$ emission. In particular we find that the $1/m_Q$
correction $\hat{\epsilon}_{3/2}$ included in the auxiliary $\tau^{eff}_{3/2}$, is of the order of  $-0.2 \pm 0.1$, the minus sign confirming the agreement with a quark model calculation. The three other quantities we obtain are of the order of $\bar{\Lambda}$, as expected in a $1/m_Q$  expansion, which is encouraging in view of the roughness of the method.
For $D_{1/2}$ mesons, estimates of the parameters  are more uncertain but this does not change our conclusion on the smallness of the contribution.

To explain measurements of exclusive $D^{(*)} \pi$ broad hadronic final states in $B$ meson semi-leptonic decays we have evaluated the contribution from $D_V^{(*)}$ components, in addition to $D_{1/2}$ decays. These components can be normalized by using $\bar{B} \to D^{(*)} \ell^- \bar{\nu}_{\ell}$ measurements but the mass dependence of the $D^{(*)}\pi$ mass distribution remains highly undetermined. 

We  propose a model  which accounts for $\bar{B} \to D^{(*)}\pi \ell^- \bar{\nu}_{\ell}$ measurements by adding $D_V^{(*)}$ and $D^{**}$ contributions. We have, at present, not considered contributions from higher mass hadronic states. Results have been compared with the $\sdce$  model  in which $D_{1/2}$ mesons alone explain the broad mass spectra.

These two models give very different expectations for the broad  $D^{(*)}\pi$ mass and $q^2$ distributions regarding light leptons. The $D\pi$ final state, in particular can provide clear informations on the relative importance of the 
$D^*_V$ and $\Dostar$ components. The two models have also very different expectations for semi-leptonic decays with a $\tau$ lepton and in Class I non-leptonic transitions with $D_s^{(*)}$ emission.

\appendix

\section{Expressions for Class I non-leptonic decays}
\label{app:nl}

Expressions for Lorentz invariant form factors are those of $LLSW$ \cite{ref:leibo}, therefore, for the IW functions, one has to use the correspondance $\tau(w)=\sqrt{3}\, \tau_{3/2}(w)$ and $\zeta(w)=2\, \tau_{1/2}(w)$. In following formulas, Lorentz invariant form-factors are evaluated at $w_D=(m_B^2-m_{P(V)}^2+m_{D^{**}}^2)/(2 m_B m_{D^{**}})$ while the quantity $w$, which appears in the rest of the expressions, is computed at the running mass value of the $D^{**}$ resonance.
Expressions for the traditional Blatt-Weisskopf damping factors in  $\overline{B}\to D^{**}X$ decays in which $X$ is a stable particle and which occur in an angular momentum $L=1,\,2,\,\rm{and}\,3$ are the following:
\begin{eqnarray}
F_{B,1}(p^{\prime}) &=& \frac{1}{\sqrt{1+z}}\\
F_{B,2}(p^{\prime})&=& \frac{1}{\sqrt{9+3\,z+z^2}}\nonumber\\
F_{B,3}(p^{\prime})&=&\frac{1}{\sqrt{225 + 45\,z +6\,z^2+z^3}}\nonumber 
\end{eqnarray}
with $z=(r_{BW}\,p^{\prime})^2$, in which $p^{\prime}$ is the decay momentum evaluated in the $\overline{B}$ rest-frame. For the damping term, we use $r_{BW}=3\,(GeV)^{-1}$. 
We define the quantities:
\begin{equation}
B_{B,L}(w)=\left ( \frac{F_{B,L}(p^{\prime})}{F_{B,L}(p_0^{\prime})}\right )^2
\end{equation}
that are the ratios of the previous functions evaluated at the running mass of the $D^{**}$ resonance and at its nominal mass, $m_{D^{**}}$. 
\\

{%\color{red}
\subsection{$\Bob \to D^{**,+} P^-$ decays}

We write the following expressions  for a generic pseudo-scalar meson denoted as "$P$". $a_{1,eff}^{{D_X},P}$ is an effective parameter describing the deviation from strict factorization %\patrick
{ which includes also possible contributions from exchange, annihilation or penguin amplitudes,}
%(where it should be $1$) \red{(check BBNS)},  
$V_{{q_1}{q_2}}$ is the relevant light-quarks $CKM$ matrix element and $f_P$ the annihilation constant. Those parameters are to be adapted to the case under consideration.

The damping factor $B_{B,2}(w)$ is used for $\overline{B} \to D_{3/2}$ decays, that occur in a D-wave and, as usual, no damping is considered for the decays into $D_{1/2}$ states, that are S-wave.  Each expression is followed by its $m_Q=\infty$ limit (between parentheses). This is valid also in the case of a final vector meson.

}
\begin{widetext}

\vspace{4mm}
$\Gamma_{\Bob \to \Ddestarp P^-}=\vert  a_{1,eff}^{{ {D^*_2}},P}\vert^2 \vert V_{cb}V_{{q_1}{q_2}}^* \vert^2\frac {G_F^2}{24\pi}  f_{P}^2 m_B m_{D^*_2}^2   \vert k_{A_1}+k_{A_2} (1-r w)+k_{A_3}(w-r) \vert^2 (w^2-1)^{5/2}%\red
{B_{B,2}(w)}$\\ \strut \hspace {1cm}($\vert  a_{1,eff}^{{ {D^*_2}},P}\vert^2 \vert V_{cb}V_{{q_1}{q_2}}^* \vert^2\frac {G_F^2}{24 \pi}   f_P^2 m_B m_{D^*_2}^2    \tau(w)^2 (1-r)^2 ( 1+w)^2  (w^2-1)^{3/2}$  in the $m_Q \to \infty $ limit)

\vspace{4mm}

$ \Gamma_{\Bob \to \Dunp P^-}=\vert  a_{1,eff}^{{ {D_1}},P}\vert^2 \vert V_{cb}V_{{q_1}{q_2}}^* \vert^2\frac {G_F^2}{16 \pi}   f_P^2 m_B m_{D_1}^2 \vert  f_{V_1}  +     f_{V_2}    (1  - r w)   +     f_{V_3}  ( w- r)\vert^2 (w^2-1)^{3/2}%\red
{B_{B,2}(w)}$
\\ \strut \hspace {1cm}($\vert  a_{1,eff}^{{ {D_1}},P}\vert^2 \vert V_{cb}V_{{q_1}{q_2}}^* \vert^2 \frac {G_F^2}{24 \pi}   f_P^2 m_B m_{D_1}^2    \tau^2 (1  - r )^2 (1+w)^2  (w^2-1)^{3/2} $)

\vspace{4mm}

$ \Gamma_{\Bob \to \Dostarp P^-}= \vert  a_{1,eff}^{{ {D^*_0}},P}\vert^2 \vert V_{cb}V_{{q_1}{q_2}}^* \vert^2\frac {G_F^2}{16 \pi}   f_P^2 m_B m_{D^*_0}^2 \vert   g_+ (1  - r)(1+ w)   +     g_- (1+r)(1- w)\vert^2 (w^2-1)^{1/2} $
\\ \strut   \hspace {1cm}($\vert  a_{1,eff}^{{ {D^*_0}},P}\vert^2 \vert V_{cb}V_{{q_1}{q_2}}^* \vert^2\frac {G_F^2}{16 \pi}   f_P^2 m_B m_{D^*_0}^2      \zeta^2 (1+r)^2( 1- w)^2 (w^2-1)^{1/2}$)
\vspace{2mm}

$ \Gamma_{\Bob \to  D_1^{\ast +} P^-}= \vert  a_{1,eff}^{{ {D^*_1}},P}\vert^2 \vert V_{cb}V_{{q_1}{q_2}}^* \vert^2\frac {G_F^2}{16 \pi}  f_P^2 m_B m_{D^*_1}^2 (w^2-1)^{3/2} \vert g_{V1} + g_{V2 } (1 - r w) + g_{V3 } (w -r)\vert^2 $
\\  \strut \hspace {1cm}( $ \vert  a_{1,eff}^{{ {D^*_1}},P}\vert^2 \vert V_{cb}V_{{q_1}{q_2}}^* \vert^2\frac {G_F^2}{16 \pi}  f_P^2 m_B m_{D^*_1}^2 (1-r)^2 \zeta^2 (w^2-1)^{3/2})$

\end{widetext}

\vspace{2mm}
\subsection{$\Bob \to D^{**,+} V^-$ decays}
{These expressions are obtained by a direct calculation. Several partial waves are contributing in each decay channel and the appropriate Blatt and Weisskopf damping factor has to be used for each contribution. It can be identified from the power of the momentum dependence.}
\begin{widetext}
\vspace{4mm}
\noindent $ \Gamma_{\Bob \to \Ddestarp V^-}=  \vert  a_{1,eff}^{{ {D_2}},V}\vert^2 \vert V_{cb}V_{{q_1}{q_2}}^* \vert^2\frac{G_F^2}{48 \pi} f_{V}^2 \frac{m_V^2 m_{D^*_2}^2 }{ m_B}  (w^2-1)^{3/2} \left[3\vert k_V\vert^2 (w^2-1){B_{B,2}(w)}        \right.$\\$\left.   +\vert k_{A_1}\vert^2 \left(5 {\,B_{B,1}(w)}+2\frac{m_B^2}{m_{V}^2}(w^2-1){B_{B,2}(w)}\right) 
+  2 \vert k_{A_2}+\frac{1}{r} k_{A_3}\vert^2    (w^2-1)^2 \frac{m_{D^*_2}^2}{m_{V}^2}{B_{B,3}(w)} \right.$\\$\left.+ 4 Re( k_{A_1 }^*(k_{A_2}+\frac{1}{r} k_{A_3})) (w^2-1)(w-r)  \frac{m_B m_{D^*_2} }{m_{V}^2}{\sqrt{B_{B,1}(w)}}{\sqrt{B_{B,3}(w)}}\right]$\\ \strut ($\vert  a_{1,eff}^{{ {D_2}},V}\vert^2 \vert V_{cb}V_{{q_1}{q_2}}^* \vert^2\frac {G_F^2}{48 \pi}  f_{V}^2\frac{ m_V^2 m_{D^*_2}^2}{m_B }  \tau^2 (5(1+w)^2+(w^2-1)(3+4 \frac{m_B m_D}{m_V^2}(1+w))  ) (w^2-1)^{3/2}$)

\vspace{4mm}

\noindent $\Gamma_{\Bob \to \Dunp V^-}=\vert  a_{1,eff}^{{ {D_1}},V}\vert^2 \vert V_{cb}V_{{q_1}{q_2}}^* \vert^2 \frac{G_F^2}{16 \pi}  f_{V}^2 m_V^2 m_{ D_1}^2 / m_B  (w^2-1)^{1/2} \left[\vert f_{V_1}\vert^2(3+\frac{m_B^2}{m_{V}^2} (w^2-1){B_{B,1}(w)})  \right.$\\$\left.+  2\vert f_A\vert^2 (w^2-1){B_{B,1}(w)}+ \vert f_{V_2} + \frac{1}{r} f_{V_3}\vert^2    (w^2-1)^2   \frac{m_{ D_1}^2}{m_{V}^2} {B_{B,2}(w)} \right.$\\$\left.+2 Re( f_{V_1 }^*(f_{V_2}+\frac{1}{r} f_{V_3}))   (w^2-1) (w-r) \frac{m_B m_{ D_1}}{m_V^2}{\sqrt{B_{B,1}(w)}}{\sqrt{B_{B,2}(w)}}\right]$\\ \strut \hspace {1mm}($\vert  a_{1,eff}^{{ {D_1}},V}\vert^2 \vert V_{cb}V_{{q_1}{q_2}}^* \vert^2\frac {G_F^2}{96 \pi}   f_{V}^2  \frac{m_V^2 m_{D_1}^2 }{ m_B}     \tau^2 \left[2(1+w)^2+(w^2-1)(3+\frac{m_B^2}{m_V^2}(2  r (1+w)+3(1+r)^2 ))\right] \\(w^2-1)^{3/2} $)

\vspace{4mm}

\noindent$ \Gamma_{\Bob \to \Dostarp V^-}=\vert  a_{1,eff}^{{ {D_0^*}},V}\vert^2 \vert V_{cb}V_{{q_1}{q_2}}^* \vert^2\frac{G_F^2}{16 \pi}  f_{V}^2 \frac{m_{ D_0^*}^4 }{ m_B }  (w^2-1)^{3/2} \vert g_+ (1+\frac{1}{r} )+ g_-  (1-\frac{1}{r} )\vert^2{B_{B,1}(w)}$
\\ \strut   \hspace {1cm}($\vert  a_{1,eff}^{{ {D_0^*}},V}\vert^2 \vert V_{cb}V_{{q_1}{q_2}}^* \vert^2\frac {G_F^2}{16 \pi}   f_{V}^2 m_{ D_0^*}^4/m_B      \zeta^2 (1-1/r)^2 (w^2-1)^{3/2}$)

\vspace{4mm}

\noindent$ \Gamma_{\Bob \to  D_1^{\ast +} V^-}= \vert  a_{1,eff}^{{ {D_1^*}},V}\vert^2 \vert V_{cb}V_{{q_1}{q_2}}^* \vert^2\frac{G_F^2}{16 \pi}  f_{V}^2 \frac{m_V^2 m_{D_1^*}^2 }{m_B } (w^2-1)^{1/2} \left[  \vert g_{V_1} \vert^2(3+\frac{m_B^2}{m_{V}^2} (w^2-1){B_{B,1}(w)}) +  \right.$\\$\left. 2  \vert g_A \vert^2 (w^2-1){B_{B,1}(w)} +  \vert g_{V_2} + \frac{1}{r} g_{V_3}) \vert^2    (w^2-1)^2   \frac{m_{ D_1^*}^2}{m_V^2} {B_{B,2}(w)} \right.$\\$\left.+ 2Re(g_{V_1 }^*(g_{V_2}+\frac{1}{r} g_{V_3}))   (w^2-1) (w-r) \frac{m_B m_{D_1^*}}{m_V^2}{\sqrt{B_{B,1}(w)}}{\sqrt{B_{B,2}(w)}}\right]$
\\ ($\vert  a_{1,eff}^{{ {D_1^*}},V}\vert^2 \vert V_{cb}V_{{q_1}{q_2}}^* \vert^2 \frac {G_F^2}{16 \pi}  f_{V}^2 m_B m_{ D_1^*}^2  \zeta^2 (3(m_V/m_B)^2(w-1)/(w+1)+2((1 -r)^2+r(1-w)))(w^2-1)^{3/2}   $)

\end{widetext}

 \section{Calculations of {$\eta^{b(c)}_{ke}$} and $\eta^{b}$ with NR treatment of the c.o.m motion}
\label{app:epsneg}
1) according to the standard analysis, the ${\cal O}(1/m_Q)$ corrections to form factors fall into two main categories, due to modifications of : 

\begin{itemize}
\item  the state vectors;
\item the current operators.
\end{itemize}

Here we are concerned with the first category, and the $\eta$'s
 or $\chi$'s  parameters are precisely characterizing the corrections to the form factors due to this  modification of state vectors, when the current operator is kept to the infinite mass limit. \vskip 1cm

More precisely, we want to consider {$\eta^{b(c)}_{ke}$}, which parameterize the effect of the kinetic operator {${\cal O}_{ke}$ on respectively} the initial ground state and on the $j=3/2$ final state, with specification to $w=1$.

2) in terms of the quark model, the modification of vector states at $w=1$ is identified with the one of the rest frame wave functions of the mesons, and the aim is then to evaluate the effect of this modification on form factors, i.e. on matrix elements of currents when the current {\bf operator} is kept to its infinite mass limit. 

As concerns  {$\eta^{b(c)}_{ke}$}, one can give an intuitive interpretation by assuming that they are generated by the addition of the heavy quark kinetic energy in the Schr\"odinger equation :

$$p^2/2 m_q \to p^2/2 m_q+p^2/2 m_Q$$

$m_q, m_Q$ being respectively the light and heavy quark masses. But this amounts simply to replace $m_q$ by the reduced mass $\mu=\frac {m_q m_Q} {m_q+m_Q}$ :

$$p^2/2 m_q \to p^2/2 \mu$$

This fact implies that the wave equation is the same as in the heavy quark limit, except for the change of mass. In the case of power-like potential $r^{\alpha}$, this allows to use the scale invariance to deduce the wave functions from the heavy mass limit ones $\phi^{\infty}$. Let the equation be :

$$ (p^2/2 m_q+b r^{\alpha})\phi^{\infty}(\vec{r})=E_n \phi^{\infty}(\vec{r})$$
Performing the substitution $\vec{r} \to \lambda  \vec{r}$, one has :

$$ (\frac {p^2} {2 m_q \lambda^{(2+\alpha)}}+b r^{\alpha})\phi^{\infty}(\lambda\vec{r})=\frac{E_n}{\lambda^{\alpha}} \phi^{\infty}(\lambda \vec{r})$$
Therefore if one chooses $\lambda=(\mu/m_q)^{(1/(2+\alpha))}=(1+(m_q/m_Q))^{(-1/(2+\alpha))}$, one sees that  $\phi^{\infty}(\lambda \vec{r})$
is the solution of the finite mass equation. The normalised wave function is :

$$\lambda^{3/2} \phi^{\infty}(\lambda \vec{r}).$$

Now, the non relativistic (NR) (in fact the familiar dipolar expression) for $\tau$ is :

                $$\tau(w=1)=m (\phi^{\infty}_f|z|\phi^{\infty}_i)$$

{ Note that j=1/2 and j=3/2 are degenerate at $m_Q=\infty$ in the NR approach (no Wigner rotation) if one disregards any spin-orbit force.
In fact, all the spin dependent forces are relativistic effects in the sense that they are  ${\cal O}(v^2/c^2)$ with respect to internal velocities. Therefore, even at finite $m_Q$, if one considers the fully NR approach,
there are no spin independent forces. This is what is done in the following paragraphs 1),2),3) below where we mean only to evaluate {\bf kinetic energy} corrections to $B$ and $D^{(**)}$.

On the other hand, for $\eta^b$, which combines all the type of $\eta$'s relative to the $B$, one has to consider the spin-dependent forces, since this contains the magnetic contributions generating the $\eta$'s (1,2,3). This is done in paragraph 4), where we use the GI model which contains all the relevant forces. An important contribution is obviously the one of spin-spin force, although it is $1/m_Q$. It lifts the degeneracy with the $B^*$.}
 
We now define :  $$\tau^{b,c}(w=1)=m (\phi^{c}_f|z|\phi^{b}_i),$$ with the substitution by the finite mass wave functions,      
whence the corrections are represented by \footnote{One must note the difference of definition of the Isgur-Wise functions in $LLSW$. This difference disappears in corrections with a hat since they represent the quotient by the Isgur-Wise functions themselves}:
\begin{eqnarray}
\tau^{b,c}(w=1)-\tau(w=1)
\end{eqnarray}
or, assuming the proportionality to the $\tau$ and defining reduced corrections as $\hat{\eta}$,
{
\begin{eqnarray}
\tau^{b,c}(w=1)/\tau(w=1)=1+\frac {\hat{\eta}^b_{ke}}{2 m_b} + \frac {\hat{\eta}^c_{ke}}{2 m_c}+...\nonumber \\
=\lambda^{3/2}_b \lambda^{3/2}_c \frac {\int{d_3\vec{r}~z~ \phi^{\infty}_1(\lambda_c \vec{r})\phi^{\infty}_0(\lambda_b \vec{r})} }{\int{d_3\vec{r}~z~\phi_1^{\infty}( \vec{r})\phi^{\infty}_0( \vec{r})}}
\end{eqnarray} 
}

\noindent where $0,1$ denote the orbital angular momentum $L$. If one passes to radial wave functions by means of an angular integration : 
{\small\begin{eqnarray}
\tau^{b,c}(w=1)/\tau(w=1)=\lambda^{3/2}_b \lambda^{3/2}_c \frac {\int{r^3 dr~ \phi^{\infty}_1(\lambda_c r)\phi^{\infty}_0(\lambda_b r)} }{\int{r^3 dr ~\phi_1^{\infty}( r)\phi^{\infty}_0( r)}}\nonumber
\end{eqnarray}}
If one makes $m_b=m_c=m_Q$, one finds : 
{
\begin{eqnarray}
\tau^{b,c}(w=1)/\tau(w=1)&=1+\frac {\hat{\eta}^b_{ke}}{2 m_Q} + \frac {\hat{\eta}^c_{ke}}{2 m_Q}+...\\ =1/\lambda_Q\ &=1+\frac{m_q/m_Q}{2+\alpha}+...
\end{eqnarray}
$$\hat{\eta}^b_{ke}+\hat{\eta}^c_{ke}=\frac{2 m_q} {2+\alpha}$$
}
{To go further and separate $\hat{\eta}^b_{ke}$ and $\hat{\eta}^c_{ke}$, one needs explicit wave functions, which is possible for $\alpha=2 $ (harmonic oscillator)  or $ -1 $  (Coulomb).
}

Harmonic oscillator ($\alpha=2$): $\phi^{\infty}_0(r) \propto e^{(-r^2/(2 R^2))}$ , $\phi^{\infty}_1( r) \propto r~e^{-r^2/(2 R^2)}$
\begin{eqnarray}
\tau^{\infty,c}(w=1)/\tau(w=1)=\left(\frac {2 \lambda_c}{\lambda_c^2+1}\right)^{5/2}
\end{eqnarray}
{whence $\hat{\eta}^c_{ke}=0$, and $\hat{\eta}^b_{ke}=(2 m_q)/(2+2)=m_q/2$}

Coulomb ($\alpha=-1$) :$\,\phi^{\infty}_0(r) \propto e^{(-r/r_0)}$ , $\phi^{\infty}_1( r) \propto r~e^{(-r/2 r_0)}$
\begin{eqnarray}
\tau^{\infty,c}(w=1)/\tau(w=1)=\left(\frac {3 \lambda^{1/2}_c}{\lambda_c^2+2}\right)^{5}
\end{eqnarray}
{whence $\hat{\eta}^c_{ke}=-10/6\, m_q$, and $\hat{\eta}^b_{ke}=(2 \,m_q)/(2-1)+10/6 \,m_q=11/3\, m_q$.}
\vskip 1cm

3) numerical calculation for a linear+Coulomb potential

Of course, the physical potential is rather of linear+Coulomb type. The result may be expected to lay between the HO and the Coulomb one. But one has no analytical solution. Therefore we perform a numerical calculation, using the particular wave functions of reference \cite{BC}
with a potential close to linear+Coulomb and find, with  $m_Q$ respectively infinite or equal to $5 \, GeV$ masses (light mass : $m_q=0.45  \, GeV$)\footnote{The numerical calculations can  be performed for fictitious heavy quark masses because we need only the coefficient of the dependence.}:
%\small 
{\small
\begin{eqnarray}
\tau^{m_b=5,m_c=\infty}(w=1)/\tau(w=1)=1+&\frac{\hat{\eta}^b_{ke}}{2 m_b}=1.053\,  \\
\tau^{m_b=\infty,m_c=5}(w=1)/\tau(w=1)=1+&\frac{\hat{\eta}^c_{ke}}{2 m_c}=0.9898\,  \\
\tau^{m_b=5,m_c=5}(w=1)/\tau(w=1)=1+&\frac{\hat{\eta}^b_{ke}+\hat{\eta}^c_{ke}}{2 m_Q}=1.045\,
\end{eqnarray}}

\noindent whence approximately: $\hat{\eta}^b_{ke} \simeq 0.5   \, GeV$, $\hat{\eta}^c_{ke} \simeq -0.1 \, GeV$, which is indeed intermediate between the results from HO and Coulomb potentials. In fact $\hat{\eta}^b_{ke}+\hat{\eta}^c_{ke}\simeq 0.5  \, GeV$ corresponds roughly to what is expected from $\alpha=0$ (Eq. above), i.e. $m_q$, and it is indeed well known that such a power potential $\alpha \simeq 0$ or a log one approximate roughly the linear+Coulomb one (e.g. Martin potentials).

{
The conclusion up to now is that $\hat{\eta}^c_{ke}<0$. In addition, $\hat{\eta}^b_{ke}>0$ but what must be estimated is $\hat{\eta}^b$}, a common combination appearing in all the form factors, and which can therefore be interpreted as the effect of the full Lagrangian contribution, i.e., intuitively, one needs to include spin dependent forces, which are not present in the potentials which have been considered up to now.

\vskip 1cm
4) numerical calculation of {$\hat{\eta}^b$} for the GI model.

We consider then the Godfrey and Isgur spectroscopic model with all relevant forces, and moreover a relativistic kinetic energy. To obtain {$\hat{\eta}^b$}, we calculate the variation of $\tau$ with the initial $B$ respectively at infinite and finite mass. We find finally {$\hat{\eta}^b\simeq -0.26  \, GeV$} : this indicates that the effect of spin-spin force is large, dominating the kinetic energy effect.

\vspace{3mm}

\section{ {Repeating the $\sdcep$ analysis of \cite{ref:ligeti1,ref:ligeti2}}
\label{app:bl}}

This comparison is intended to show that, using the same input data and constraints we find, using our own code, the same values for fitted parameters as in
\cite{ref:ligeti1,ref:ligeti2}, with the $\sdcep$ approach.

For this purpose measurements of $B$-meson semi-leptonic decays and $\Bob \to D_{3/2}^+ \pi^-$ non-leptonic decays, reported in the third column of Table \ref{tab:constraint}, are used to constrain parameterizations of hadronic form factors. In addition,
measurements from Belle \cite{ref:belle_dsstarl} which provide, respectively, 4 and 5 values for the production fractions
of $\Ddestar$ and $\Dostar$ mesons, in different bins of the $w$ variable are used. The $w$ dependence of the two IW functions is assumed to be linear.
The validity of  factorization , with $a_1=1$, is assumed, to relate semi and non-leptonic decays
in which $D_{3/2}$ mesons, only, are emitted.

Therefore $\Bob \to D_{1/2}^+ \pi^-$ non-leptonic decays are not included in the analysis. 
$D^{**}$ mesons are assumed to be stable (no mass distribution is considered).

We have modified accordingly our analysis { but some differences remain:}
\begin{itemize}
\item for the fractions measured in different $w$ bins, we have not used one of the measurements in each of the two samples because these quantities are not independent (their sum is equal to one);
\item the parameterization of the different form factors is derived from the original article of \cite{ref:leibo}, without using different approximations;
\item in addition to $\alpha_s$ corrections, we have also included, for $D_{3/2}$ mesons, those at order $1/m_Q \times \alpha_s$, provided in \cite{ref:leibo}; 

\end{itemize}

\subsection{{Numerical aspects }}

Production of $D_{3/2}$ and $D_{1/2}$ mesons are evaluated separately. In addition to the normalization and slope of the IW functions, the same parameters which determine $1/m_Q$ corrections, as in \cite{ref:ligeti2}, are fitted.

Considering $D_{3/2}$ mesons only, values of fitted parameters are compared in Table \ref{tab:ligeti32}.

We obtain very similar results. 
The numbers of degree of freedom differ by one unit, in the two analyses, because we have not used one of the measurements for the $w$ dependence of $\Ddestar$ production in semi-leptonic decays. A similar comparison is done, see Table \ref{tab:ligeti12}, fitting only data relative to $D_{1/2}$ mesons, measured in semi-leptonic decays (in \cite{ref:ligeti2} measurements of non-leptonic transitions are not used). We have also modified our code to use the zero width formulation of our expressions as done in \cite{ref:ligeti2}.\\ \\\\

\begin{widetext}\begin{center}\begin{table}[!htb]
  \begin{tabular}{|c|c|c|c|c|c|}
    \hline
 analysis & $\tau_{3/2}^{eff.}$ & $\sigma_{3/2}^2$&$\hat{\tau}_1 (CeV)$ & $\hat{\tau}_2 (GeV)$& $\chi^2/NDF$\\
\hline
 \cite{ref:ligeti2}  & $0.40 \pm 0.04$ & $1.6 \pm 0.2$ & $-0.5 \pm 0.3\, $& $2.9 \pm 1.4\, $ &$2.4/4 $\\
\hline
 our code & $0.41 \pm 0.06$ & $1.60 \pm 0.25$ & $-0.66 \pm 0.42\,$&$5. \pm 2.\, $ & $1.8/3 $\\
\hline
  \end{tabular}
  \caption[]{\it {Comparison between the values of fitted parameters obtained in \cite{ref:ligeti2} and with our own code, modified to be similar to the previous analysis %\patrick{
  and using the same input measurements.}%.}
  \label{tab:ligeti32}}
%}
\end{table}
\end{center}%\end{widetext}
%\begin{widetext}
\begin{center}\begin{table}[!htb]
\begin{center}
{\small
  \begin{tabular}{|c|c|c|c|c|}
    \hline
 analysis & $\tau_{1/2}^{eff.}$ & $\sigma_{1/2}^2$&$\hat{\zeta}_1\small$ & $\chi^2/NDF$\\
\hline
 \cite{ref:ligeti2}  & $0.35 \pm 0.11$ & $0.2 \pm 1.4$ & $0.6 \pm 0.3$ & $9.1/4 $\\
\hline
 our code ($\Gamma(D_{1/2}) = 0$)& $0.37 \pm 0.11$ & $0.26 \pm 1.23$ & $0.23 \pm 0.31$& $7.0/3 $\\
\hline\hline
 our  code ($\Gamma(D_{1/2}) \neq 0$)& $0.30 \pm 0.18$ & $-1.6 \pm 3.2$ & $0.45 \pm 0.28$& $6.0/3 $\\
\hline
  \end{tabular}}
  \caption[]{\it {Comparison between the values of fitted parameters obtained in \cite{ref:ligeti2} and with our own code, modified to be similar to the previous analysis. In the third line, values are obtained using the physical widths for the broad $D_{1/2}$ resonances.}
  \label{tab:ligeti12}}
\end{center}
\end{table}
\end{center}\end{widetext}

Very similar results are obtained in the two analyses when considering zero width resonances. In the third line, obtained using the physical resonance widths, the central value of the IW function slope comes out negative, which is unexpected and inconclusive because the corresponding uncertainty is large. From all fits, with or without a finite resonance width, it can be concluded that data are not able to measure really this slope.

\subsection{The main difficulty of the $\sdcep$ analysis}

Values obtained in this way, for  $\tau_{1/2}^{eff.}(1)$ and $\tau_{3/2}^{eff.}(1)$ are compatible. This comes simply from the fact that the measurement $ {\cal B}(\Bob \to \Dostarp \pi^-)$ is not included in \cite{ref:ligeti1,ref:ligeti2} analyses.
Meanwhile, expectations from relativistic quark models and LQCD \cite{ref:lqcd-tau}, obtained in the  $m_{b,\,c}\to \infty$ limit are very different with  $\tau_{1/2}(1) \ll  \tau_{3/2}(1)$.

If one assumes factorization ($a_1=1$), the  branching fractions of the NL $\bar B \to D_{1/2}$ Class I decays predicted in $\sdce $ are, at present,  higher than the measured values by at least four standard deviations (when including the two channels)

{This is illustrated in Section \ref{sec:sc2_nl}, Table \ref{tab:sc2dpi}, where we compare expectations from our model and from the $\sdce$ analysis, for non-leptonic Class I $\overline{B} \to D_{1/2}$ decays.}

{  Then we show that} it is possible to fit data in a way which satisfies
factorization for narrow and broad states and which is compatible with present theoretical expectations.

\section{{A summary of how the $\sdce$  analysis   differs from the $\sdcep$ one and from our model}}
\label{app:second}

{
%\red
The $\sdce$ %second 
analysis uses  the same hypotheses as $\sdcep$ does but, to make it directly comparable with our model we use the same input measurements and the same constraints from theory, when possible. 

$\sdce$ differs from our model by the following points:

\begin{itemize}
\item constraints from factorization are ignored in the production of $D_{1/2}$ mesons;
\item possible contributions from $D^{(*)}_V$ decays are ignored;
\item semi-leptonic branching fractions, ${\cal B}(\Bob \to D_{1/2}^+ \ell^-\bar{\nu}_{\ell})$, are taken from the last column of Table \ref{tab:constraint}. It can be noted that the uncertainty taken for ${\cal B}(\Bob \to \Dunstarp \ell^-\bar{\nu}_{\ell})$ is three times larger than the one assumed in \cite{ref:ligeti1,ref:ligeti2} to account for the fact that the corresponding central value is obtained from an average of not compatible experimental results (the factor three is evaluated using the usual PDG recipe to scale uncertainties in this situation);
\item no constraint is used on $\tau^{eff.}_{1/2}$;
\item the measured fractions, in several $w$ bins, attributed by Belle \cite{ref:belle_dsstarl}, to the  $\Bob \to \Dostarp \ell^-\bar{\nu}_{\ell}$ decay distribution are used;
\end{itemize}

Because $\sdce$ violates  factorization  in $\Dostar$ production and theoretical expectations for $\tau_{1/2}(1)$ it cannot be considered as a possible alternative to our model. We simply mean to illustrate
the large expected differences between our model and previous analyses that can be confronted with data, when available. 
}

\subsection{{Expected values for ${\cal B}(\Bob \to D_{i}^{**,\,+} \ell^- \bar{\nu}_{\ell})$}}
\label{app:sl_sc2}

{Expected values for semi-leptonic branching fractions, with a light or the $\tau$ lepton, in $\sdce$, are given in Table \ref{tab:brsl_dsstar_el_sc2}.  As expected, they differ mainly from those quoted in Table \ref{tab:brsl_dsstar_el} on the production of $D_{1/2}$ mesons and are now similar to those obtained for $D_{3/2}$ mesons.  Values obtained in this approach, are essentially identical with the input values given in Table \ref{tab:constraint}. Therefore estimates for the $\Dunstar$ are quite inaccurate.%\end{widetext}

\begin{widetext}
 \begin{center}\begin{table}[htb]%{
{%\color{blue}
{
  \begin{tabular}{|c|c|c|c|}
    \hline
 channel & $e$ or $\mu$ & $\tau$ & ${\cal R}_{D^{**}}$ \\
         & $\times 10^3$ & $\times 10^4$ & $(\%)$ \\
\hline
$ {\cal B}(\Bob \to \Ddestarp \ell^-\bar{\nu}_{\ell})$ & $3.16 \pm 0.30$ & $1.90\pm0.27\pm.07 $ & $6.01 \pm0.49 \pm 0.19$ \\
$ {\cal B}(\Bob \to \Dunp \ell^-\bar{\nu}_{\ell})$ & $6.40 \pm 0.44$ & $6.19\pm0.56\pm0.15 $ & $9.67 \pm0.62 \pm 0.24$\\
\hline
         & $\times 10^4$ & $\times 10^5$ & $(\%)$\\
\hline
$ {\cal B}(\Bob \to \Dostarp \ell^-\bar{\nu}_{\ell})$ & $39.1 \pm 7.0 \pm 0.2$ & $31.9\pm 7.9\pm 2.0 $ & $8.2 \pm 1.5 \pm 0.5$\\
$ {\cal B}(\Bob \to \Dunstarp \ell^-\bar{\nu}_{\ell})$ & $17. \pm 15. \pm 2.$ & $12.8\pm 11.5\pm 5.0 $ & $7.6 \pm 1.0 \pm 2.0$\\
\hline
  \end{tabular}}
  \caption[]{\it {$\sdce$ model: expected semi-leptonic branching fractions with a light or a $\tau$ lepton, and their ratio, for the individual $D^{**}$ mesons. Only model systematic uncertainties are quoted.}
  \label{tab:brsl_dsstar_el_sc2}}
}
\end{table}\end{center}\end{widetext}

%\begin{widetext}
% \begin{center}\begin{table}[!htb]{
{%\color{blue}
%{
%  \begin{tabular}{|c|c|c|c|}
%    \hline
% channel & $e$ or $\mu$ & $\tau$ & ${\cal R}_{D^{**}}$ \\
%         & $\times 10^3$ & $\times 10^4$ & $(\%)$ \\
%\hline
%$ {\cal B}(\Bob \to \Ddestarp \ell^-\bar{\nu}_{\ell})$ & $3.16 \pm 0.30$ & $1.90\pm0.27\pm.07 $ & $6.01 \pm0.49 \pm 0.19$ \\
%$ {\cal B}(\Bob \to \Dunp \ell^-\bar{\nu}_{\ell})$ & $6.40 \pm 0.44$ & $6.19\pm0.56\pm0.15 $ & $9.67 \pm0.62 \pm 0.24$\\
%\hline
%         & $\times 10^4$ & $\times 10^5$ & $(\%)$\\
%\hline
%$ {\cal B}(\Bob \to \Dostarp \ell^-\bar{\nu}_{\ell})$ & $39.1 \pm 7.0 \pm 0.2$ & $31.9\pm 7.9\pm 2.0 $ & $8.2 \pm 1.5 \pm 0.5$\\
%$ {\cal B}(\Bob \to \Dunstarp \ell^-\bar{\nu}_{\ell})$ & $17. \pm 15. \pm 2.$ & $12.8\pm 11.5\pm 5.0 $ & $7.6 \pm 1.0 \pm 2.0$\\
%\hline
%  \end{tabular}}
 % \caption[]{\it {$\sdce$ model: expected semi-leptonic branching fractions with a light or a $\tau$ lepton, and their ratio, for the individual $D^{**}$ mesons. %Only model systematic uncertainties are quoted.}
%  \label{tab:brsl_dsstar_el_sc2}}
%}}
%\end{table}\end{center}\end{widetext}
\subsection{{Expected values for ${\cal B}(\Bob \to D_{i}^{**,\,+} \Dsm)$ }}
\label{app:ds_sc2}
Values for $ \Bob \to D_{3/2} \Dsm$ branching fractions  are essentially identical with those obtained in our analysis. On the contrary, for $D_{1/2}$ mesons, they differ by about an order of~{magnitude, as expected.}
\hspace{1mm}

%\begin{widetext}
 \begin{center}\begin{table}[!htb]
{
{
  \begin{tabular}{|c|c|c|}
    \hline
 channel & value $\pm$ fit & model \\
\hline
$ {\cal B}(\Bob \to \Ddestarp \Dsm)\times 10^4$ & $5.7 \pm 0.7$ & $0.7$ \\
${\cal R}_{\Ddestar}^{\tau,\,D_s}$& $ 0.34\pm 0.06$ & $0.03$ \\
\hline
$ {\cal B}(\Bob \to \Dunp \Dsm)\times 10^4$ & $12.3 \pm 3.2$ & $1.3$ \\
${\cal R}_{\Dun}^{\tau,\,D_s}$& $0.50 \pm 0.12$ & $0.01$\\
\hline\hline
$ {\cal B}(\Bob \to \Dostarp \Dsm)\times 10^4$ & $16.0 \pm 4.1$ & $3.0$ \\
${\cal R}_{\Dostar}^{\tau,\,D_s}$& $0.20 \pm 0.03$ & $0.04$ \\
\hline
$ {\cal B}(\Bob \to \Dunstarp \Dsm)\times 10^4$ & $5.3 \pm 4.5$ & $3.2$\\
${\cal R}_{\Dunstar}^{\tau,\,D_s}$& $0.24 \pm 0.04$ & $^{+0.54}_{-0.12}$ \\
\hline
  \end{tabular}}
  \caption[]{\it %\color{orange}
 $\sdce$ model : ${\cal B}(\Bob \to D^{**,\,+}_i \Dsm)$ branching fractions, and their ratio, to corresponding semi-leptonic decays with a $\tau$ lepton. Only model systematic uncertainties are quoted.
  \label{tab:brds_dsstar_el_sc2}}
}
\end{table}\end{center}%\end{widetext}

\section{Acknowledgements} {We would like to thank colleagues who have provided interest and help on various aspects of this analysis, namely, D. Becirevic, I. Bigi,  B. Blossier, S. Descotes-Genon and S. Fajfer. This study has been triggered by G. Wormser who asked us some advice about the interest of  measuring $\bar{B} \to D^{**} D_s^-$  decays to improve our knowledge on $D^{**}$ production in $\bar B$ decays.}

  \end{document}